\documentclass[aps,preprint]{revtex4}%
\usepackage{amsfonts}
\usepackage{amsmath}
\usepackage{amssymb}
\usepackage{subfigure}
\usepackage{graphicx}%
\begin{document}
\title{Stability of horizon with pressure and volume of d-dimensional charged AdS
black holes with cloud of strings and quintessence}
\author{Rui Yin$^{a,b}$}
\email{2016141221093@stu.scu.edu.cn}
\author{Jing Liang$^{a,b}$}
\email{jingliang@stu.scu.edu.cn}
\author{Benrong Mu$^{a,b}$}
\email{benrongmu@cdutcm.edu.cn}
\affiliation{$^{a}$ Physics Teaching and Research section, College of Medical Technology,
Chengdu University of Traditional Chinese Medicine, Chengdu, 611137, PR China}
\affiliation{$^{b}$Center for Theoretical Physics, College of Physics, Sichuan University,
Chengdu, 610064, PR China}

\begin{abstract}
In this paper, the thermodynamics and the stability of horizon of the charged
AdS black hole surrounded by quintessence and cloud of strings in
d-dimensional spacetime are studied via the scalar field scattering and the
charged particle absorption. The cosmological constant is interpreted as a
thermodynamics variable. During the study, we consider the case where the
energy of the particle(scalar field) is related to the internal energy of the
black hole. Furthermore, we also consider another assumption, which is
proposed in [Phys. Rev. D 100, no.10, 104022 (2019)]. This assumption
considers that the energy of the particle(scalar field) is related to the
internal energy of the black hole. In addition, we compare and discuss the
results obtained under these two assumptions. At the same time, we also
considered the effect of the dimension. The thermodynamics of black holes in
different dimensions has also been studied and compared.

\end{abstract}
\maketitle
\tableofcontents

{}

\bigskip{}



\section{Introduction}

Usually, black holes are defined as one of the compact objects formed by the
concentration of matter in a small space, which exhibits various features of
gravity. A significant feature of a black hole is its event horizon, through
which no particle can escape from its gravity, even if the particles are
photons. Therefore, the event horizon plays a foremost role in preventing the
observer from viewing the inside of a black hole. A particle going through the
outside region horizon cannot be seen, but its physical quantities affect the
black hole through a back-reaction. The recent theoretical developments are in
favor of a scenario that represents black holes energies were divided into two
parts: irreducible mass, reducible energy
\cite{Christodoulou:1970wf,Bardeen:1970zz}. The irreducible mass increases in
an irreversible process, even if a Penrose process extracts energy from the
black hole. As energy, the irreducible mass is considered to be distributed on
the horizon's surface area and is proportional to the square-root of the
horizon surface area. However, the mass of a black hole can decrease, such as
the Penrose process, and the reduced mass is the reducible energy among the
energy of a black hole. This reducible energy includes electric and rotational
energies, and external fields or particles can reduce it. In thermodynamics,
the irreducible property of entropy is similar to that of irreducible mass.
The Bekenstein-Hawking entropy of a black hole is proportional to the square
of the irreducible mass \cite{Bekenstein:1973ur,Bekenstein:1974ax}. According
to these definitions of the temperature and entropy of a black hole, the laws
of thermodynamics are defined. Furthermore, black holes can be regarded as
thermodynamic systems with the Hawking temperature
\cite{Hawking:1974sw,Hawking:1976de}, for the reason that there is an energy
radiated from the black hole that to do with the quantum effects near the
horizon \cite{Hawking:1974sw}.

High precision astronomical observations have shown that the universe is
undergoing a phase of accelerated expansion
\cite{Riess:1998cb,Perlmutter:1998np}. Formation of a singularity with
infinite matter density is inevitable during the gravitational collapse
\cite{Penrose:1964wq}. The existence of a singularity will destroy the
deterministic nature of general relativity. Since a naked singularity without
a horizon causes problems in terms of causality, the weak cosmic censorship
conjecture states that the singularity should be hidden to an observer in the
spacetime of a black hole owing to the horizon. Hence, the horizon should be
stable. There is not a concrete proof of the weak cosmic censorship
conjecture, whose validity should be checked in different spacetimes. Wald
proposed firstly a gedanken experiment to check this conjecture by examining
whether the black hole horizon can be destroyed by absorbing a point particle.
Until now, there are some debates on the test particle mode. When it comes to
the higher order terms in the energy, angular momentum, and charge of the test
particle are taken into account, the weak cosmic censorship conjecture was
found to be violated too even for an extremal Kerr-Newman black hole
\cite{Jacobson:2009kt}. Later, it was claimed that in all of these situations,
the test particle assumption was not perfect since they did not take into
account the self force \cite{Barausse:2010ka,Barausse:2011vx,Colleoni:2015ena}
and back reaction effects \cite{Hubeny:1998ga,Isoyama:2011ea}. As these
effects were considered, the weak cosmic censorship conjecture was found to be
valid for both the extremal and near-extremal black holes. Especially, by
applying the Wald formalism rather than matter, a new version of Gedanken
experiment has been designed recently. Over the years, the validity of the
weak universe censorship conjecture has received extensive attention and a lot
of research work has been carried out under particle absorption
\cite{Shaymatov:2019pmn,Shaymatov:2020tna,Ying:2020bch,Han:2020fsa,Hod:2020lgt,Zhang:2020txy,Duztas:2020xnl,Li:2020dnc,Zeng:2019baw,He:2019kws,Zeng:2019huf,Wang:2020osg,Hu:2020biz,Hu:2020lkg,Liu:2020cji,Isoyama:2012mgk,Gim:2018axz,Chen:2019pdj,Han:2019kjr,Han:2019lfs,Wang:2019dzl,Zeng:2019aao,Mu:2019bim,He:2019fti}%
.  The weak cosmic censorship conjecture was found to be valid for the
non-extremal black holes. In this framework, the second order variation of the
mass of the black hole emerges, which somehow incorporates both the self force
and back reaction effects. Then, this study was also generalised to scalar
field \cite{Semiz:2005gs,Toth:2011ab}. Semiz first proposed a way of
destroying the event horizon of a black hole to test the validity of the weak
cosmic censorship conjecture, which is the scattering of a classical test
field. Others have extended this approach further
\cite{Duztas:2013wua,Semiz:2015pna,Duztas:2014sga}. Recently, Gwak divided the
scattering process into a series of in-finitesimal time interval and
considered an infinitesimal process only, the result shows that Kerr-(anti) de
Sitter black holes cannot be overspun by a test scalar field
\cite{Gwak:2018akg}. It is important that the time interval for particles
crossing the event horizon for the weak cosmic censorship conjecture
\cite{Gwak:2016gwj,Yang:2020iat,Liang:2018wzd}. And further developed by
others
\cite{Wang:2020vpn,Mu:2020szg,Li:2020smq,Ding:2020zgg,Duztas:2019mxr,Shaymatov:2019del,Jiang:2019ige,Jiang:2019vww,Gwak:2019rcz,He:2019mqy,Bai:2020ieh,Hong:2020zcf,VandeMoortel:2020olr,Goncalves:2020ccm}%

In addition, various investigations have been conducted on the conjecture for
not only black holes of Einstein's theory of gravity, but also anti-de Sitter
(AdS), lower-dimensional, and higher-dimensional black holes
\cite{Colleoni:2015afa,Hod:2016hqx,Gwak:2017icn,Rocha:2011wp,Gao:2012ca,Rocha:2014gza,Rocha:2014jma,Natario:2016bay,Horowitz:2016ezu}%
. From the research results, this conjecture and the laws of thermodynamics
have great relevance. If the entropy of the black hole increases, as ensured
by the second law for an irreversible process, the horizon can cover the
inside of a black hole, and the variation of a black hole is consistent with
the first law of thermodynamics under particle absorption
\cite{Gwak:2015fsa,Gwak:2016gwj}. In recent years, black hole thermodynamics
related issues received much attention with the discoveries of the Bekenstein
Hawking entropy and Hawking radiation. This has changed our understanding of
black holes ever since, opening up vast areas of research including phase
transitions and holography \cite{Wall:2018ydq}. For an black hole, the usual
first law of black hole thermodynamics takes the form
\begin{equation}
dM=TdS+\phi dQ.
\end{equation}
where $M$ denotes the Arnowitt-Deser-Misner (ADM) mass of the black hole, $T$
is the Hawking temperature, $S$ is the Bekenstein-Hawking entropy, $\phi$ is
the electrostatic potential and $Q$ is the electrical charge. Compared with
ordinary thermodynamics, there is no an absence of a $VdP$ term. This term in
the context of black hole spacetime was eventually introduced and requires an
anti-de Sitter (AdS) background \cite{Dolan:2012jh,Kastor:2009wy}. Then the
first law is generalized to
\begin{equation}
dM=TdS+VdP+\phi dQ.
\end{equation}
Where $M$ is now reinterpreted as the enthalpy, $V$ is the volume of the black
hole and is defined as the thermodynamic conjugate to the pressure. The
relationship between $M$, the internal energy $U$ and $PV$ of the black hole
is
\begin{equation}
M=U+PV.
\end{equation}

The d-precision observations confirmed the existence of a gravitationally
repulsive interaction at a global scale (cosmic dark energy) recently
\cite{intro-Ade:2013sjv}. It is founded that one type of dark energy models
produces some gravitational effect when it surrounds black holes. For this
type of dark energy, the equation of state parameters is in the interval
$[-1,-\frac{1}{3}]$ \cite{intro-Saleh:2011zz}. This type of dark energy models
is called quintessence dark energy or quintessence for short. In this case,
the first law of thermodynamics is given by \cite{intro-Li:2014ixn}
\begin{equation}
dM=TdS+VdP+\phi dQ-\frac{1}{2r_{h}^{3\omega_{q}}}d\alpha,\label{eqn:dM3}%
\end{equation}
where $\alpha$ is a positive normalization factor. There has been much
interest in studying the physics of black holes surrounded by quintessence
\cite{intro-Singh:2020tkf,intro-Haldar:2020jmt,intro-Chen:2020rov,intro-Hong:2019yiz,intro-Moinuddin:2019mzf,intro-Toledo:2019mlz,intro-Chabab:2017xdw,intro-Liu:2017baz,intro-Ghaderi:2016dpi,intro-Fernando:2014rsa,intro-Fernando:2014wma,intro-Guo:2019hxa,intro-Nandan:2016ksb,intro-Malakolkalami:2015cza,intro-WangChun-Yan:2012tcg,intro-Xi:2008ce,intro-Harada:2006dv,
intro-Kiselev:2002dx}.

According to string theory, nature can be represented by a set of extended
objects (such as one-dimensional strings) rather than point particles.
Therefore, understanding the gravitational effects caused by a set of strings
is necessary. This can be achieved by solving Einstein's equations with a
finite number of strings. The results obtained by the Letelier show that the
existence of cloud of strings will produce a global origin effect that related
to a solid deficit angle. Moreover, the solid deficit angle depends on the
parameters that determine the existence of the cloud
\cite{intro-Letelier:1979ej}. Therefore, the existence of cloud of strings
will have an impact on black holes. When we consider the existence of cloud of
strings, the first law of thermodynamics takes on the form as
\begin{equation}
dM=TdS+VdP+\phi dQ-\frac{r_{h}}{2}da,\label{eqn:dM4}%
\end{equation}
where $a$ is the state parameter of cloud of strings. The effect of cloud of
strings on black holes have been explored for various black holes
\cite{intro-Li:2020zxi,intro-Cai:2019nlo,intro-Toledo:2019szg,intro-Ghaffarnejad:2018tpr,intro-Graca:2016cbd,intro-Ghosh:2014dqa}%
. As noted in \cite{intro-Toledo:2019amt}, considered that the parameters
related to the cloud of string and quintessence are extensive thermodynamic
parameters, the first law of thermodynamics of black hole is modified as
\begin{equation}
dM=TdS+VdP+\phi dQ-\frac{1}{2r_{h}^{3\omega_{q}}}d\alpha-\frac{r_{h}}{2}da.\label{eqn:dM5}%
\end{equation}
There has been much interest in deducing and discussing the physical
properties of various black holes when they are surrounded by cloud of strings
and quintessence
\cite{Chabab:2020ejk,Liang:2020hjz,intro-Sakti:2019iku,intro-Ma:2019pya}.

The rest is organized as follows. In section \ref{sec:A}, we present a
generalized solution corresponding to charged AdS black holes surrounded by
quintessence and cloud of strings in higher dimensional spacetime. In section
\ref{sec:A}, we have to study the problem from four aspects of the black hole
with particle absorption. In section \ref{sec:Ba}, we investigate the
absorptions of the scalar particle and fermion by the black hole. The relation
between energy and charge of the particle is gotten. In section \ref{sec:Bb},
the thermodynamics in the extended phase space are investigated by the
absorptions of the particles. In section \ref{sec:Bc}, the overcharging
problem is tested by throwing a particle in the near-extremal and extremal
black holes. In section \ref{sec:Bd}, the first and second laws of
thermodynamics and the stability of the horizon are discussed under a new
assumption. In section \ref{sec:C}, We also describe the related problems of
black holes Under scalar field scattering from four aspects. In section
\ref{sec:Ca}, we get the changes in the internal energy and charge of the
black hole during the time interval $dt$. In section \ref{sec:Cb}, the laws of
thermodynamics through scalar field scattering are discussed. In section
\ref{sec:Cc}, we tested the stability of horizon by evaluating the minimum of
function $f$ in the final state. In section \ref{sec:Cd}, We use the
scattering of a scalar field to investigate thermodynamics and the stability
of horizon under a new assumption. The last section is devoted to our
discussion and conclusion.

\section{Quintessence surrounding d-dimensional RN-AdS black holes with a
cloud of strings}

\label{sec:A} It was recently considered a metric for AdS asymptotically
spacetime in $d$-dimension, which generated by a charged static black hole and
surrounded by cloud of strings and quintessence. Assuming that the cloud of
strings and quintessence do not interact \cite{Toledo:2018pfy}, the energy
momentum tensor of the two sources can be seen as a linear superposition. The
solution corresponding to a black hole immersed in quintessence with cloud of
strings, in a $d$-dimensional spacetime, is given by the general form
\cite{Chabab:2020ejk}
\begin{equation}
dS_{d}^{2}=-f(r)dt^{2}+f(r)^{-1}dr^{2}+r^{2}d\varOmega_{d-2}^{2}.
\end{equation}
\label{eqn:ds1} Where $d\varOmega_{d-2}^{2}$ denotes the metric on unit
$(d-2)$-sphere, which can eliminate $\rho_{q}$ from the Einstein equation. The
following equations can be obtained from the metric ansatza above and the
Einstein equation
\begin{equation}
G_{\mu}^{\nu}+\Lambda g_{\mu}^{\nu}=\sum T_{\mu}^{\nu},
\end{equation}%
\begin{equation}
-\frac{d-2}{2r}f^{'}(r)-\frac{(d-2)(d-3)}{2r^{2}}(f(r)-1)-\Lambda=\sum T_{t}^{t}=\sum T_{r}^{r},
\end{equation}%
\begin{equation}
-\frac{f^{^{\prime\prime}}(r)}{2}-\frac{d-3}{2r}f^{^{\prime}}(r)-\frac
{(d-3)(d-4)}{2r^{2}}(f(r)-1)-\Lambda=\sum T_{\theta_{1}}^{\theta_{1}}=\sum
T_{\theta_{d-2}}^{\theta_{d-2}},
\end{equation}
which gives the following master equation
\begin{equation}
r^{2}f^{^{\prime\prime}}(r)+F_{1}rf^{^{\prime}}(r)+F_{2}(f(r)-1)+F_{3}%
r^{2}+F_{4}r^{-2(d-3)}+F_{5}r^{-(d-4)}=0,
\end{equation}
with
\begin{equation}
\begin{aligned} &F_{1}=((d-1)\omega_{q}+2d-5),\\ &F_{2}=(d-3)((d-1))\omega_{q}+d-3),\\ &F_{3}=\Lambda\frac{2(d-1)(\omega_{q}+1)}{d-2},\\ &F_{4}=q^{2}(d-3)((d-1)\omega_{q}-d+3),\\ &F_{5}=\frac{2((d-1)\omega_{q}+1)a}{d-2}.\\ \end{aligned}
\end{equation}
It is important to note that the required advertising space cosmological
constant $\Lambda$ is negative, and then we use Maxwell equations ($\nabla_{\nu}(\sqrt{-g}F^{\mu\nu})=0$) to evaluate the potential
\begin{equation}
A=-\sqrt{\frac{d-2}{2(d-3)}}\frac{q}{r^{d-3}}dt.
\end{equation}
The solution of the main equation is given by
\begin{equation}
f(r)=1-\frac{m}{r^{d-3}}+\frac{q^{2}}{r^{2(d-3)}}-\frac{2\Lambda r^{2}%
}{(d-2)(d-1)}-\frac{\alpha}{r^{(d-1)\omega_{q}+d-3}}-\frac{2a}{(d-2)r^{d-4}%
},\label{eqn:fr}%
\end{equation}
where $m$ is the integral constant proportional to the mass, and $q$ is
proportional to the integral constant black holes, which are given by the
following equation \cite{Chamblin:1999tk,Gunasekaran:2012dq}
\begin{equation}
M=\frac{(d-2)}{16\pi}\Omega_{d-2}m,Q=\frac{\sqrt{2(d-2)(d-3)}\varOmega_{d-2}%
q}{8\pi},
\end{equation}
where $\Omega_{d-2}$ is the volume of unit $(d-2)$-sphere, $\alpha$ is a
positive normalization factor related to the quintessence, whose relationship
with density $\rho_{q}$ is \cite{Caldwell:1997ii},
\begin{equation}
\rho_{q}=-\frac{\alpha\omega_{q}(d-1)(d-2)}{4r^{(d-1)(\omega_{q}+1)}}.
\end{equation}
In addition, the asymptotic effect of the quintessence term may be different
due to the existence of power $[\frac{\alpha}{r^{(d-1)\omega_{q}+d-3}}]$ in
Eq. $\left(  \ref{eqn:fr}\right)  $ . When only the quintessential
contribution is considered, the above formula can be modified as
\begin{equation}
f_{\alpha}(r)=1-\frac{m}{r^{d-3}}-\frac{\alpha}{r^{(d-1)\omega_{q}+d-3}}.
\end{equation}
Where the spacetime becomes asymptotically $dS$-like for $\omega_{q}%
<-\frac{d-3}{d-1}$, otherwise it becomes asymptotically flat. In this paper we
consider only the asymptotically $dS$ behavior and set $\omega_{q}$ to the
value $\omega_{q}=-\frac{d-2}{d-1}$ in numerical analysis. \begin{figure}[tbh]
\begin{center}
\subfigure[{$\alpha=0.01$, $a=0.01$.}]{
\includegraphics[width=0.45\textwidth]{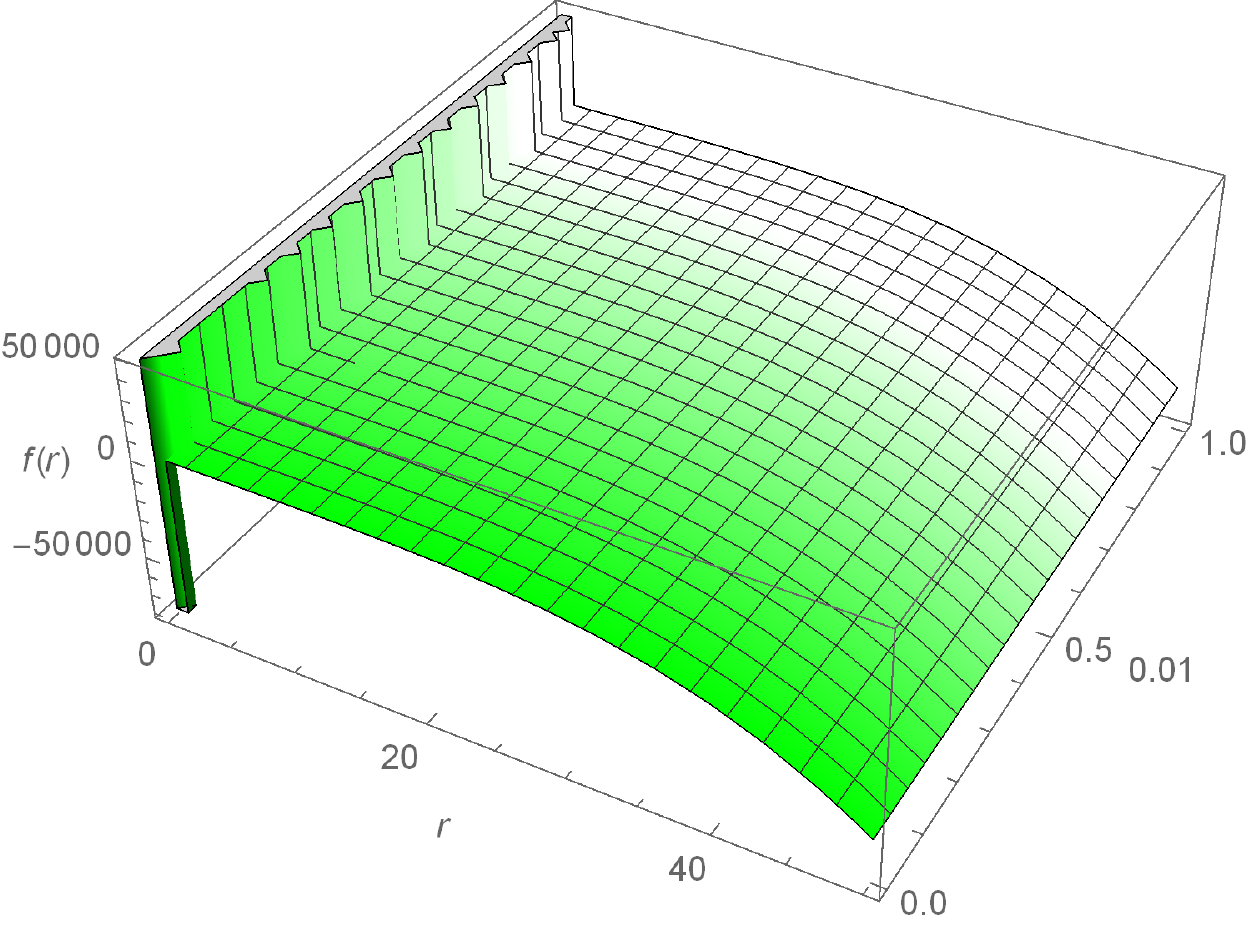}\label{fig:f1}}
\subfigure[{$\alpha=0.01$, $a=20$.}]{
\includegraphics[width=0.45\textwidth]{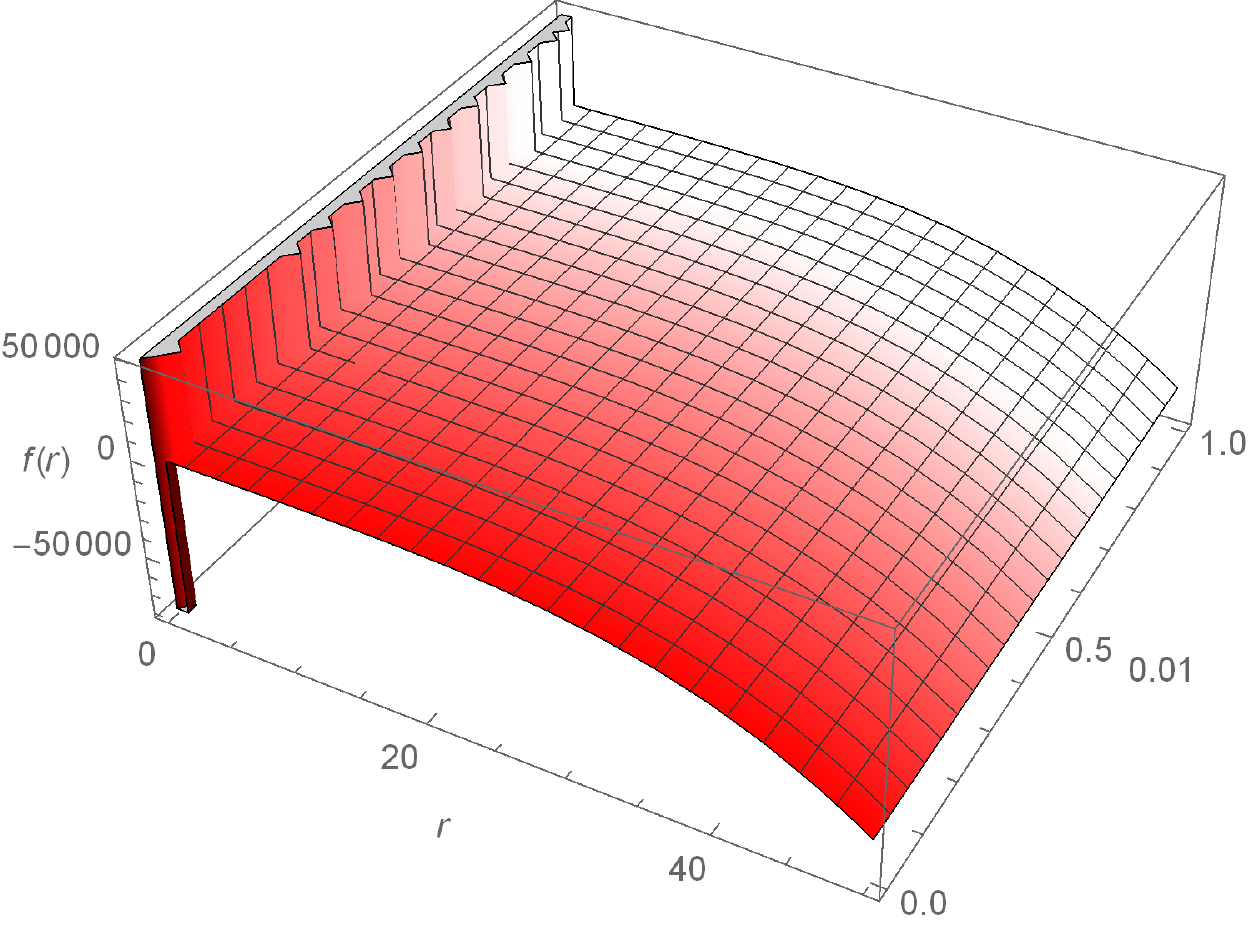}\label{fig:f2}}
\subfigure[{$\alpha=0.01$, $a=20$.}]{
\includegraphics[width=0.45\textwidth]{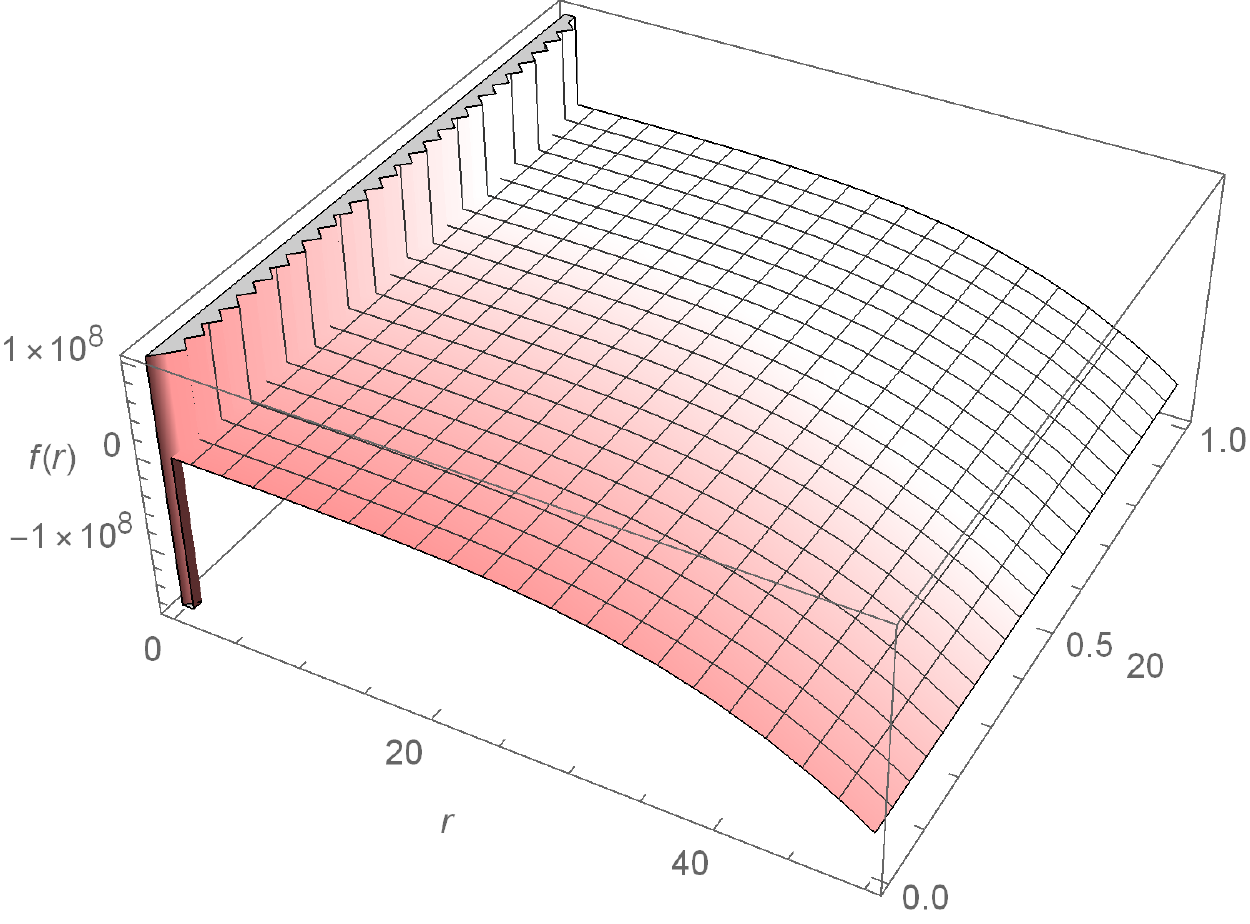}\label{fig:f3}}
\end{center}
\caption{The relationship between $f(r)$, $Q$ and $r_{h}$ for different values
of $a$ and $\alpha$. We choose $M=1$,$l=1,\omega_{q}=-\frac{d-2}{d-1}$,$d=5$
and $\varOmega_{d-2}=1$.}%
\label{fig:F1}%
\end{figure}

\begin{figure}[tbh]
\begin{center}
\subfigure[{ }]{
\includegraphics[width=0.45\textwidth]{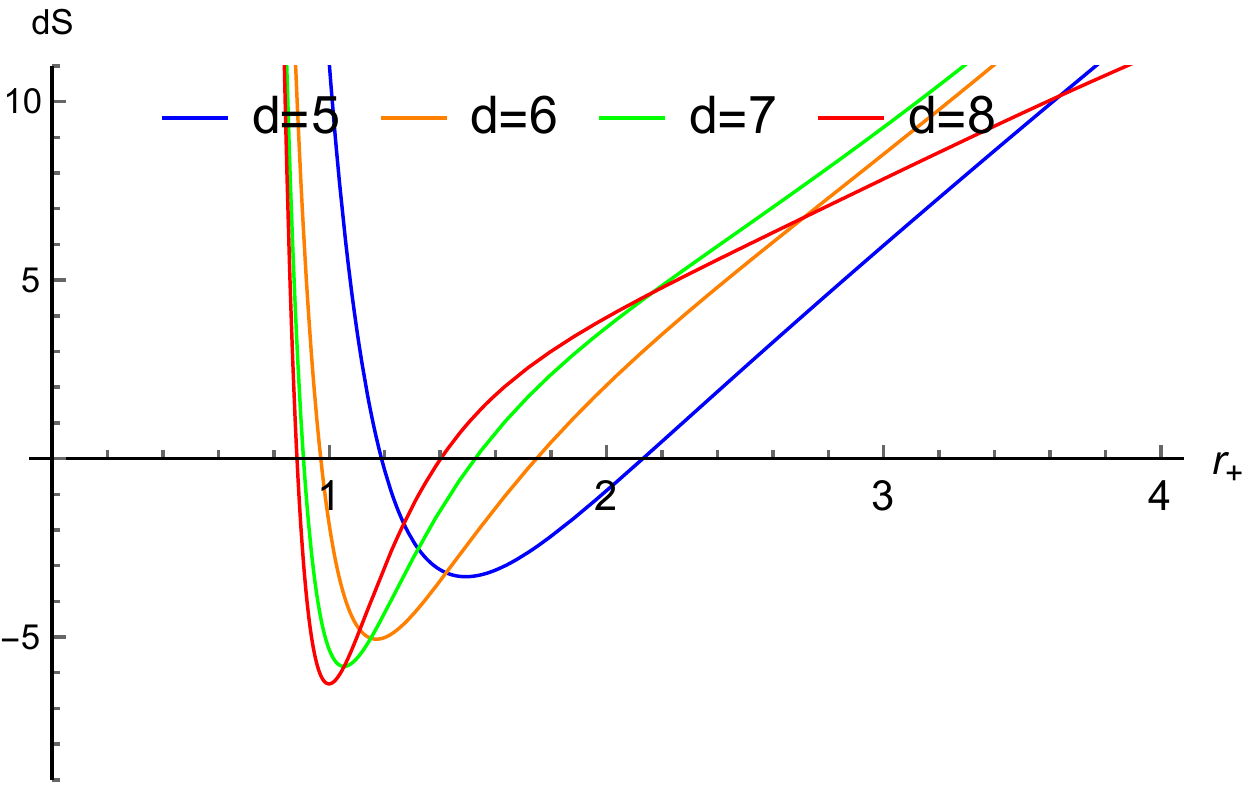}\label{fig:f4}}
\end{center}
\caption{The relationship between $f(r)$ and $r_{h}$ with parameter values
$M=2,l=1,\omega_{q}=-\frac{d-2}{d-1},Q=0.7,a=0.08,\alpha=0.01$ and
$\varOmega_{d-2}=1$. }%
\label{fig:F2}%
\end{figure}In Fig. \ref{fig:F1}, the graphs of the function $f(r)$ are shown
for different values of the parameters $a$, $\alpha$ when $d=5$. In Fig.
\ref{fig:F2}, the graphs of the function $f(r)$ are shown for different values
of the parameters $a$, $\alpha$ and $d$, when it is the non-extremal black
hole, the equation $f(r)$ = 0 has two positive real roots $r_{h}$ and $r_{-}$.
The $r_{h}$ represents the radius of the event horizon. When it is the
extremal black hole, $f(r)=0$ has only one root. The mass of the black hole
is
\begin{equation}
M=\frac{(d-2)\Omega_{d-2}r_{h}^{d-3}}{16\pi}+\frac{(d-2)\Omega_{d-2}q^{2}%
}{16\pi r_{h}^{d-3}}+\frac{\Omega_{d-2}Pr_{h}^{d-1}}{(d-1)}-\frac
{\alpha(d-2)\Omega_{d-2}}{16\pi r_{h}^{(d-1)\omega_{q}}}-\frac{ar_{h}%
\Omega_{d-2}}{8\pi}.\label{eqn:mmm}%
\end{equation}
Where the mass of the black hole $M$ is defined as its enthalpy. Therefore,
the relationship between enthalpy, internal energy and pressure can be
expressed by the following equation
\begin{equation}
M=U+PV.
\end{equation}

\section{Particle absorption}

\label{sec:B}

\subsection{Scalar particle's absorption}

\label{sec:Ba} In this subsection, we discuss the absorption of the scalar
particle in the d-dimensional spacetime and the motion of scattered particles
satisfy the Klein-Gordon equation \cite{Wang:2019jzz} of curved spacetime,
which is
\begin{equation}
-\frac{1}{\sqrt{-g}}(\frac{\partial}{\partial x^{\mu}}-\frac{iq}{\hbar}A_{\mu
})[\sqrt{-g}g^{\mu\nu}(\frac{\partial}{\partial x^{\nu}}-\frac{iq}{\hbar
}A_{\nu})]\phi-\frac{m^{2}}{\hbar^{2}}\phi=0 .\label{eqn:s1}%
\end{equation}
Where $m$ and $q$ are the particle's mass and charge, respectively, $\phi$ is
the scalar field, and $A_{\mu}$ is the electro magnetic potential. Assuming
the WKB ansatz for $\phi$
\begin{equation}
\phi=A\exp(\frac{iI}{\hbar}).
\end{equation}
Where $A$ is a slowly varying amplitude. In a semiclassical approximation, the
Hamilton-Jacobi equation for a scalar particle is the lowest order of the WKB
expansion of the corresponding Klein-Gordon equation. We can expand Eq.
$\left( \ref{eqn:s1}\right) $ in powers of $\hbar$ and find that the lowest
order term is
\begin{equation}
g^{\mu\nu}(p_{\mu}-qA_{\mu})(p_{\nu}-qA_{\nu})+m^{2}=0,\label{eqn:s3}%
\end{equation}
with
\begin{equation}
p_{\mu}=\partial_{\mu}I.
\end{equation}
Which is the Hamilton-Jacobi equation. Where $p_{\mu}$ is the momentum of the
particle, and $I$ is the Hamilton action of the particle. Considering the
symmetry of space and time, the role of the Hamiltonian motion of the
particles can be divided into
\begin{equation}
I=-\omega t+I_{r}(r)+\sum_{i=1}^{d-3}I_{\theta_{i}}(\theta_{i})+L\varPsi.
\end{equation}
And where the conserved quantities $\omega$ and $L$ are energy and angular
momentum of particle, based on the formula $\left( \ref{eqn:s3}\right) $ of
symmetry and translational regulatory moderate, which is the amount of time
and space conservation in gravitational systems. In addition, $I_{r}(r)$ and
$I_{\theta_{i}}(\theta_{i})$ are the radial directional component and $\theta
$-directional component of the action respectively. The black hole includes a
$(d-2)$ -dimensional sphere $\Omega_{d-2}$ because of d-dimensional solution,
whose the translation symmetry of the last angle coordinate corresponding to
the angular momentum $L$. Then, the $(d-2)$ -dimensional sphere can be written
as
\begin{equation}
h_{ij}dx^{i}dx^{j}=\sum_{i=1}^{d-2}(\prod_{j-1}^{i}\sin^{2}\theta
_{j-1})d\theta_{i}^{2},\theta_{d-2}=\varPsi.
\end{equation}

To solve the Hamilton-Jacobi equation, we inserting the above ansatz and the
contravariant metric of the black hole into the Klein-Gordon equation and
yields
\begin{equation}
\begin{aligned} g^{\mu\nu}\partial_{\mu}\partial_{\nu}=&-f(r)^{-1}(\partial_{t})^{2}+f(r)(\partial_{r})^{2}+r^{-2}\sum_{i=1}^{d-2}(\prod_{j=1}^{i}\sin^{2}\theta_{j-1})(\partial_{\theta_{i}})^{2}.\\ \end{aligned}
\end{equation}
Substituting the above equations into Eq. $\left( \ref{eqn:s3}\right) $, we
obtain
\begin{equation}
\begin{aligned} -m^{2}=&-f(r)^{-1}r^{2}(-\omega-qA_{t})^{2}+f(r)(\partial_{r}I(r))^{2}+r^{-2}\sum_{i=1}^{d-3}(\prod_{j=1}^{i}sin^{-2}\theta_{j-1})(\partial_{\theta_{i}}I(\theta_{i}))^{2}\\ &+r^{-2}(\prod_{j=1}^{d-2}sin^{-2}\theta_{j-1})L^{2}.\\ \end{aligned}
\end{equation}
We carry out the separation of variables by introducing a variable $\kappa$
and $R_{i}$, Therefore, the radial and angular components are
\begin{equation}
\kappa=-m^{2}r^{2}+\frac{r^{2}}{f(r)}(-\omega-qA_{t})^{2}-r^{2}f(r)(\partial
_{r}I(r))^{2},
\end{equation}
with
\begin{equation}
R_{i}=\sum_{i=1}^{d-3}(\prod_{j=1}^{i}sin^{-2}\theta_{j-1})(\partial
_{\theta_{i}}I(\theta_{i}))^{2}+(\prod_{j=1}^{d-2}sin^{-2}\theta_{j-1})L^{2}.
\end{equation}
Then, we can determine entire equations of motion. The radial and $\theta
$-directional are sufficient to obtain the relationship between the equations
and the energy of the charged particles. The momenta of the particle are
\begin{equation}
p^{r}\equiv g^{rr}\partial_{r}I(r)=f(r)\sqrt{\frac{-m^{2}r^{2}-\kappa}%
{r^{2}f(r)}+\frac{1}{f(r)^{2}}(-\omega-qA_{t})^{2}}.
\end{equation}
We take into account the case of the absorbed particle near the event horizon.
This implies $f(r)\rightarrow0$ and the above equation is simplified to
\begin{equation}
p^{r}=\omega-qA_{t}=\omega-q\phi,
\end{equation}
where $\phi=\sqrt{\frac{d-2}{2(d-3)}}\frac{q}{r_{h}^{d-3}}$ represents the
electric potential at the event horizon. The condition of the super radiation
is that the boundary condition of the scalar field should be in the asymptotic
region and $\omega<q\phi$. Then, at the limit of the outer horizon, the energy
relation between conserved quantities and momenta is obtained as
\begin{equation}
E=\sqrt{\frac{d-2}{2(d-3)}}\frac{q^{2}}{r_{h}^{d-3}}+p^{r}.
\end{equation}
The particle enters the black hole in the positive flow of time. At this
moment, the energy of the particle should be defined as a positive value thus
that the signs of $E$ and $\mid p^{r}\mid$ are both positive. Therefor a
positive sign is required in front of the $\mid p^{r}\mid$ term
\begin{equation}
E=\sqrt{\frac{d-2}{2(d-3)}}\frac{q^{2}}{r_{h}^{d-3}}+\mid p^{r}\mid,
\end{equation}
in which various dependencies between variables are reduced to this simple relation.

\subsection{Fermion absorption}

In curved spacetime, a spin-1/2 fermion of the mass $m$ and the charge $q$
obeys the Dirac equation
\begin{equation}
i\gamma_{\mu}(\partial^{\mu}+\Omega^{\mu}-\frac{iqA^{\mu}}{\hbar
})\varPsi-\frac{m}{\hbar}\varPsi=0.\label{eqn:f1}%
\end{equation}
where $\Omega_{\mu}\equiv\frac{i}{2}\omega_{\mu}{}^{ab}\varSigma_{ab}$ is the
Lorentz spinor generator, $\varSigma_{ab}$ is the Lorentz spinor generator,
$\omega_{\mu}{}^{ab}$ is the spin connection and $\left\{  \gamma_{\mu}%
,\gamma_{\nu}\right\}  =2g_{\mu\nu}$. The Greek indices are raised and lowered
by the curved metric $g_{\mu\nu}$, while the Latin indices are governed by the
flat metric $\eta_{ab}$. In order to obtain the fermions Hamilton - Jacobi
equation, assuming that the WKB ansatz $\varPsi$ is
\begin{equation}
\varPsi=\exp(\frac{iI}{\hbar})u,\label{eqn:f2}%
\end{equation}
where $u$ is a slowly varying spinor amplitude. Substituting Eq. $\left(
\ref{eqn:f1}\right) $ into Eq. $\left( \ref{eqn:f2}\right) $, we find that the
lowest order term of $\hbar$ is
\begin{equation}
\gamma_{\mu}(\partial^{\mu}I-qA^{\mu})u=-mu,\label{eqn:f3}%
\end{equation}
which is the Hamilton-Jacobi equation for the fermion. Multiplying both sides
of Eq. $\left( \ref{eqn:f3}\right) $ from the left by $\ensuremath{\gamma_{\nu}}(\partial^{\nu}I+qA^{\nu})$ and then using Eq. $\left( \ref{eqn:f3}\right) $
to simplify the RHS, one obtains
\begin{equation}
\gamma_{\nu}(\partial^{\nu}I-qA^{\nu})\gamma_{\mu}(\partial^{\mu}I-qA^{\mu
})u=m^{2}u.
\end{equation}
Using $\left\{  \gamma_{\mu},\gamma_{\nu}\right\}  =2g_{\mu\nu}$, we have
\begin{equation}
\left[ (\partial^{\mu}I-qA^{\mu})(\partial_{\mu}I-qA_{\mu})-m^{2}\right] u=0.
\end{equation}
Since $u$ is nonzero, the Hamilton-Jacobi equation reduces to
\begin{equation}
(\partial^{\mu}I-qA^{\mu})(\partial_{\mu}I-qA_{\mu})=m^{2}.
\end{equation}
which is the same as the Hamilton-Jacobi equation for a scalar. And then, the
following formula can be obtained by using the same way
\begin{equation}
E=\sqrt{\frac{d-2}{2(d-3)}}\frac{q^{2}}{r_{h}^{d-3}}+\mid p^{r}\mid.
\end{equation}

\subsection{The first and second laws of Thermodynamics}

\label{sec:Bb} \begin{table}
\caption{The relation between $dS$, $Q$ and $r_{h}$ for $d = 5$ in the
extended phase space via particle absorption .}%
\label{tab:dsa1}%
\begin{tabular}
[c]%
{p{0.6in}|p{0.6in}|p{0.6in}|p{0.6in}|p{0.6in}|p{0.6in}|p{0.6in}|p{0.6in}|p{0.6in}}%
\hline
\multicolumn{3}{c|}{$a=0.01$} & \multicolumn{3}{|c|}{$a=10$} &
\multicolumn{3}{|c}{$a=20$}\\\hline
$Q$ & $r_{h}$ & $dS$ & $Q$ & $r_{h}$ & $dS$ & $Q$ & $r_{h}$ & $dS$\\\hline
0.640747 & 1.43509 & 1.720080 & 0.965855 & 1.75208 & 3.231130 & 1.33654 &
2.01130 & 5.074160\\
0.64 & 1.40537 & 1.562300 & 0.96 & 1.67481 & 2.512370 & 0.99 & 1.32019 &
0.726566\\
0.6 & 1.19173 & 0.797021 & 0.9 & 1.46858 & 1.367870 & 0.9 & 1.20661 &
0.539207\\
0.55 & 1.04937 & 0.504545 & 0.8 & 1.26515 & 0.773806 & 0.8 & 1.08407 &
0.384662\\
0.5 & 0.93213 & 0.338204 & 0.7 & 1.09707 & 0.476048 & 0.7 & 0.96294 &
0.268491\\
0.45 & 0.82613 & 0.228518 & 0.6 & 0.94171 & 0.292462 & 0.6 & 0.84135 &
0.180201\\
0.4 & 0.72648 & 0.152164 & 0.5 & 0.79128 & 0.171589 & 0.5 & 0.71771 &
0.113574\\
0.35 & 0.63078 & 0.098106 & 0.4 & 0.64168 & 0.091637 & 0.4 & 0.59042 &
0.064739\\
0.3 & 0.53766 & 0.060084 & 0.3 & 0.49007 & 0.041264 & 0.3 & 0.45777 &
0.031139\\
0.2 & 0.35598 & 0.017191 & 0.2 & 0.33413 & 0.013315 & 0.2 & 0.31756 &
0.010813\\
0.1 & 0.17741 & 0.002109 & 0.1 & 0.17163 & 0.001847 & 0.1 & 0.166670 &
0.001639\\\hline
\end{tabular}
\end{table}

\begin{table}
\caption{The relation between $dS$, $Q$ and $r_{h}$ for $d = 6$ in the
extended phase space via particle absorption.}%
\label{tab:dsa2}%
\begin{tabular}
[c]%
{p{0.6in}|p{0.6in}|p{0.6in}|p{0.6in}|p{0.6in}|p{0.6in}|p{0.6in}|p{0.6in}|p{0.6in}}%
\hline
\multicolumn{3}{c|}{$a=0.01$} & \multicolumn{3}{|c|}{$a=10$} &
\multicolumn{3}{|c}{$a=20$}\\\hline
$Q$ & $r_{h}$ & $dS$ & $Q$ & $r_{h}$ & $dS$ & $Q$ & $r_{h}$ & $dS$\\\hline
0.751095 & 1.26380 & 1.339280 & 1.09831 & 1.43654 & 2.198080 & 1.48119 &
1.57562 & 2.624600\\
0.75 & 1.24297 & 1.210610 & 0.99 & 1.23344 & 0.813316 & 0.99 & 1.07695 &
0.341080\\
0.7 & 1.10288 & 0.631245 & 0.9 & 1.12473 & 0.517795 & 0.9 & 1.01152 &
0.262104\\
0.65 & 1.02277 & 0.437442 & 0.8 & 1.02853 & 0.345687 & 0.8 & 0.93810 &
0.192757\\
0.6 & 0.95369 & 0.316940 & 0.7 & 0.93622 & 0.231018 & 0.7 & 0.86278 &
0.138147\\
0.55 & 0.88937 & 0.232461 & 0.6 & 0.84395 & 0.150458 & 0.6 & 0.78436 &
0.095181\\
0.5 & 0.82727 & 0.170035 & 0.5 & 0.74900 & 0.092995 & 0.5 & 0.70140 &
0.061812\\
0.4 & 0.70437 & 0.086572 & 0.4 & 0.64873 & 0.052578 & 0.4 & 0.61198 &
0.036642\\
0.3 & 0.57726 & 0.038301 & 0.3 & 0.53976 & 0.025506 & 0.3 & 0.51316 &
0.018679\\
0.2 & 0.43870 & 0.012623 & 0.2 & 0.41655 & 0.009229 & 0.2 & 0.39967 &
0.007159\\
0.1 & 0.27580 & 0.001955 & 0.1 & 0.26666 & 0.001595 & 0.1 & 0.25905 &
0.001337\\\hline
\end{tabular}
\end{table}

\begin{table}
\caption{The relation between $dS$, $Q$ and $r_{h}$ for $d = 7$ in the
extended phase space via particle absorption.}%
\label{tab:dsa3}%
\begin{tabular}
[c]%
{p{0.6in}|p{0.6in}|p{0.6in}|p{0.6in}|p{0.6in}|p{0.6in}|p{0.6in}|p{0.6in}|p{0.6in}}%
\hline
\multicolumn{3}{c|}{$a=0.01$} & \multicolumn{3}{|c|}{$a=10$} &
\multicolumn{3}{|c}{$a=20$}\\\hline
$Q$ & $r_{h}$ & $dS$ & $Q$ & $r_{h}$ & $dS$ & $Q$ & $r_{h}$ & $dS$\\\hline
0.818145 & 1.16885 & 1.0788 & 1.17528 & 1.28263 & 1.459720 & 1.56137 &
1.37345 & 1.795660\\
0.8 & 1.10675 & 0.728925 & 0.99 & 1.07803 & 0.443659 & 0.99 & 0.99049 &
0.219661\\
0.75 & 1.03766 & 0.480032 & 0.9 & 1.01703 & 0.316912 & 0.9 & 0.94394 &
0.170971\\
0.7 & 0.98535 & 0.351321 & 0.8 & 0.95202 & 0.220464 & 0.8 & 0.89072 &
0.127292\\
0.65 & 0.93833 & 0.264973 & 0.7 & 0.88699 & 0.151683 & 0.7 & 0.83506 &
0.092296\\
0.6 & 0.89360 & 0.201726 & 0.6 & 0.81988 & 0.101251 & 0.6 & 0.77593 &
0.064373\\
0.5 & 0.80579 & 0.115536 & 0.5 & 0.74876 & 0.064123 & 0.5 & 0.71197 &
0.042409\\
0.4 & 0.71509 & 0.061968 & 0.4 & 0.67125 & 0.037275 & 0.4 & 0.64122 &
0.025614\\
0.3 & 0.61620 & 0.028949 & 0.3 & 0.58382 & 0.018751 & 0.3 & 0.56042 &
0.013418\\
0.2 & 0.50162 & 0.010235 & 0.2 & 0.47991 & 0.007168 & 0.2 & 0.46334 &
0.005381\\
0.1 & 0.35417 & 0.001781 & 0.1 & 0.34299 & 0.001376 & 0.1 & 0.33386 &
0.001105\\\hline
\end{tabular}
\end{table}\begin{table}
\caption{The relation between $dS$, $Q$ and $r_{h}$ for $d=8$ in the extended
phase space via particle absorption.}%
\label{tab:dsa4}%
\begin{tabular}
[c]%
{p{0.6in}|p{0.6in}|p{0.6in}|p{0.6in}|p{0.6in}|p{0.6in}|p{0.6in}|p{0.6in}|p{0.6in}}%
\hline
\multicolumn{3}{c|}{$a=0.01$} & \multicolumn{3}{|c|}{$a=10$} &
\multicolumn{3}{|c}{$a=20$}\\\hline
$Q$ & $r_{h}$ & $dS$ & $Q$ & $r_{h}$ & $dS$ & $Q$ & $r_{h}$ & $dS$\\\hline
0.861699 & 1.11151 & 0.900437 & 1.22409 & 1.19494 & 1.15260 & 1.611 &
1.26111 & 1.369570\\
0.8 & 1.01782 & 0.438211 & 0.99 & 1.01482 & 0.31313 & 0.99 & 0.95131 &
0.162462\\
0.75 & 0.97711 & 0.324878 & 0.9 & 0.96949 & 0.22958 & 0.9 & 0.91486 &
0.127323\\
0.7 & 0.94082 & 0.249153 & 0.8 & 0.91979 & 0.16287 & 0.8 & 0.87272 &
0.095452\\
0.65 & 0.90635 & 0.193416 & 0.7 & 0.86901 & 0.11381 & 0.7 & 0.82812 &
0.069675\\
0.6 & 0.87253 & 0.150415 & 0.6 & 0.81565 & 0.07703 & 0.6 & 0.78013 &
0.048944\\
0.5 & 0.80409 & 0.089039 & 0.5 & 0.75808 & 0.04948 & 0.5 & 0.72747 &
0.032514\\
0.4 & 0.73116 & 0.049179 & 0.4 & 0.69412 & 0.02924 & 0.4 & 0.66824 &
0.019851\\
0.3 & 0.64919 & 0.023723 & 0.3 & 0.62028 & 0.01502 & 0.3 & 0.59919 &
0.010559\\
0.2 & 0.55070 & 0.008747 & 0.2 & 0.52978 & 0.00593 & 0.2 & 0.51382 &
0.004342\\
0.1 & 0.41685 & 0.001632 & 0.1 & 0.40457 & 0.00121 & 0.1 & 0.39465 &
0.000939\\\hline
\end{tabular}
\end{table}

\begin{figure}[tbh]
\begin{center}
\subfigure[{$d=5$ .}]{
\includegraphics[width=0.45\textwidth]{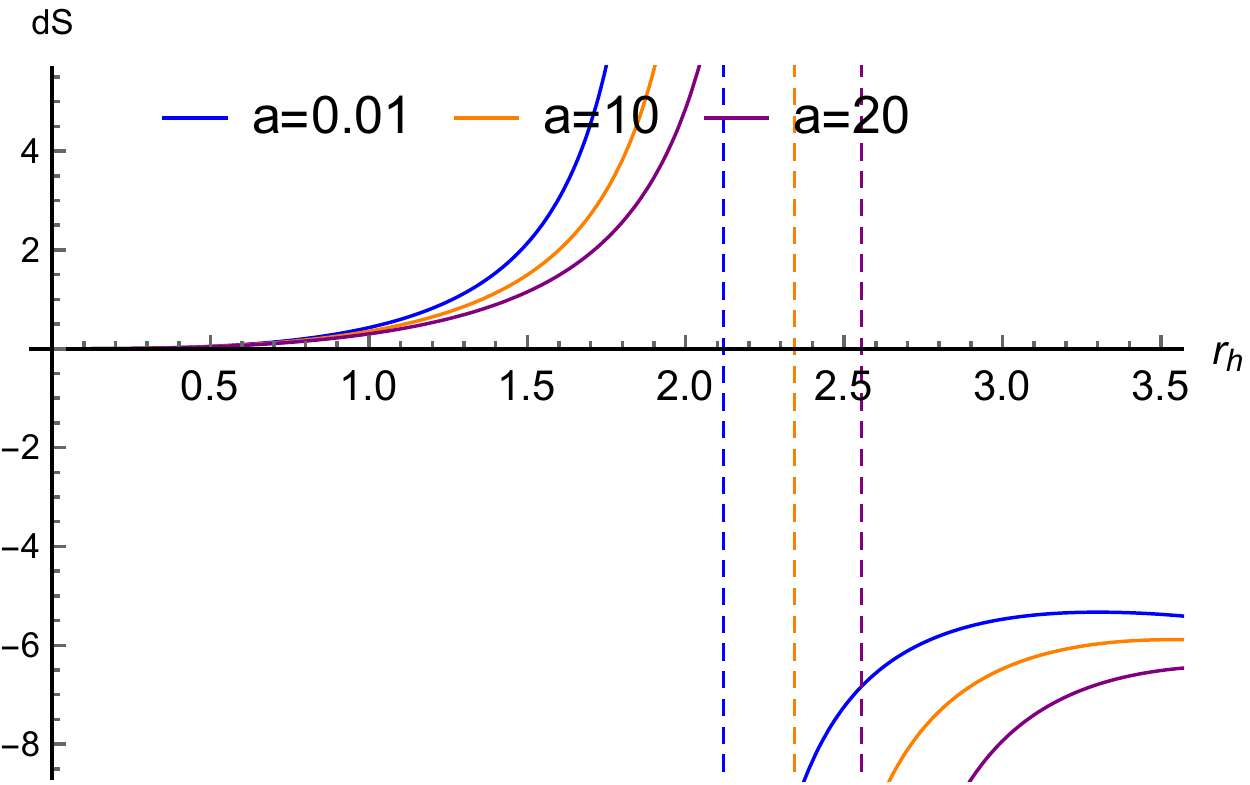}\label{ldsa5}}
\subfigure[{$d=6$ .}]{
\includegraphics[width=0.45\textwidth]{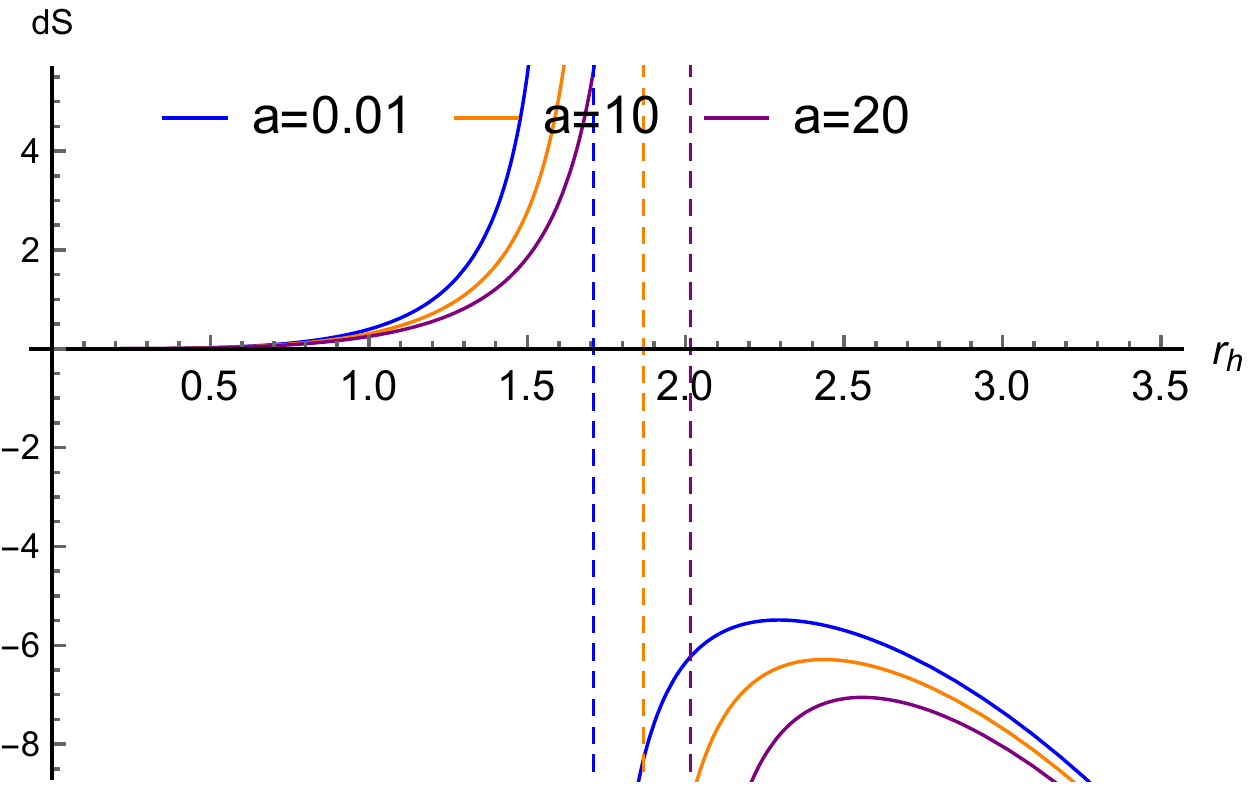}\label{ldsa6}}
\subfigure[{$d=7$ .}]{
\includegraphics[width=0.45\textwidth]{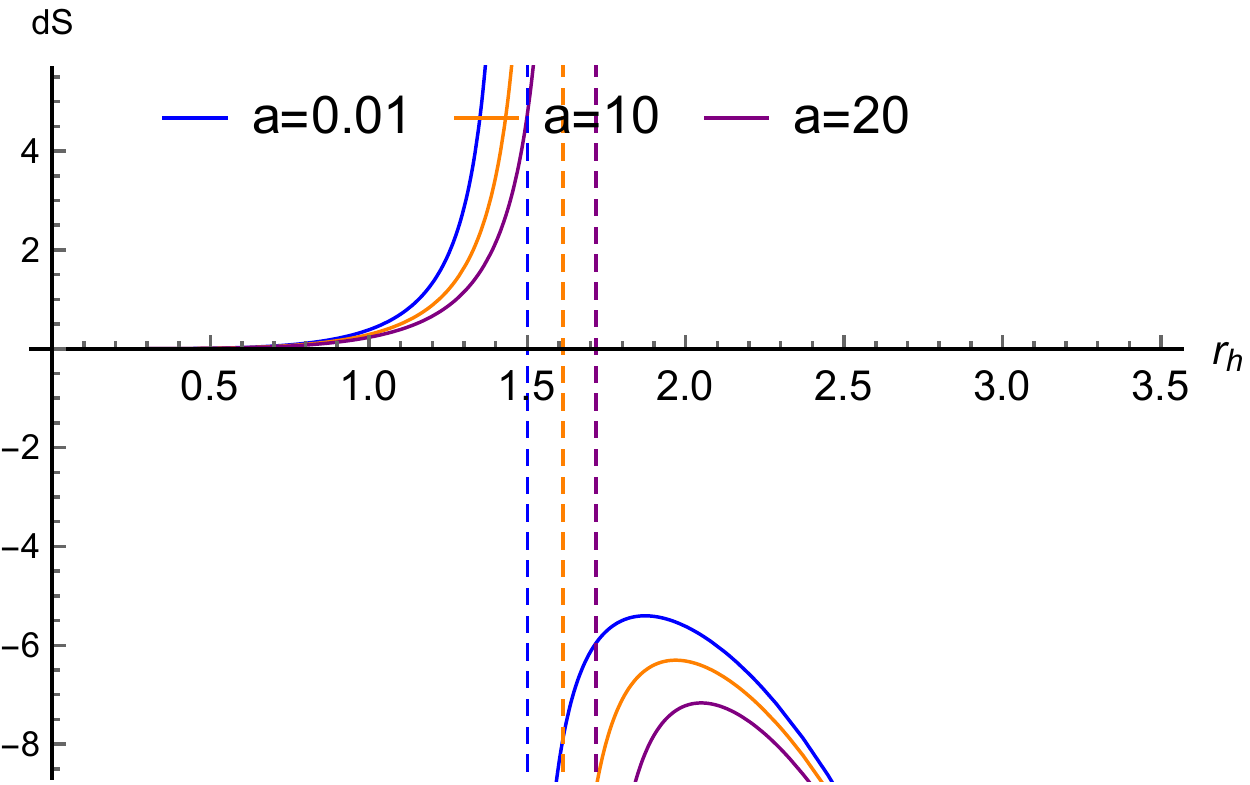}\label{ldsa7}}
\subfigure[{$d=8$ .}]{
\includegraphics[width=0.45\textwidth]{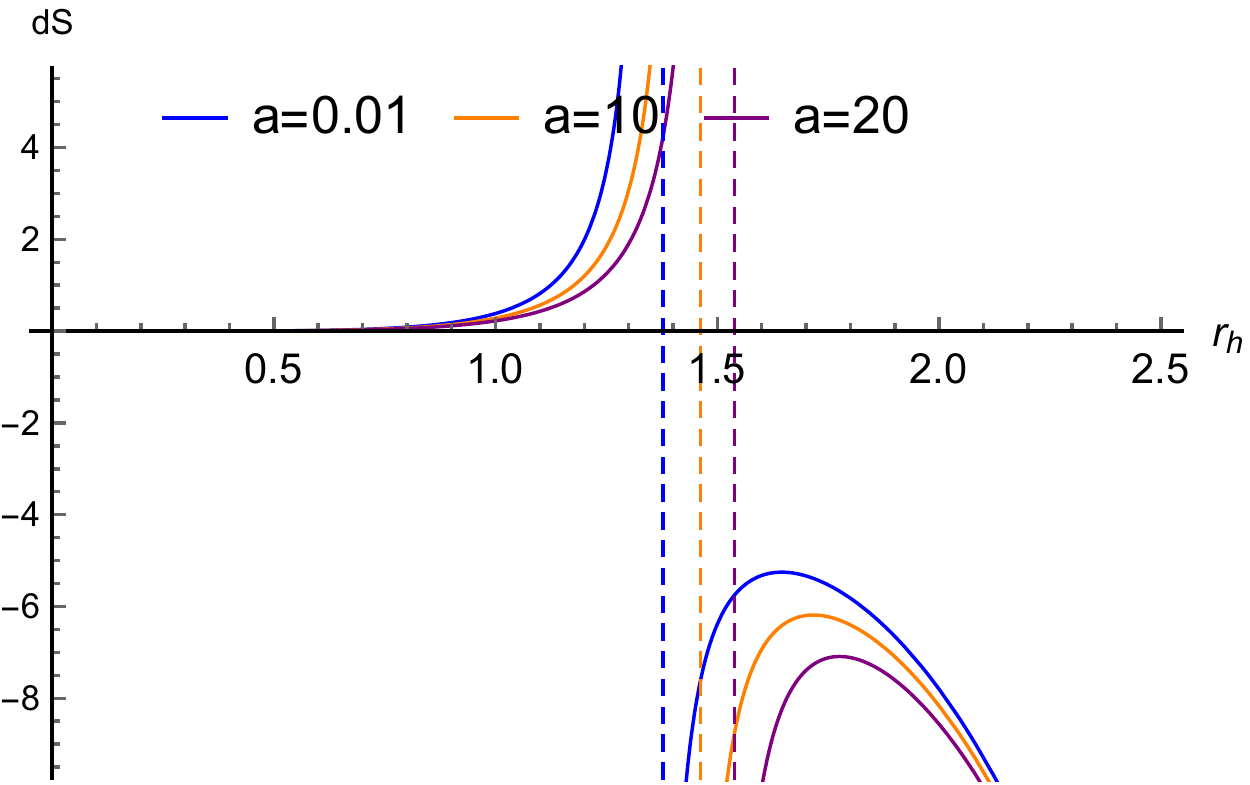}\label{ldsa8}}
\end{center}
\caption{The relationship between $dS$,$Q$ and $r_{h}$ for $a=0.1,10,20$. }%
\label{fig:ds1}%
\end{figure}

\begin{table}
\caption{The relation between $dS$, $Q$ and $r_{h}$ for $d = 5$ in the
extended phase space via particle absorption.}%
\label{tab:dspha1}%
\begin{tabular}
[c]%
{p{0.6in}|p{0.6in}|p{0.6in}|p{0.6in}|p{0.6in}|p{0.6in}|p{0.6in}|p{0.6in}|p{0.6in}}%
\hline
\multicolumn{3}{c|}{$\alpha=0.01$} & \multicolumn{3}{|c|}{$\alpha=10$} &
\multicolumn{3}{|c}{$\alpha=20$}\\\hline
$Q$ & $r_{h}$ & $dS$ & $Q$ & $r_{h}$ & $dS$ & $Q$ & $r_{h}$ & $dS$\\\hline
0.640747 & 1.43509 & 1.72008 & 0.99 & 1.25816 & 3.276210 & 0.99 & 1.11358 &
6.888790\\
0.6 & 1.19173 & 0.797021 & 0.9 & 1.18899 & 1.840720 & 0.9 & 1.05811 &
2.220120\\
0.55 & 1.04937 & 0.504545 & 0.8 & 1.10560 & 1.105600 & 0.8 & 0.99118 &
1.087460\\
0.5 & 0.93212 & 0.338204 & 0.7 & 1.01377 & 0.671353 & 0.7 & 0.91733 &
0.619633\\
0.45 & 0.82613 & 0.228518 & 0.6 & 0.91142 & 0.411186 & 0.6 & 0.83458 &
0.367834\\
0.4 & 0.72648 & 0.152164 & 0.5 & 0.79599 & 0.240074 & 0.5 & 0.74015 &
0.214234\\
0.35 & 0.63078 & 0.098105 & 0.4 & 0.66474 & 0.126627 & 0.4 & 0.63015 &
0.115212\\
0.3 & 0.53766 & 0.060086 & 0.3 & 0.51596 & 0.055265 & 0.3 & 0.50004 &
0.052067\\
0.2 & 0.35598 & 0.017191 & 0.2 & 0.35138 & 0.016747 & 0.2 & 0.34726 &
0.016361\\
0.1 & 0.17741 & 0.002109 & 0.1 & 0.17712 & 0.002103 & 0.1 & 0.17682 &
0.002096\\\hline
\end{tabular}
\end{table}

\begin{table}
\caption{The relation between $dS$, $Q$ and $r_{h}$ for $d = 6$ in the
extended phase space via particle absorption.}%
\label{tab:dspha2}%
\begin{tabular}
[c]%
{p{0.6in}|p{0.6in}|p{0.6in}|p{0.6in}|p{0.6in}|p{0.6in}|p{0.6in}|p{0.6in}|p{0.6in}}%
\hline
\multicolumn{3}{c|}{$\alpha=0.01$} & \multicolumn{3}{|c|}{$\alpha=10$} &
\multicolumn{3}{|c}{$\alpha=20$}\\\hline
$Q$ & $r_{h}$ & $dS$ & $Q$ & $r_{h}$ & $dS$ & $Q$ & $r_{h}$ & $dS$\\\hline
0.751095 & 1.2638 & 1.339280 & 0.99 & 1.04990 & 0.778412 & 0.99 & 0.96842 &
0.689644\\
0.7 & 1.10288 & 0.631251 & 0.9 & 1.00783 & 0.585157 & 0.9 & 0.93329 &
0.506712\\
0.6 & 0.95369 & 0.316940 & 0.8 & 0.95630 & 0.422232 & 0.8 & 0.89028 &
0.361227\\
0.55 & 0.88937 & 0.232461 & 0.7 & 0.89852 & 0.297847 & 0.7 & 0.84200 &
0.254615\\
0.5 & 0.82727 & 0.170035 & 0.6 & 0.83281 & 0.201932 & 0.6 & 0.78685 &
0.174284\\
0.45 & 0.76594 & 0.122736 & 0.5 & 0.75694 & 0.128363 & 0.5 & 0.72250 &
0.113016\\
0.4 & 0.70437 & 0.086573 & 0.4 & 0.66796 & 0.073522 & 0.4 & 0.64545 &
0.066696\\
0.3 & 0.57726 & 0.038301 & 0.3 & 0.56214 & 0.035340 & 0.3 & 0.55061 &
0.033263\\
0.2 & 0.43871 & 0.012623 & 0.2 & 0.43457 & 0.012268 & 0.2 & 0.43085 &
0.011959\\
0.1 & 0.27580 & 0.001955 & 0.1 & 0.27538 & 0.001946 & 0.1 & 0.27496 &
0.001938\\\hline
\end{tabular}
\end{table}

\begin{table}
\caption{The relation between $dS$, $Q$ and $r_{h}$ for $d = 7$ in the
extended phase space via particle absorption.}%
\label{tab:dspha3}%
\begin{tabular}
[c]%
{p{0.6in}|p{0.6in}|p{0.6in}|p{0.6in}|p{0.6in}|p{0.6in}|p{0.6in}|p{0.6in}|p{0.6in}}%
\hline
\multicolumn{3}{c|}{$\alpha=0.01$} & \multicolumn{3}{|c|}{$\alpha=10$} &
\multicolumn{3}{|c}{$\alpha=20$}\\\hline
$Q$ & $r_{h}$ & $dS$ & $Q$ & $r_{h}$ & $dS$ & $Q$ & $r_{h}$ & $dS$\\\hline
0.818145 & 1.16885 & 1.078800 & 0.99 & 0.97520 & 0.441467 & 0.99 & 0.918038 &
0.366732\\
0.8 & 1.12844 & 0.833796 & 0.9 & 0.94455 & 0.348482 & 0.9 & 0.891935 &
0.289174\\
0.7 & 0.98534 & 0.351322 & 0.8 & 0.90670 & 0.263045 & 0.8 & 0.859722 &
0.219253\\
0.65 & 0.93833 & 0.264973 & 0.7 & 0.86385 & 0.192861 & 0.7 & 0.823226 &
0.823226\\
0.6 & 0.89360 & 0.20173 & 0.6 & 0.81459 & 0.135414 & 0.6 & 0.781100 &
0.116055\\
0.5 & 0.80578 & 0.115536 & 0.5 & 0.75695 & 0.089096 & 0.5 & 0.731328 &
0.078253\\
0.4 & 0.71509 & 0.061968 & 0.4 & 0.68812 & 0.053003 & 0.4 & 0.67080 &
0.048053\\
0.3 & 0.61620 & 0.028949 & 0.3 & 0.60401 & 0.026718 & 0.3 & 0.594569 &
0.025125\\
0.2 & 0.50162 & 0.010235 & 0.2 & 0.49782 & 0.009929 & 0.2 & 0.494408 &
0.009664\\
0.1 & 0.35416 & 0.001786 & 0.1 & 0.35368 & 0.001771 & 0.1 & 0.353200 &
0.001762\\\hline
\end{tabular}
\end{table}\begin{table}
\caption{The relation between $dS$, $Q$ and $r_{h}$ for $d = 8$ in the
extended phase space via particle absorption.}%
\label{tab:dspha4}%
\begin{tabular}
[c]%
{p{0.6in}|p{0.6in}|p{0.6in}|p{0.6in}|p{0.6in}|p{0.6in}|p{0.6in}|p{0.6in}|p{0.6in}}%
\hline
\multicolumn{3}{c|}{$\alpha=0.01$} & \multicolumn{3}{|c|}{$\alpha=10$} &
\multicolumn{3}{|c}{$\alpha=20$}\\\hline
$Q$ & $r_{h}$ & $dS$ & $Q$ & $r_{h}$ & $dS$ & $Q$ & $r_{h}$ & $dS$\\\hline
0.861699 & 1.11151 & 0.900437 & 0.99 & 0.941476 & 0.314099 & 0.99 & 0.89711 &
0.256923\\
0.8 & 1.01782 & 0.438211 & 0.9 & 0.91718 & 0.178334 & 0.9 & 0.87616 &
0.207588\\
0.75 & 0.97711 & 0.324878 & 0.8 & 0.88701 & 0.194437 & 0.8 & 0.85019 &
0.161139\\
0.7 & 0.94082 & 0.249153 & 0.7 & 0.85267 & 0.145112 & 0.7 & 0.82059 &
0.121929\\
0.6 & 0.87253 & 0.150415 & 0.6 & 0.81291 & 0.103680 & 0.6 & 0.78620 &
0.088816\\
0.5 & 0.80409 & 0.089039 & 0.5 & 0.76597 & 0.069489 & 0.5 & 0.74525 &
0.061066\\
0.4 & 0.73116 & 0.049179 & 0.4 & 0.70928 & 0.042241 & 0.4 & 0.69494 &
0.038315\\
0.3 & 0.64919 & 0.023723 & 0.3 & 0.63877 & 0.021899 & 0.3 & 0.63064 &
0.020586\\
0.2 & 0.55071 & 0.0087473 & 0.2 & 0.54719 & 0.008477 & 0.2 & 0.54403 &
0.008241\\
0.1 & 0.41685 & 0.0016326 & 0.1 & 0.41633 & 0.001623 & 0.1 & 0.41583 &
0.001613\\\hline
\end{tabular}
\end{table}\begin{figure}[tbh]
\begin{center}
\subfigure[{$d=5$ .}]{
\includegraphics[width=0.45\textwidth]{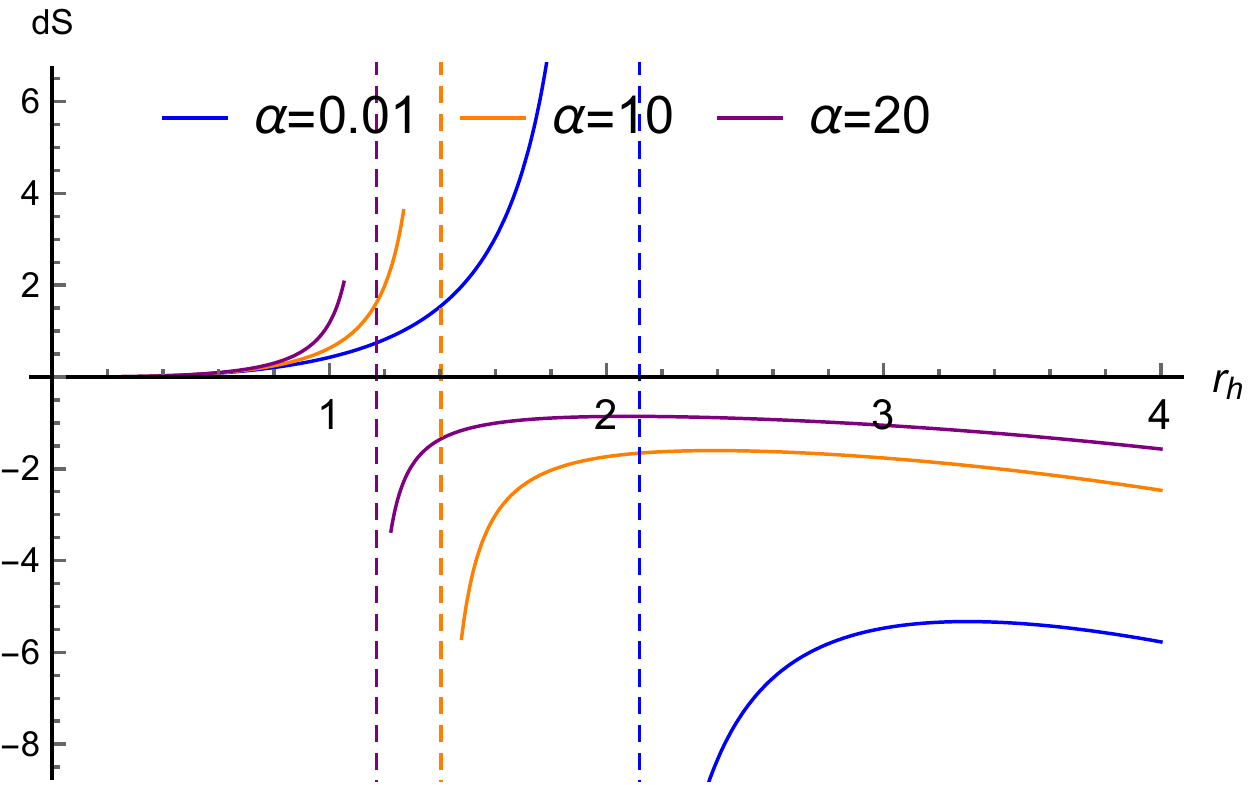}\label{ldsph5}}
\subfigure[{$d=6$ .}]{
\includegraphics[width=0.45\textwidth]{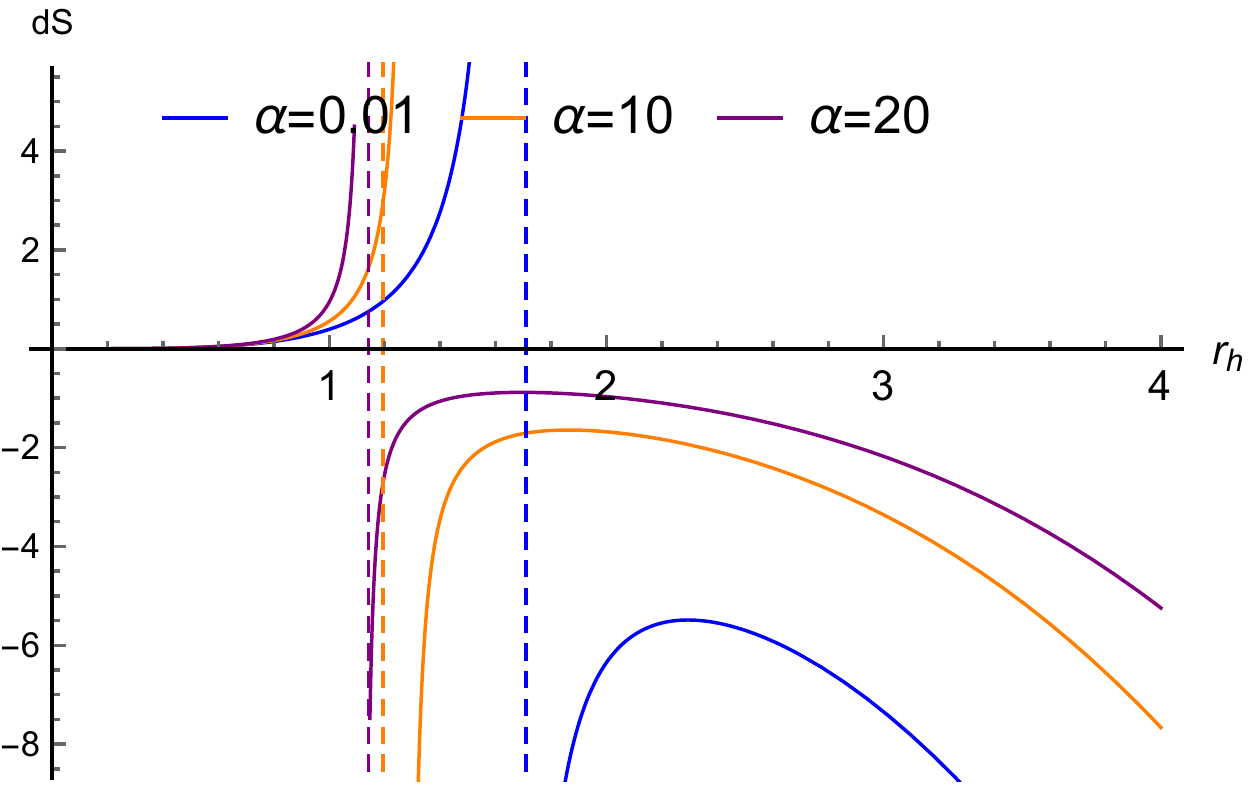}\label{ldsph6}}
\subfigure[{$d=7$ .}]{
\includegraphics[width=0.45\textwidth]{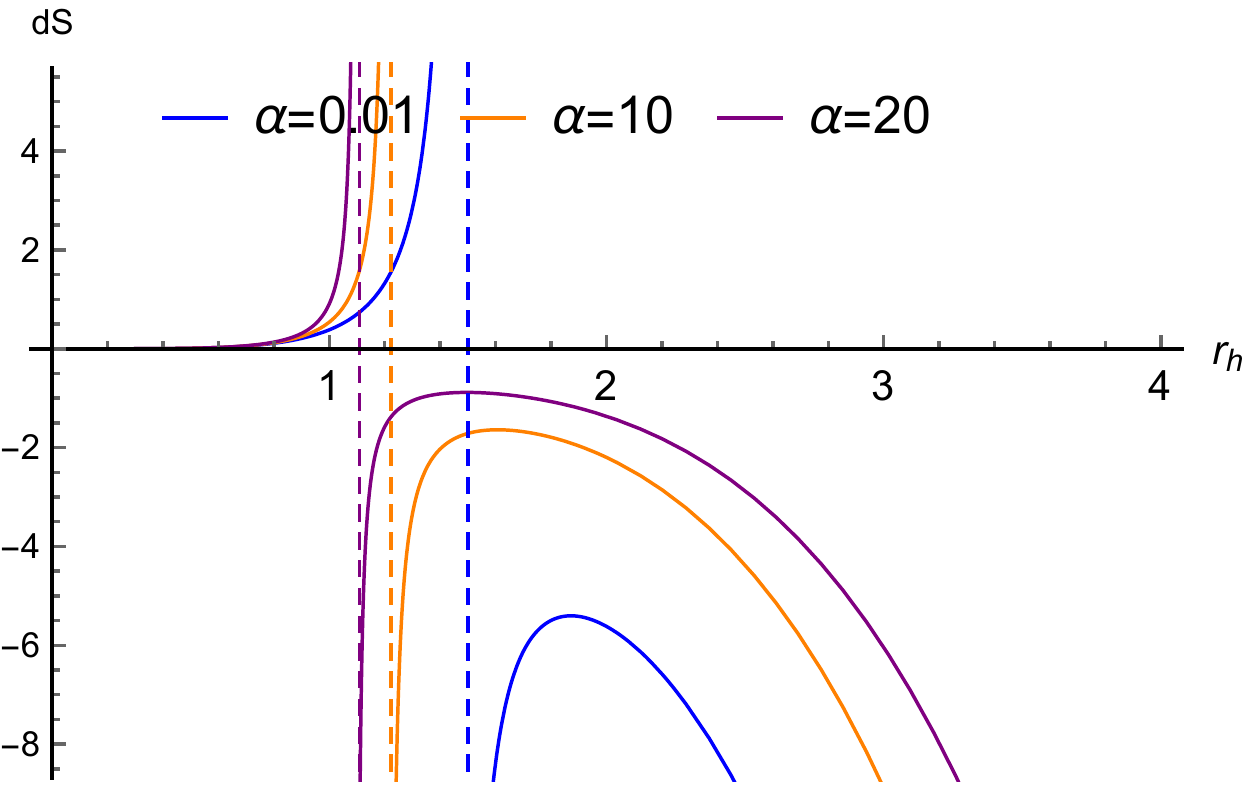}\label{ldsph7}}
\subfigure[{$d=8$ .}]{
\includegraphics[width=0.45\textwidth]{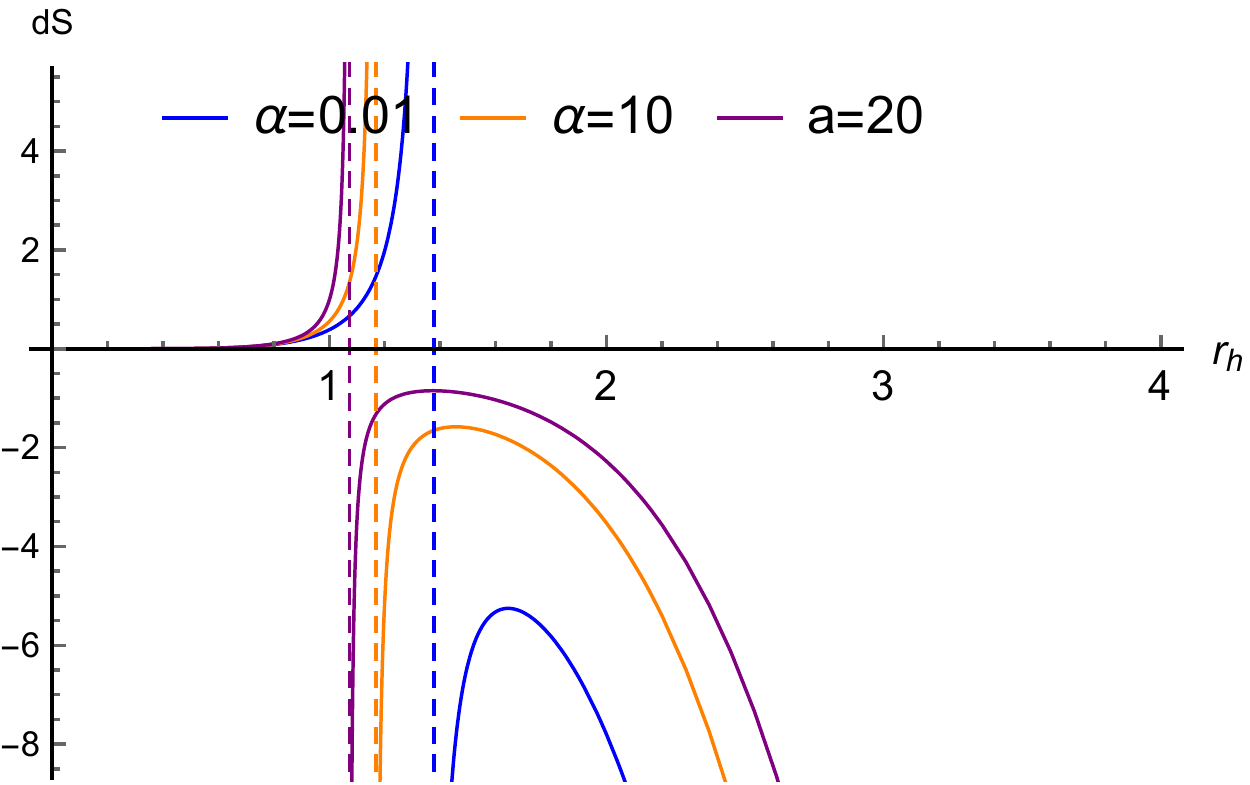}\label{ldsph8}}
\end{center}
\caption{The relationship between $dS$,$Q$ and $r_{h}$ for $\alpha=0.1,10,20$.
}%
\label{fig:ds2}%
\end{figure}

\begin{table}
\caption{The relation between $dS$, $Q$ and $r_{h}$ for $d = 5$ in the
extended phase space via particle absorption.}%
\label{tab:dsda1}%
\begin{tabular}
[c]%
{p{0.6in}|p{0.6in}|p{0.65in}|p{0.6in}|p{0.6in}|p{0.65in}|p{0.6in}|p{0.6in}|p{0.6in}}%
\hline
\multicolumn{3}{c|}{$d\alpha=0.6,da=0.9$} & \multicolumn{3}{|c|}{$d\alpha
=0.6,da=-0.9$} & \multicolumn{3}{|c}{$d\alpha=-0.6,da=0.9$}\\\hline
$Q$ & $r_{h}$ & $dS$ & $Q$ & $r_{h}$ & $dS$ & $Q$ & $r_{h}$ & $dS$\\\hline
0.796654 & 1.697710 & 5.7962600 & 0.796654 & 1.697710 & 5.2260700 & 0.796654 &
1.697710 & 4.152850\\
0.7 & 1.299390 & 1.1726300 & 0.7 & 1.299390 & 1.0756400 & 0.7 & 1.299390 &
1.008870\\
0.6 & 1.081420 & 0.5704530 & 0.6 & 1.081420 & 0.5296950 & 0.6 & 1.081420 &
0.522788\\
0.5 & 0.889426 & 0.2910120 & 0.5 & 0.889426 & 0.2734750 & 0.5 & 0.889426 &
0.277139\\
0.4 & 0.707405 & 0.1395270 & 0.4 & 0.707405 & 0.1327160 & 0.4 & 0.707405 &
0.136119\\
0.3 & 0.529490 & 0.0569941 & 0.3 & 0.529490 & 0.0548840 & 0.3 & 0.529490 &
0.056402\\
0.2 & 0.353027 & 0.0166699 & 0.2 & 0.353027 & 0.0162543 & 0.2 & 0.353027 &
0.016618\\
0.1 & 0.176765 & 0.0020799 & 0.1 & 0.176765 & 0.0020537 & 0.1 & 0.176765 &
0.002079\\\hline
\end{tabular}
\end{table}

\begin{table}
\caption{The relation between $dS$, $Q$ and $r_{h}$ for $d = 6$ in the
extended phase space via particle absorption.}%
\label{tab:dsda2}%
\begin{tabular}
[c]%
{p{0.6in}|p{0.6in}|p{0.65in}|p{0.6in}|p{0.6in}|p{0.65in}|p{0.6in}|p{0.6in}|p{0.6in}}%
\hline
\multicolumn{3}{c|}{$d\alpha=0.6,da=0.9$} & \multicolumn{3}{|c|}{$d\alpha
=0.6,da=-0.9$} & \multicolumn{3}{|c}{$d\alpha=-0.6,da=0.9$}\\\hline
$Q$ & $r_{h}$ & $dS$ & $Q$ & $r_{h}$ & $dS$ & $Q$ & $r_{h}$ & $dS$\\\hline
0.978391 & 1.465170 & 5.184920 & 0.978391 & 1.465170 & 4.757350 & 0.978391 &
1.465170 & 3.391810\\
0.9 & 1.267230 & 1.430560 & 0.9 & 1.267230 & 1.319450 & 0.9 & 1.267230 &
1.129080\\
0.8 & 1.137970 & 0.754340 & 0.8 & 1.137970 & 0.699487 & 0.8 & 1.137970 &
0.646563\\
0.7 & 1.025040 & 0.443271 & 0.7 & 1.025040 & 0.413400 & 0.7 & 1.025040 &
0.400375\\
0.6 & 0.916339 & 0.263653 & 0.6 & 0.916339 & 0.247428 & 0.6 & 0.916339 &
0.247008\\
0.5 & 0.806853 & 0.151338 & 0.5 & 0.806853 & 0.143002 & 0.5 & 0.806853 &
0.145500\\
0.4 & 0.693037 & 0.079978 & 0.4 & 0.693037 & 0.076146 & 0.4 & 0.693037 &
0.078278\\
0.3 & 0.571182 & 0.036246 & 0.3 & 0.571182 & 0.034801 & 0.3 & 0.571182 &
0.035887\\
0.2 & 0.435772 & 0.012159 & 0.2 & 0.435772 & 0.011786 & 0.2 & 0.435772 &
0.012118\\
0.1 & 0.274769 & 0.001913 & 0.1 & 0.274769 & 0.001875 & 0.1 & 0.274769 &
0.001916\\\hline
\end{tabular}
\end{table}

\begin{table}
\caption{The relation between $dS$, $Q$ and $r_{h}$ for $d = 7$ in the
extended phase space via particle absorption.}%
\label{tab:dsda3}%
\begin{tabular}
[c]%
{p{0.6in}|p{0.6in}|p{0.65in}|p{0.6in}|p{0.6in}|p{0.65in}|p{0.6in}|p{0.6in}|p{0.6in}}%
\hline
\multicolumn{3}{c|}{$d\alpha=0.6,da=0.9$} & \multicolumn{3}{|c|}{$d\alpha
=0.6,da=-0.9$} & \multicolumn{3}{|c}{$d\alpha=-0.6,da=0.9$}\\\hline
$Q$ & $r_{h}$ & $dS$ & $Q$ & $r_{h}$ & $dS$ & $Q$ & $r_{h}$ & $dS$\\\hline
1.09785 & 1.329700 & 4.421750 & 1.09785 & 1.329700 & 4.09676 & 1.09785 &
1.329700 & 2.728440\\
0.9 & 1.094730 & 0.684067 & 0.9 & 1.094730 & 0.636731 & 0.9 & 1.094730 &
0.570757\\
0.8 & 1.019380 & 0.429625 & 0.8 & 1.019380 & 0.401167 & 0.8 & 1.019380 &
0.378410\\
0.7 & 0.945735 & 0.274614 & 0.7 & 0.945735 & 0.257376 & 0.7 & 0.945735 &
0.251631\\
0.6 & 0.870696 & 0.172710 & 0.6 & 0.870696 & 0.162559 & 0.6 & 0.870696 &
0.162987\\
0.5 & 0.791846 & 0.103746 & 0.5 & 0.791846 & 0.098126 & 0.5 & 0.791846 &
0.100064\\
0.4 & 0.706553 & 0.057305 & 0.4 & 0.706553 & 0.054505 & 0.4 & 0.706553 &
0.056142\\
0.3 & 0.611095 & 0.027317 & 0.3 & 0.611095 & 0.026153 & 0.3 & 0.611095 &
0.027046\\
0.2 & 0.498774 & 0.009809 & 0.2 & 0.498774 & 0.009465 & 0.2 & 0.498774 &
0.009773\\
0.1 & 0.352893 & 0.001732 & 0.1 & 0.352893 & 0.001688 & 0.1 & 0.352893 &
0.001731\\\hline
\end{tabular}
\end{table}

\begin{table}
\caption{The relation between $dS$, $Q$ and $r_{h}$ for $d = 8$ in the
extended phase space via particle absorption.}%
\label{tab:dsda4}%
\begin{tabular}
[c]%
{p{0.6in}|p{0.6in}|p{0.65in}|p{0.6in}|p{0.6in}|p{0.65in}|p{0.6in}|p{0.6in}|p{0.6in}}%
\hline
\multicolumn{3}{c|}{$d\alpha=0.6,da=0.9$} & \multicolumn{3}{|c|}{$d\alpha
=0.6,da=-0.9$} & \multicolumn{3}{|c}{$d\alpha=-0.6,da=0.9$}\\\hline
$Q$ & $r_{h}$ & $dS$ & $Q$ & $r_{h}$ & $dS$ & $Q$ & $r_{h}$ & $dS$\\\hline
1.17861 & 1.244110 & 3.755540 & 1.17861 & 1.244110 & 3.500120 & 1.17861 &
1.244110 & 2.232980\\
0.99 & 1.075100 & 0.681459 & 0.99 & 1.075100 & 0.635796 & 0.99 & 1.075100 &
0.550287\\
0.9 & 1.024280 & 0.459031 & 0.9 & 1.024280 & 0.428949 & 0.9 & 1.024280 &
0.391199\\
0.8 & 0.969409 & 0.304303 & 0.8 & 0.969409 & 0.284994 & 0.8 & 0.969409 &
0.271241\\
0.7 & 0.913819 & 0.201539 & 0.7 & 0.913819 & 0.189263 & 0.7 & 0.913819 &
0.185893\\
0.6 & 0.855703 & 0.130299 & 0.6 & 0.855703 & 0.122757 & 0.6 & 0.855703 &
0.123379\\
0.5 & 0.793252 & 0.080237 & 0.5 & 0.793252 & 0.075880 & 0.5 & 0.793252 &
0.077500\\
0.4 & 0.724135 & 0.045465 & 0.4 & 0.724135 & 0.043189 & 0.4 & 0.724135 &
0.044559\\
0.3 & 0.644708 & 0.022331 & 0.3 & 0.644708 & 0.021328 & 0.3 & 0.644708 &
0.022108\\
0.2 & 0.547972 & 0.008357 & 0.2 & 0.547972 & 0.008029 & 0.2 & 0.547972 &
0.008319\\
0.1 & 0.415445 & 0.001579 & 0.1 & 0.415445 & 0.001533 & 0.1 & 0.415445 &
0.001578\\\hline
\end{tabular}
\end{table}\begin{figure}[tbh]
\begin{center}
\subfigure[{$d=5$ .}]{
\includegraphics[width=0.45\textwidth]{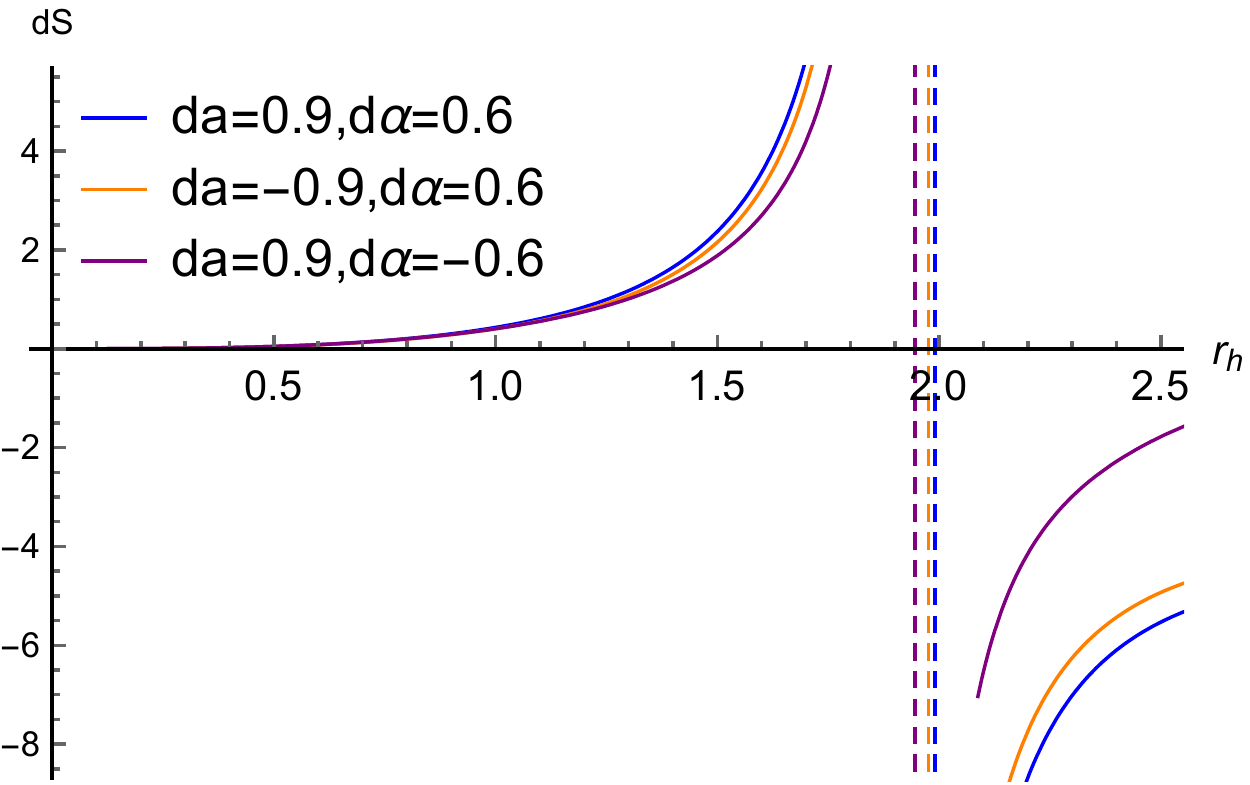}\label{ldsda5}}
\subfigure[{$d=6$ .}]{
\includegraphics[width=0.45\textwidth]{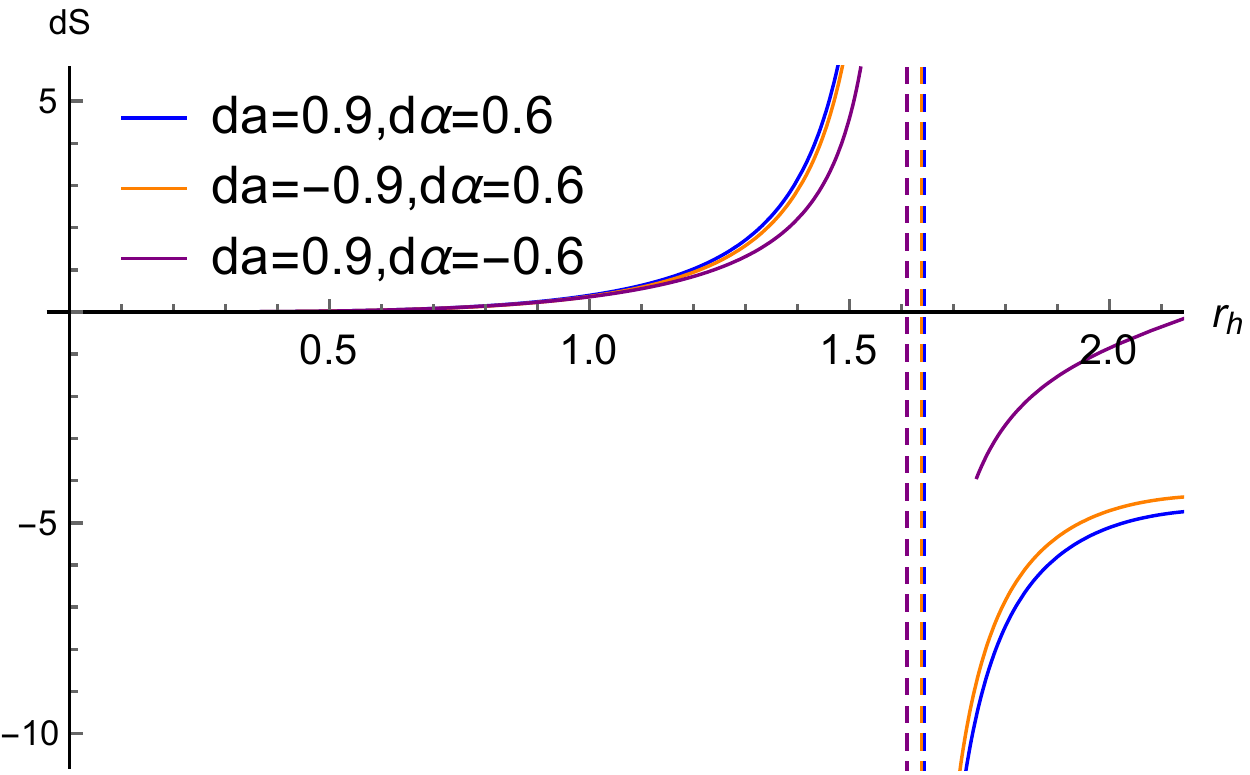}\label{ldsda6}}
\subfigure[{$d=7$ .}]{
\includegraphics[width=0.45\textwidth]{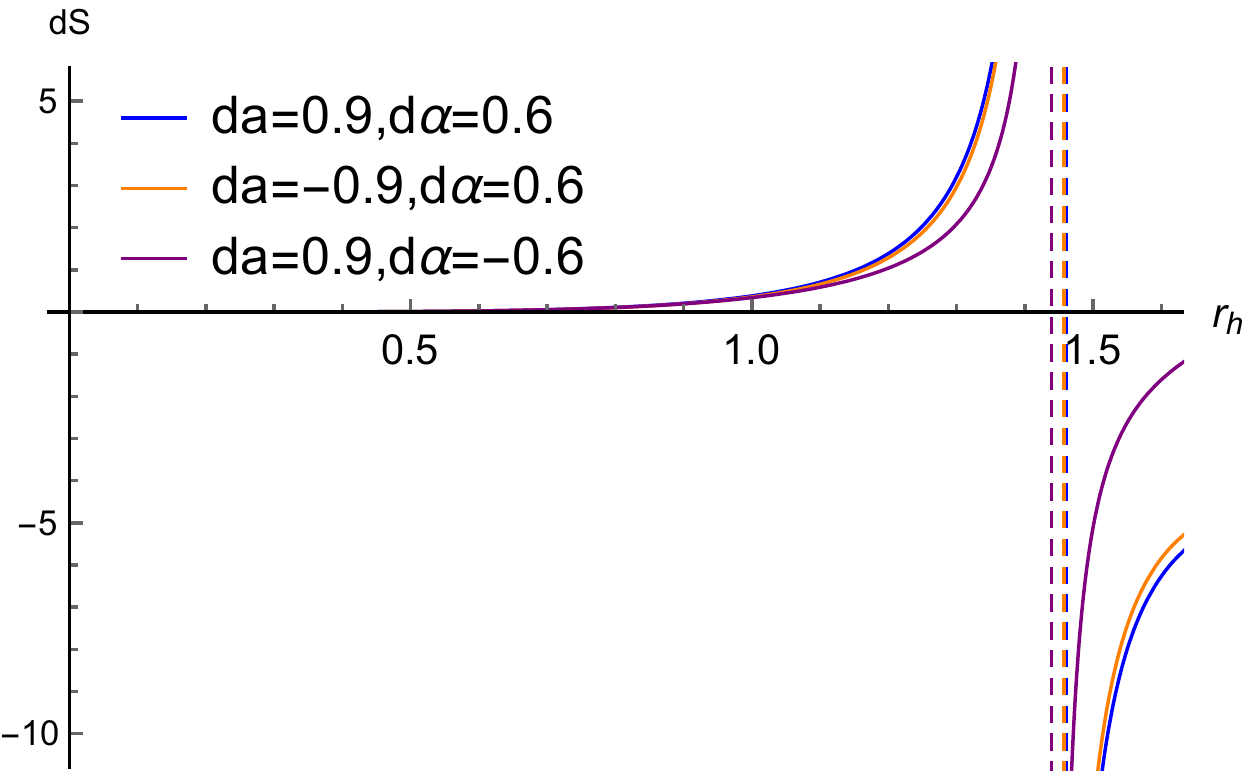}\label{ldsda7}}
\subfigure[{$d=8$ .}]{
\includegraphics[width=0.45\textwidth]{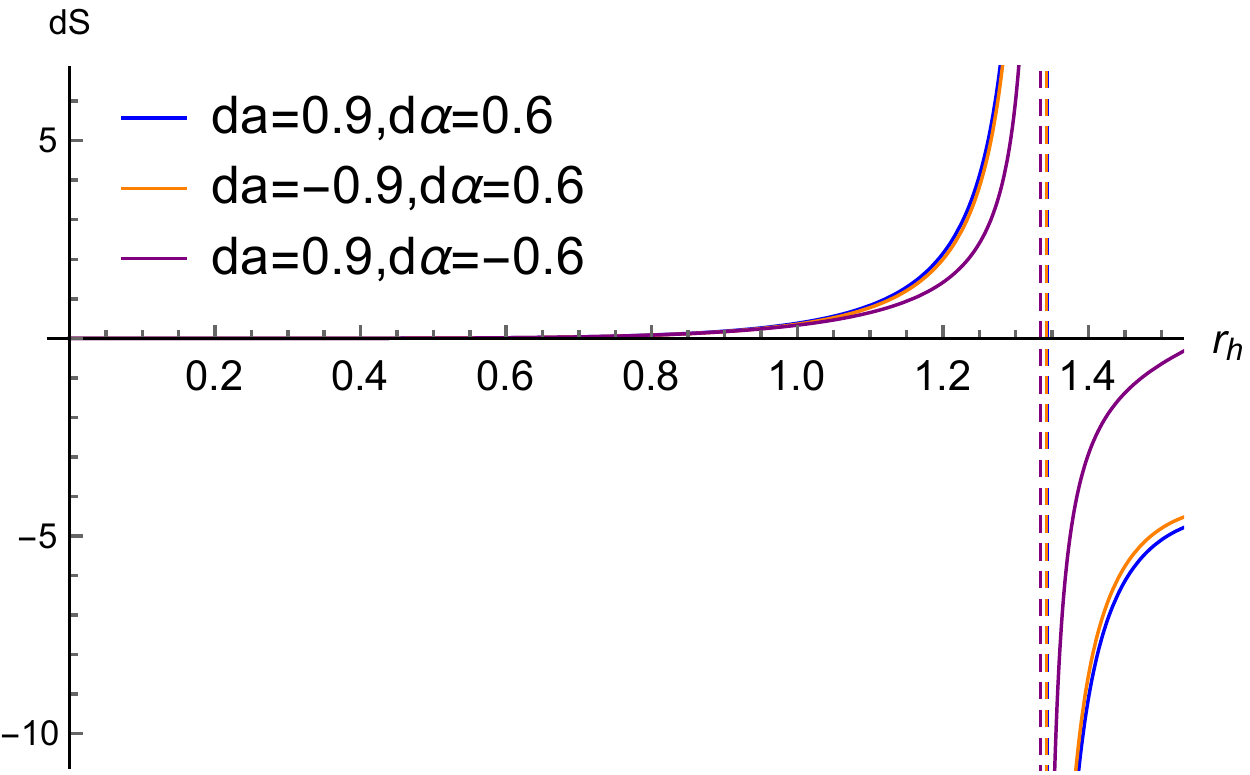}\label{ldsda8}}
\end{center}
\caption{The relationship between $dS$,$Q$ and $r_{h}$ for $da$ and $d\alpha$.
}%
\label{fig:ds3}%
\end{figure}

\begin{figure}[tbh]
\begin{center}
\subfigure[{$d=5$ .}]{
\includegraphics[width=0.45\textwidth]{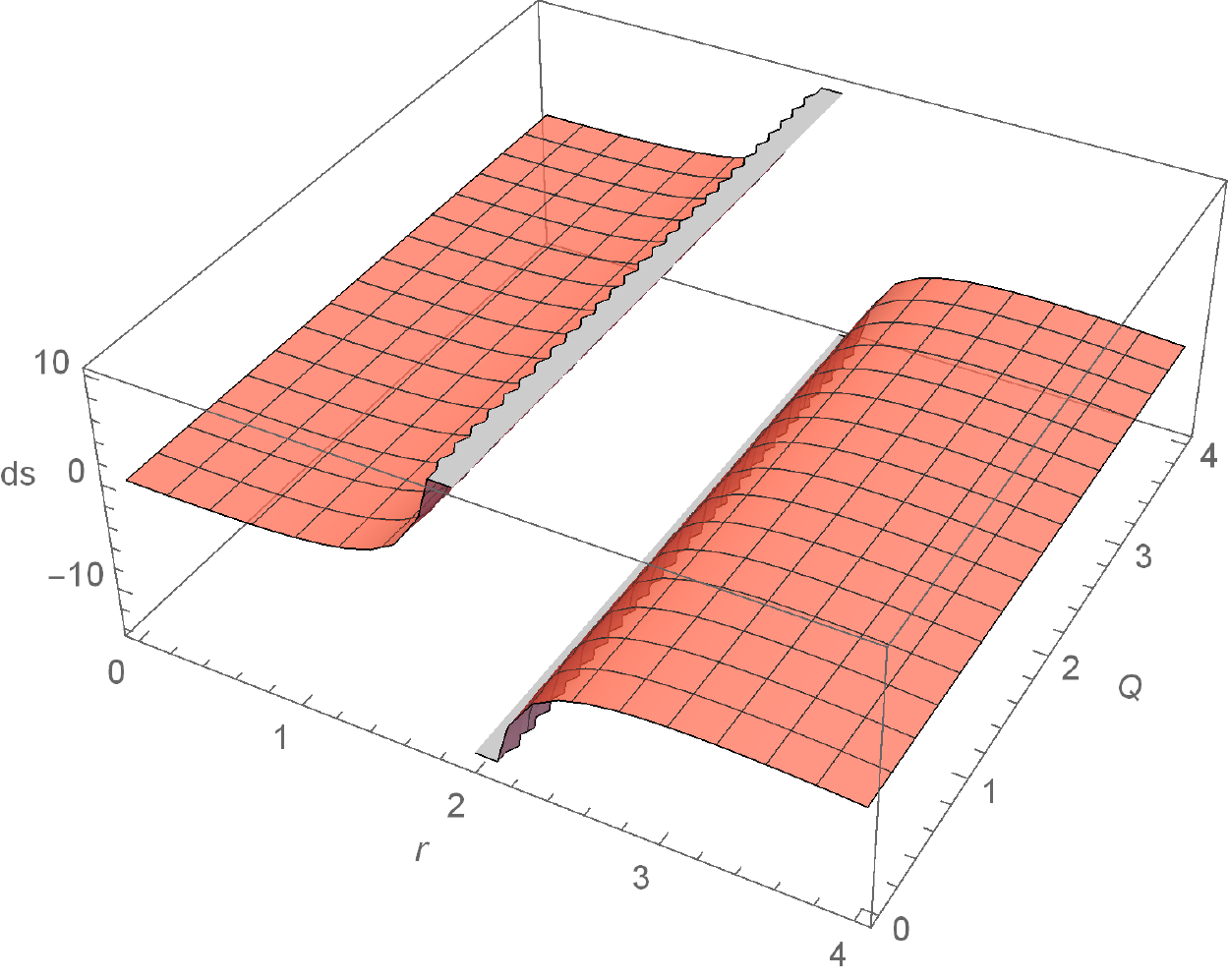}\label{fig:3Dlds5}}
\subfigure[{$d=6$ .}]{
\includegraphics[width=0.45\textwidth]{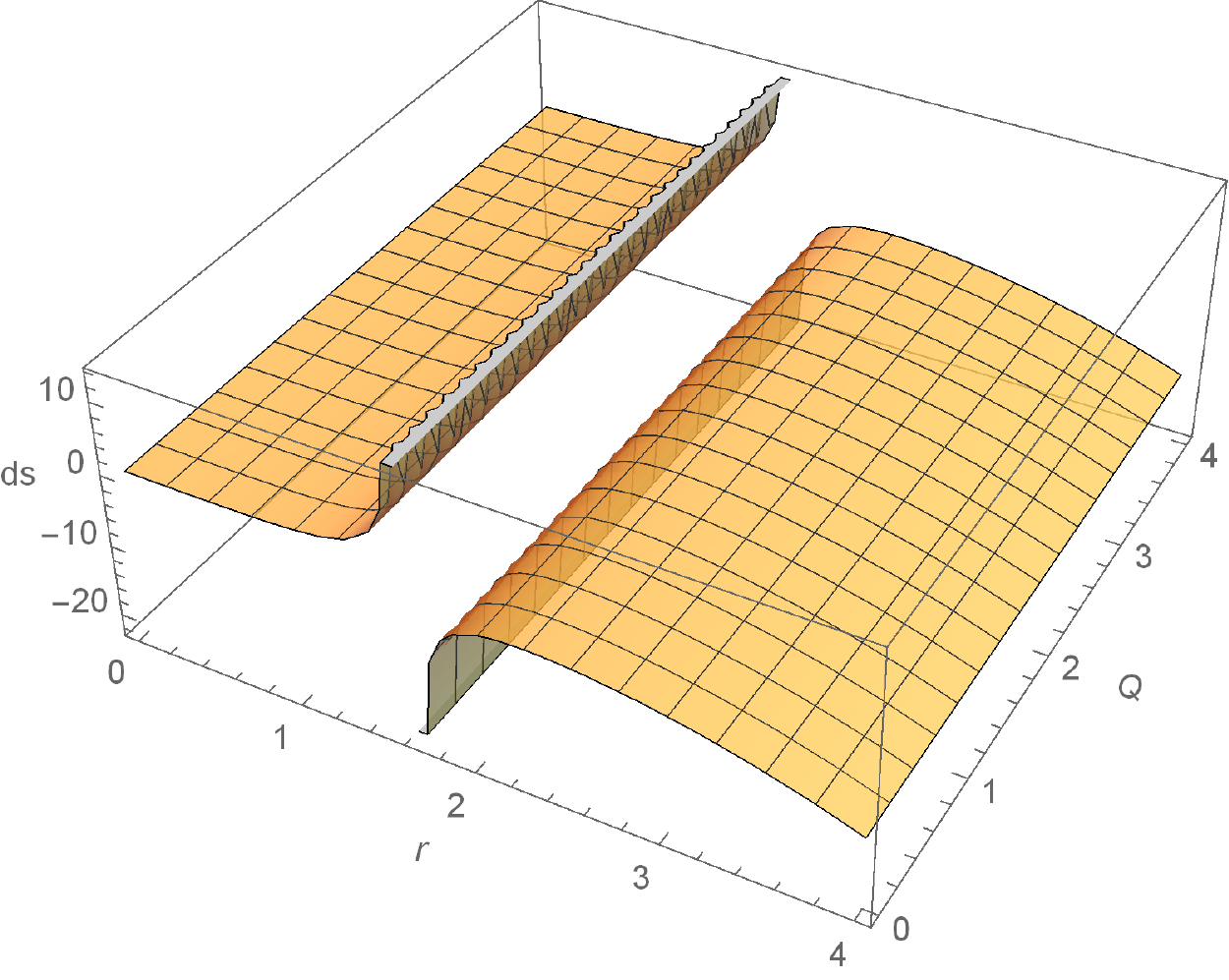}\label{fig:3Dlds6}}
\subfigure[{$d=7$ .}]{
\includegraphics[width=0.45\textwidth]{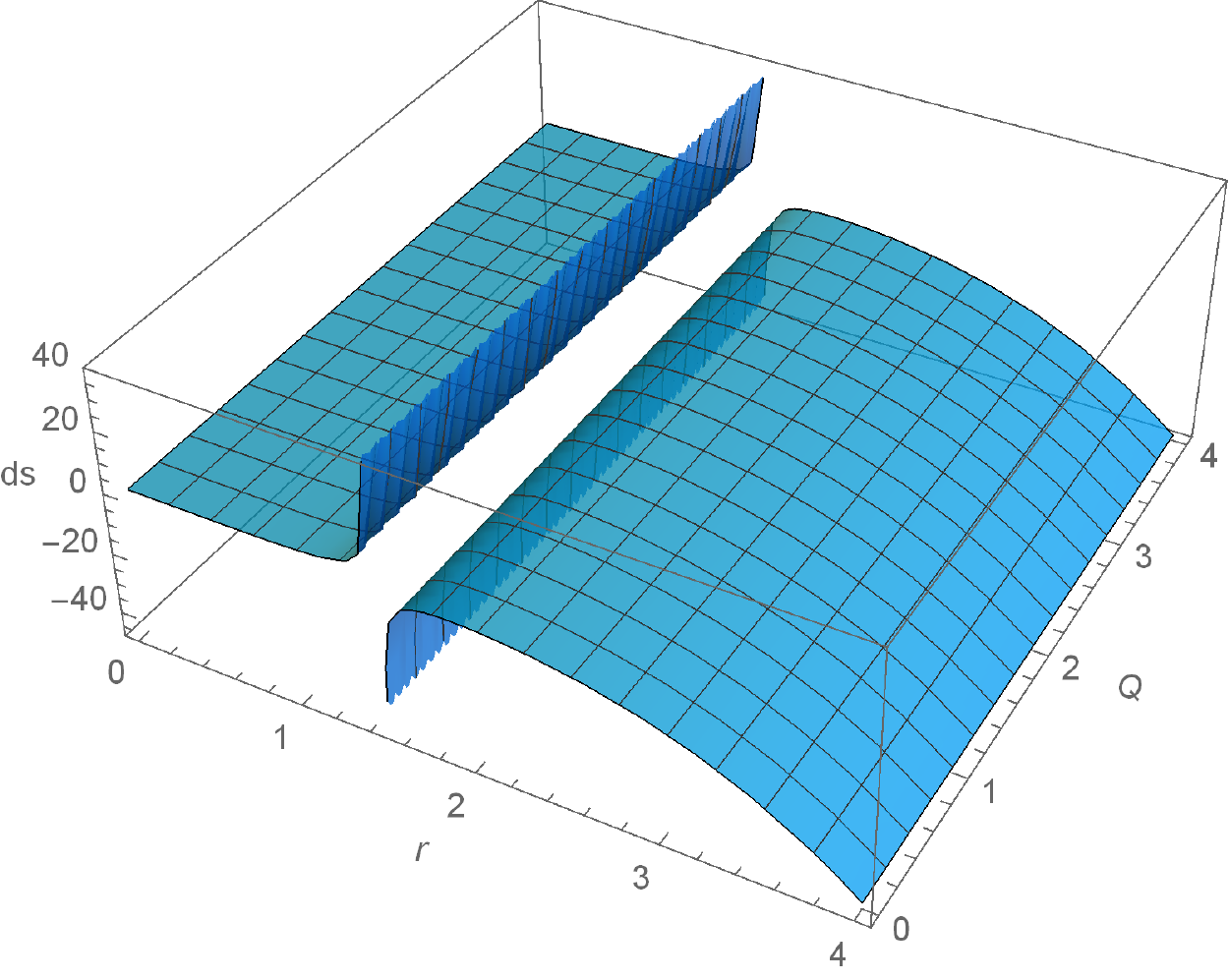}\label{fig:3Dlds7}}
\subfigure[{$d=8$ .}]{
\includegraphics[width=0.45\textwidth]{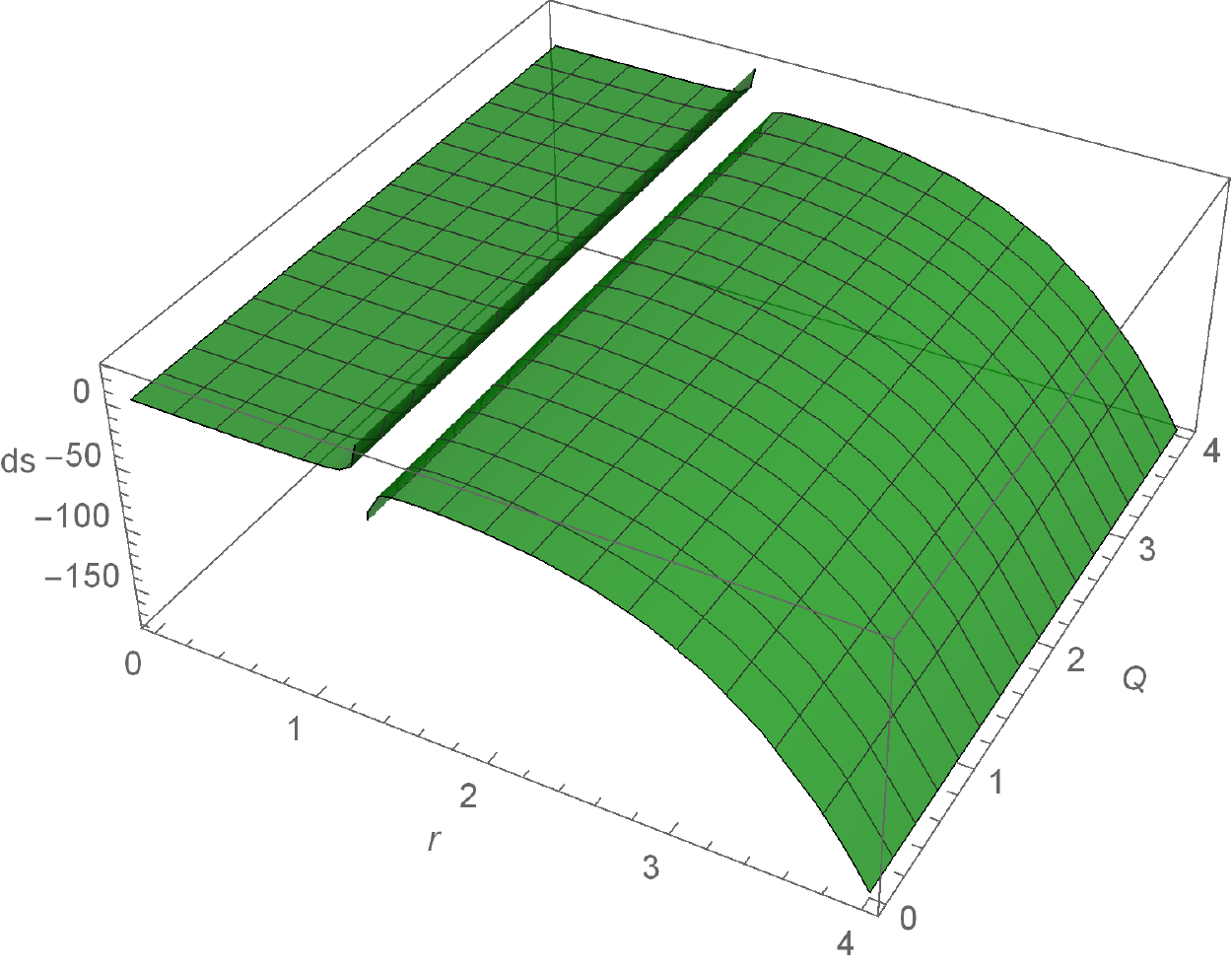}\label{fig:3Dlds8}}
\end{center}
\caption{The relationship between $dS$,$Q$ and $r_{h}$. }%
\label{fig:ds4}%
\end{figure}

\begin{figure}[tbh]
\begin{center}
\subfigure[{$d=4$ .}]{
\includegraphics[width=0.45\textwidth]{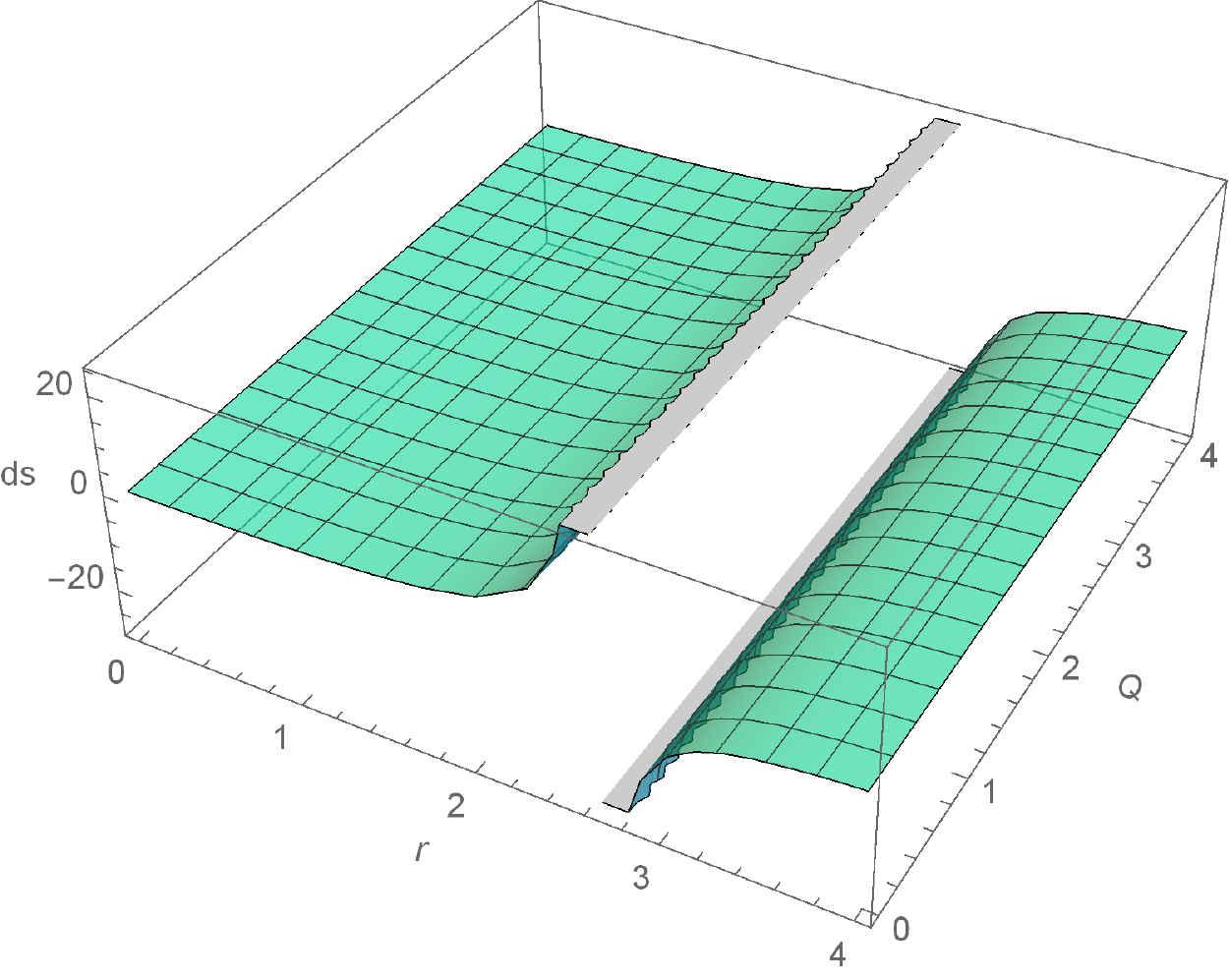}\label{fig:3Dlds4}}
\end{center}
\caption{The relationship between $dS$,$Q$ and $r_{h}$. }%
\label{fig:ds5}%
\end{figure}In this section, we will discuss the thermodynamics-related issues
of a $d$-dimensional charged AdS black hole surrounded by quintessence and
cloud of strings. As usual, we consider the cosmological constant as the
dynamical pressure of a black hole.
\begin{equation}
P=\frac{-\Lambda}{8\pi}=\frac{(d-1)(d-2)}{16\pi l^{2}}.\label{eqn:pp1}%
\end{equation}
The Hawking temperature of the black hole is expressed as
\begin{equation}
\begin{aligned} &T=\frac{f(r)^{'}}{4\pi}\mid_{r=r_{h}}=\frac{m(d-3)}{4\pi r_{h}^{d-2}}+\frac{q^{2}(3-d)}{2\pi r_{h}^{2d-5}}+\frac{8Pr_{h}}{(d-2)(d-1)}\\ &+\frac{[(d-1)\omega_{q}+d-3]\alpha}{4\pi r_{h}^{(d-1)\omega_{q}+d-2}}+\frac{(d-4)a}{2\pi(d-2)r_{h}^{d-3}}.\\ \end{aligned}
\end{equation}
With the help of the Bekenstein-Hawking formula \cite{Hawking:1974sw}, Entropy
can be obtained
\begin{equation}
S=\frac{A_{d-2}}{4}=\frac{\Omega_{d-2}}{4}r_{h}^{d-2}.
\end{equation}

After the black hole absorbs a particle, the change in the enthalpy is
connected to the change in internal energy as
\begin{equation}
E=dU=d(M-PV).
\end{equation}
with
\begin{equation}
dU=dM-PdV-VdP=\frac{8\pi Q}{2(d-3)r_{h}^{d-3}\varOmega_{d-2}}dQ+|p^{r}%
|.\label{eqn:du1}%
\end{equation}
The initial state of the black hole is represented by $(M,Q,P,a,\alpha,r_{h}%
)$, and the final state is represented by $(M+dM,Q+dQ,P+dP,a+da,\alpha
+d\alpha,r_{h}+dr_{h})$. The functions $f(M,Q,P,a,\alpha,r_{h})$ and
$f(M+dM,Q+dQ,P+dP,a+da,\alpha+d\alpha,r_{h}+dr_{h})$ satisfy
\begin{equation}
f(M,Q,P,a,\alpha,r_{h})=f\left( M+dM,Q+dQ,P+dP,a+da,\alpha+d\alpha
,r_{h}+dr_{h}\right) =0.
\end{equation}
The relation between the functions $f(M,Q,P,a,\alpha, r_{h})$ and
$f(M+dM,Q+dQ,P+dP,a+da,\alpha+d\alpha, r_{h}+dr_{h})$is
\begin{equation}
\begin{aligned} &f\left(M+dM,Q+dQ,P+dP,a+da,\alpha+d\alpha,r_{h}+dr_{h}\right)=f(M,Q,P,a,\alpha, r_{h})\\ &+\frac{\partial f}{\partial M}|_{r=r_{h}}dM+\frac{\partial f}{\partial Q}|_{r=r_{h}}dQ+\frac{\partial f}{\partial r}|_{r=r_{h}}dr_{h}+\frac{\partial f}{\partial P}|_{r=r_{h}}dP+\frac{\partial f}{\partial a}|_{r=r_{h}}da+\frac{\partial f}{\partial\alpha}|_{r=r_{h}}d\text{\ensuremath{\alpha}}.\\ \end{aligned}\label{eqn:ff1}%
\end{equation}
Where
\begin{equation}
\begin{aligned} &\frac{\partial f}{\partial M}|_{r=r_{h}}=-\frac{16\pi}{r_{h}^{d-3}(d-2)\Omega_{d-2}},\\ &\frac{\partial f}{\partial Q}|_{r=r_{h}}=\frac{16\pi q}{r_{h}^{2\text{(}d-3)}\Omega_{d-2}\sqrt{2(d-3)(d-2)}},\\ &\frac{\partial f}{\partial P}|_{r=r_{h}}=\frac{16\text{\ensuremath{\pi}}r_{h}^{2}}{(d-2)(d-1)},\\ &\frac{\partial f}{\partial r}|_{r=r_{h}}=4\pi T,\\ &\frac{\partial f}{\partial\alpha}|_{r=r_{h}}=-\frac{1}{r_{h}^{(d-1)\omega_{q}+d-3}},\\ &\frac{\partial f}{\partial a}|_{r=r_{h}}=\frac{-2}{(d-2)r^{d-4}}.\\ \end{aligned}
\end{equation}
Combining Eq. $\left( \ref{eqn:du1}\right) $ with Eq. $\left( \ref{eqn:ff1}%
\right) $, we get
\begin{equation}
dr_{h}=\frac{\frac{2}{(d-2)r_{h}^{d-4}}da+\frac{1}{r_{h}^{(d-1)\omega_{q}%
+d-3}}d\alpha+\frac{16\pi}{r_{h}^{d-3}(d-2)\varOmega_{d-2}}|p^{r}|}{4\pi
T-\frac{16\pi Pr_{h}}{d-2}}.
\end{equation}
Then the variations of entropy and thermodynamic volume of the black hole are
obtained as
\begin{equation}
dS=\frac{\frac{\Omega_{d-2}r_{h}}{2}da+\frac{\Omega_{d-2}(d-2)}%
{4r^{(d-1)\omega_{q}}}d\alpha+4\pi|p^{r}|}{4\pi T-\frac{16\pi Pr_{h}}{d-2}%
},\label{eqn:ds}%
\end{equation}
and
\begin{equation}
dV=\frac{\frac{2\varOmega_{d-2}r_{h}^{2}}{(d-2)}da+\frac{\varOmega_{d-2}%
}{r_{h}^{(d-1)\omega_{q}-1}}d\alpha+\frac{16\pi r_{h}}{(d-2)}|p^{r}|}{4\pi
T-\frac{16\pi Pr_{h}}{d-2}}.\label{eqn:dv}%
\end{equation}
Using Eqs. $\left( \ref{eqn:ds}\right) $ and $\left( \ref{eqn:dv}\right) $
yields
\begin{equation}
TdS-PdV=\frac{\frac{T\Omega_{d-2}r_{h}}{2}-\frac{2P\varOmega_{d-2}r_{h}^{2}}{(d-2)}}{4\pi T-\frac{16\pi Pr_{h}}{d-2}}da+\frac{\frac{T\Omega_{d-2}(d-2)}{4r_{h}^{(d-1)\omega_{q}}}-\frac{P\varOmega_{d-2}}{r_{h}^{(d-1)\omega_{q}-1}}}{4\pi T-\frac{16\pi Pr_{h}}{d-2}}d\alpha+\frac{4T\pi-\frac{16\pi r_{h}P}{(d-2)}}{4\pi T-\frac{16\pi Pr_{h}}{d-2}}|p^{r}|.
\end{equation}
The generalized first law in the extended phase space which account for the
cosmological constant effect, the cloud of strings and the quintessence
contributions is then expressed as
\begin{equation}
dM=TdS+VdP+\phi dQ+\mathcal{A}da+\mathcal{Q}d\alpha.\label{eqn:fcc}%
\end{equation}
Where $\mathcal{Q}$ and $\mathcal{A}$ are the physical quantity conjugated to
the parameter $\alpha$ and $a$ respectively, they satisfy
\begin{equation}
\mathcal{Q}=\left( \frac{\partial M}{\partial\alpha}\right) _{S,P}%
=\frac{(2-d)\Omega_{d-2}}{16\pi r_{h}^{(d-1)\omega_{q}}},\mathcal{A}=\left(
\frac{\partial M}{\partial a}\right) _{S,P}=\frac{-\varOmega_{d-2}r_{h}}{8\pi
}.
\end{equation}

According to the above, the first law of thermodynamics proved to be
satisfied. However, the effectiveness of the first law does not mean that the
second law is also effective. The second law of thermodynamics needs to be
tested in extended phase space, which states that the entropy of the black
hole never decreases. In other words, as the particle is absorbed, the entropy
of the final state is always greater than the initial state according to the
second law of thermodynamics.

When it is the extremal black hole, the temperature is zero. Then Eq. $\left(
\ref{eqn:ds}\right) $ is modified as
\begin{equation}
dS=\frac{\frac{\Omega_{d-2}r_{h}}{2}da+\frac{\Omega_{d-2}(d-2)}{4r_{h}^{(d-1)\omega_{q}}}d\alpha+4\pi|p^{r}|}{-\frac{16\pi Pr_{h}}{d-2}}.
\end{equation}
It is negative, which means that the second law is invalid for the extremal
black hole. Next, we focus on investigating the non-extremal black hole by
analyzing Eq. $\left( \ref{eqn:ds}\right) $ numerically to represent the
changes of entropy intuitively. We set $M=1,\Omega_{d-2}=1,\mid p^{r}%
\mid=1,l=1$ to discuss the influence of other parameters on the change of
entropy for given values of $d$. First, the object of our explore is the
behaviour of the function $\left( \ref{eqn:ds}\right) $, for different values
of $a$ in the case of $da=0.9,d\alpha=0.6,\alpha=0.01$, which are represented
by Fig. \ref{fig:ds1} and Table \ref{tab:dsa1}, Table \ref{tab:dsa2}, Table
\ref{tab:dsa3} and Table \ref{tab:dsa4}. When the charge is less than the
extremal charge, it can be obtained that the event horizon of the black hole
and the variation of entropy decreases when the charge of the black hole
decreases. While for $dS$, there is a divergent point, which divides the
variation of entropy into the positive and negative region. We also find that
as the values of $a$ decrease, the values of the critical horizon where $dS$
is divergent become smaller. And as the values of $d$ decrease, the values of
the divergent point become greater. So the second law of thermodynamics is
violated in extended phase space. This conclusion is independent of the values
of $d$ and $a$.

We also can set $a=0.01$ investigate how $d$ and $\alpha$ affect the values of
$dS$. For different values of $\alpha$ and $d$, the Eq. $\left( \ref{eqn:ds}\right) $ is represented in Fig.
\ref{fig:ds2} and Table \ref{tab:dspha1}, Table \ref{tab:dspha2}, Table
\ref{tab:dspha3} and Table \ref{tab:dspha4}. From these tables, it can be seen
that the event horizon of the black hole and the variation of entropy decrease
when the charge of the black hole decreases. From Figs above, it is evident
that there exists a phase transition point that divides the value of $dS$ into
positive and negative regions. The values of the divergent point decreases as
$d$ increases. At the same time, the value of divergence point also decreases
with the increase of $\alpha$. The invalidity of the second law for the
near-extremal black holes thus is universal, independent of the values of
$\alpha$ and $d$.

From Ref. \cite{Liang:2020uul} we know that the value of the state parameter
of the cloud of strings or quintessence affects the second law of
thermodynamics. Still, the parameters do not determine whether the second law
of thermodynamics is ultimately violated, which is consistent with our conclusion.

In fact, the relation between $dS$ and $r_{h}$ also can be effected by $da$
and $d\alpha$. We fix $a=0.001$ and $\alpha=0.001$ to investigate entropy in
different dimensions. From Fig. \ref{fig:ds3}, there is a phase transition
point which divides $dS$ into two branches. By comparing the data in the Table
\ref{tab:dsda1}, Table \ref{tab:dsda2}, Table \ref{tab:dsda3} and Table
\ref{tab:dsda4}, we find that $d\alpha$ has more obvious influence on the
change of entropy. While for the values of the divergent point, it decreases
as $d$ increases.  In order to explore the difference of entropy change in
high and low dimensional cases, the function graph is used to express the
relationship between $dS$, $Q$ and $r_{h}$ in different situations, which is
shown in Fig. $\left( \ref{fig:ds4}\right) $ and Fig. $\left( \ref{fig:ds5}%
\right) $. It is clear that there is indeed a phase change point that divides
$dS$ into positive and negative values. This conclusion is independent of
dimension $d$. From the above discussion, it can be concluded that the second
law of thermodynamics is not always valid for near-extremal black holes in the
extended phase space.

\subsection{Stability of horizon}

\label{sec:Bc}

In this section, we consider whether the horizons continue exist in the final
state because the horizons are significant in defining a black hole. The outer
horizon not only divides the inside and outside of the black hole, but is also
the location where the thermodynamic variables are defined. We will examine
the stability of horizon. The mass and charge of the black hole will change
after absorbing particles, which will lead to changes in the field of horizon
inside and outside the black hole. The event horizon is determined by the
metric component $f(r)$. If the minimal value of $f(r)$ is negative or zero,
the horizon exists. Otherwise, the horizon does not exist.

The sign of the minimum value in the final of $f(r)$ state can be obtained in
term of the initial state. Assuming $(M,Q,P,r_{0},a,\alpha)$ and
$(M+dM,Q+dQ,P+dP,r_{0}+dr_{0},a+da,\alpha+d\alpha)$ represent the initial
state and the finial state, respectively. At $r=r_{0}+dr_{0}$,
$f(M+dM,Q+dQ,P+dP,r_{0}+dr_{0},a+da,\alpha+d\alpha)$ is written as
\begin{equation}
\begin{aligned} &f\left(M+dM,Q+dQ,P+dP,a+da,\alpha+d\text{\ensuremath{\alpha}},dr_{0}+r_{0}\right)\\ &=\delta+\frac{\partial f}{\partial M}|_{r=r_{0}}dM+\frac{\partial f}{\partial Q}|_{r=r_{0}}dQ+\frac{\partial f}{\partial P}|_{r=r_{0}}dP\\ &+\frac{\partial f}{\partial a}|_{r=r_{0}}da+\frac{\partial f}{\partial\alpha}|_{r=r_{0}}d\text{\ensuremath{\alpha}}+\frac{\partial f}{\partial r}|_{r=r_{0}}dr,\\ \end{aligned}\label{eqn:wc1}%
\end{equation}
where
\begin{equation}
\begin{aligned} &\frac{\partial f}{\partial r}|_{r=r_{0}}=0,\\ &\frac{\partial f}{\partial M}|_{r=r_{o}}=-\frac{16\pi}{r_{0}^{d-3}(d-2)\Omega_{d-2}},\\ &\frac{\partial f}{\partial Q}|_{r=r_{0}}=\frac{16\pi q}{r_{0}^{2\text{(}d-3)}\Omega_{d-2}\sqrt{2(d-3)(d-2)}}, \\&\frac{\partial f}{\partial P}|_{r=r_{0}}=\frac{16\text{\ensuremath{\pi}}r_{0}^{2}}{(d-2)(d-1)},\\ &\frac{\partial f}{\partial\alpha}|_{r=r_{o}}=-\frac{1}{r_{0}^{(d-1)\omega_{q}+d-3}},\\ &\frac{\partial f}{\partial a}|_{r=r_{0}}=\frac{-2}{(d-2)r_{0}^{d-4}}.\\ \end{aligned}\label{eqn:wc2}%
\end{equation}
Therefore, we have
\begin{equation}
f\left( M,Q,P,a,\alpha,r_{0}\right) \equiv f_{0}=\delta\leq0,\label{eqn:wc3}%
\end{equation}
and
\begin{equation}
\partial_{r}f\left( M,Q,P,a,\alpha,r_{0}\right) \equiv f_{min}^{\prime
}=0.\label{eqn:wc4}%
\end{equation}
From Eqs.$\left( \ref{eqn:wc1}\right) $, $\left( \ref{eqn:wc2}\right) $,
$\left( \ref{eqn:wc3}\right) $ and $\left( \ref{eqn:wc4}\right) $ we obtain
\begin{equation}
\begin{aligned} &f\left(M+dM,Q+dQ,P+dP,a+da,\alpha+d\text{\ensuremath{\alpha}},r_{0}+dr_{0}\right)=\\ &\delta+\frac{16\pi q}{\varOmega_{d-2}r_{0}^{d-3}\sqrt{2(d-2)(d-3)}}(\frac{1}{r_{0}^{d-3}}-\frac{1}{r_{h}^{d-3}})dQ\\ &-\frac{16\pi}{r_{0}^{d-3}\Omega_{d-2}}[\frac{16\pi P r_{h}}{(d-2)(4\pi T-\frac{16\pi Pr_{h}}{d-2})}+1]|p^{r}|\\ &-\frac{2}{(d-2)r_{0}^{d-3}}[r_{0}+\frac{16\pi Pr_{h}^{2}}{(d-2)4\pi T-16\pi Pr_{h})}]da\\ &-r_{0}^{3-d}[\frac{16\pi Pr_{h}}{r_{h}^{(d-1)\omega_{q}}[4\pi T(d-2)-16\pi Pr_{h}]}+\frac{1}{r_{0}^{(d-1)\omega_{q}}}]d\alpha.\\ \end{aligned}\label{eqn:wc5}%
\end{equation}
When the initial black hole is the extremal black hole, $r_{0}=r_{h}$, $T=0$
and $\delta=0$. Then we can obtain $f_{min}=\delta=0$ and $f_{min}^{\prime}%
=0$. Hence, Eq. $\left( \ref{eqn:wc5}\right) $ is written as
\begin{equation}
f\left( M+dM,Q+dQ,P+dP,a+da,\alpha+d\text{$\alpha$},dr_{0}+r_{0}\right)
=0.\label{eqn:wc6}%
\end{equation}
This implies that the horizon of the extremal black hole is still exists at
the final state. When the initial black hole is the near-extremal black hole,
$r_{0}$ and $r_{h}$ do not coincide. Two locations $(r_{h}, r_{0})$ are very
close for the near-extremal black holes. Thus, we assume the condition
$r_{h}=r_{0}+\epsilon$, Using this condition, the Eq. $\left( \ref{eqn:wc5}%
\right) $ can be expand at the location $r_{0}$. To the first order, it
yields
\begin{equation}
\begin{aligned} &f\left(M+dM,Q+dQ,P+dP,a+da,\alpha+d\text{\ensuremath{\alpha}},r_{0}+dr_{0}\right)=\\ &\delta+\frac{16\pi q}{\varOmega_{d-2}\sqrt{2(d-2)(d-3)}}\frac{1}{r_{0}^{d-3}}[\frac{(d-3)\epsilon}{r_{0}^{d-2}}+O(\epsilon)^{2}]dQ\\ &-\frac{16\pi}{\Omega_{d-2}}\frac{1}{r_{0}^{d-3}}[\frac{16\pi P(r_{0}+\epsilon)}{(d-2)(4\pi T)-16\pi P(r_{0}+\epsilon)}+1]|p^{r}|\\ &-\frac{2}{(d-2)}\frac{1}{r_{0}^{d-3}}[r_{0}+\frac{16\pi P[r_{0}^{2}+2r_{0}\epsilon+O(\epsilon)^{2}]}{(d-2)4\pi T-16\pi P(r_{0}+\epsilon)}]da\\ &-\frac{1}{r_{0}^{d-3}}\{\frac{16\pi P(r_{0}+\epsilon)}{[4\pi T(d-2)-16\pi P(r_{0}+\epsilon)]}[\frac{1}{r_{0}^{(d-1)\omega_{q}}}-\frac{(d-1)\omega_{q}\epsilon}{r_{0}^{(d-1)\omega_{q}-1}}+O(\epsilon)^{2}]+\frac{1}{r_{0}^{(d-1)\omega_{q}}}\}d\alpha,\\ \end{aligned}\label{eqn:wccc7}%
\end{equation}
where $\delta$ and $\epsilon$ are all the very small quantity. We set
$dQ\sim\epsilon$, $d\alpha\sim\epsilon$, $da\sim\epsilon$. Thus, Eq. $\left(
\ref{eqn:wccc7}\right) $ is modified as
\begin{equation}
\begin{aligned} &f\left(M+dM,Q+dQ,P+dP,a+da,\alpha+d\text{\ensuremath{\alpha}},r_{0}+dr_{0}\right)\\ &=\delta_{\epsilon}-\frac{16\pi}{\Omega_{d-2}}\frac{1}{r_{0}^{d-3}}[\frac{16\pi P(r_{0}+\epsilon)}{(d-2)(4\pi T)-16\pi P(r_{0}+\epsilon)}+1]|p^{r}|\\ &-\frac{2}{(d-2)}\frac{1}{r_{0}^{d-3}}[r_{0}+\frac{16\pi Pr_{0}^{2}\epsilon}{(d-2)4\pi T-16\pi P(r_{0}+\epsilon)}]\\ &-\frac{1}{r_{0}^{(d-1)\omega_{q}+d-3}}\frac{16\pi P(r_{0}+r_{0}\epsilon+\epsilon)}{[4\pi T(d-2)-16\pi P(r_{0}+\epsilon)]}+O(\epsilon)^{2}.\\ \end{aligned}\label{eqn:wc7}%
\end{equation}
Therefore, at the minimum point, we have
\begin{equation}
f\left( M+dM,Q+dQ,P+dP,a+da,\alpha+d\text{$\alpha$},r_{0}+dr_{0}\right) \leq0.
\end{equation}
Where the term is negative, which implies that the minimum value is always
negative. Hence, the stability of horizons exists in spacetime. The
near-extremal black hole can not be overcharged, which stays near-extremal
after absorbing a particle. The WCCC is satisfied for both the extremal and
near-extremal black holes in the extended phase space.

\subsection{A new assumption: $E = dM$}

\label{sec:Bd} In this section, by dropping particles into the black hole, we
have employed the recently new assumption \cite{Gwak:2017kkt,Liu:2020cji} that
the change of the black hole mass(enthalpy) should be the same amount as the
energy of an infalling particle $(E=dM)$, to test the laws of thermodynamics
and stability of horizon of a black hole in extended phase spaces under this
assumption. Where the energy-momentum relation near the event horizon can be
simplified as
\begin{equation}
E=\phi dQ+|p^{r}|,\label{eqn:ne1}%
\end{equation}
In Eq. $\left( \ref{eqn:ne1}\right) $, we choose the positive sign in front of
the $|p^{r}|$ term to ensure the positive flow of time direction of a particle
when it fell into the black hole. If it is assumed that the changes in the
black hole parameters are not lost in this process, the changes in the black
hole parameters should be same as the changes in the falling particles. In
this sense, the relationship between the infalling particle changes the
enthalpy of the black hole is
\begin{equation}
E=dM,\label{eqn:ne2}%
\end{equation}
In this case, Eq. $\left( \ref{eqn:ne1}\right) $ change into
\begin{equation}
dM=\phi dQ+p^{r}.\label{eqn:ne3}%
\end{equation}
As a charged particle dropped into the black hole, the configurations of the
black hole will be changed. This progress will lead to a shift for the
horizon. The relation between the functions $f(M,Q,P,a,\alpha, r_{h})$ and
$f(M+dM,Q+dQ,P+dP,a+da,\alpha+d\alpha, r_{h}+dr_{h})$is
\begin{equation}
\begin{aligned} &f\left(M+dM,Q+dQ,P+dP,a+da,\alpha+d\alpha,r_{h}+dr_{h}\right)\\ &=f(M,Q,P,a,\alpha, r_{h})+\frac{\partial f}{\partial M}|_{r=r_{h}}dM+\frac{\partial f}{\partial Q}|_{r=r_{h}}dQ+\frac{\partial f}{\partial r}|_{r=r_{h}}dr_{h}\\ &+\frac{\partial f}{\partial P}|_{r=r_{h}}dP+\frac{\partial f}{\partial a}|_{r=r_{h}}da+\frac{\partial f}{\partial\alpha}|_{r=r_{h}}d\text{\ensuremath{\alpha}}. \label{eqn:ne4} \end{aligned}
\end{equation}
By substituting Eq. $\left( \ref{eqn:ne3}\right) $ into Eq. $\left(
\ref{eqn:ne4}\right) $, we can obtain the value of the $dr_{h}$, which is
\begin{equation}
dr_{h}=\frac{\frac{2}{(d-2)r_{h}^{d-4}}da+\frac{1}{r_{h}^{(d-1)\omega_{q}%
+d-3}}d\alpha+\frac{16\pi}{r_{h}^{d-3}(d-2)\varOmega_{d-2}}p^{r}-\frac{16\pi
r_{h}^{2}}{(d-2)(d-1)}dP}{4\pi T}.\label{eqn:ne5}%
\end{equation}
With the aid of Eq. $(dS=\frac{\Omega_{d-2}(d-2)r^{d-3}}{4}dr_{h})$, the
variation of entropy is given by
\begin{equation}
dS=\frac{\frac{\varOmega_{d-2}r_{h}}{2}da+\frac{\varOmega_{d-2}(d-2)}%
{4r_{h}^{(d-1)\omega_{q}}}d\alpha+4\pi p^{r}-\frac{4\pi r_{h}^{d-1}%
\varOmega_{d-2}}{(d-1)}dP}{4\pi T}.\label{eqn:ne6}%
\end{equation}
Using Eq. $\left( \ref{eqn:ne6}\right) $, it is easy to get
\begin{equation}
\begin{aligned} TdS-PdV=&\frac{\frac{T\varOmega_{d-2}r_{h}}{2}-\frac{2P\varOmega_{d-2}r_{h}^{2}}{(d-2)}}{4\pi T}da +\frac{\frac{T\varOmega_{d-2}}{r_{h}^{(d-1)\omega_{q}+1}}-\frac{\varOmega_{d-2}P}{r_{h}^{(d-1)\omega_{q}+1}}}{4\pi T}d\alpha\\ &+\frac{4T\pi-\frac{16P\pi r_{h}}{(d-2)}}{4\pi T}p^{r}-\frac{\frac{4\pi Tr_{h}^{d-1}\varOmega_{d-2}}{(d-1)}-\frac{16\pi Pr_{h}^{d}\varOmega_{d-2}}{(d-2)(d-1)}}{4\pi T}dP. \end{aligned}\label{eqn:ne7}%
\end{equation}
Then, the Eq. $\left( \ref{eqn:ne3}\right) $ is rewritten as
\begin{equation}
dM=TdS+VdP+\phi dQ+\mathcal{A}da+\mathcal{Q}d\alpha.\label{eqn:ne8}%
\end{equation}

Obviously, the Eq. $\left( \ref{eqn:ne8}\right) $ is the same as Eq. $\left(
\ref{eqn:fcc}\right) $. It is means that the first law of black hole
thermodynamics still holds. Next, we will continue to check the second law of
black hole thermodynamics when a charged particle is captured by the black
hole. As the black hole entropy increase in a clockwise direction will not be
less than zero, we can examine the second law of thermodynamics of the black
hole by studying the change in entropy. For the extremal black hole where it's
temperature is zero. Then, combining this condition and the black hole mass,
the variation of entropy finally reads
\begin{equation}
dS_{extremal}\rightarrow\infty.\label{eqn:ne9}%
\end{equation}
Therefore the second law of black hole thermodynamics is still valid for the
extremal black holes. Besides, the non-extremal black holes have temperatures
greater than zero, so the variation of entropy $dS$ always has a positive
value under certain conditions, which means the second law of black hole
thermodynamics dose not violate for the non-extremal black holes. Next, we
will further check the stability of horizon black hole with particle's
absorption. In a similar way, Eq. $\left( \ref{eqn:wc5}\right) $ is rewritten
as
\begin{equation}
\begin{aligned} &f\left(M+dM,Q+dQ,P+dP,a+da,\alpha+d\text{\ensuremath{\alpha}},r_{0}+dr_{0}\right)\\ &=\delta+\frac{(r_{0}^{3-d}-r_{h}^{3-d})16\pi q}{r_{0}^{d-3}\Omega_{d-2}\sqrt{2(d-3)(d-2)}}dQ\\ &-\frac{16\pi p^{r}}{r_{0}^{d-3}(d-2)\Omega_{d-2}}+\frac{16\text{\ensuremath{\pi}}r_{0}^{2}}{(d-2)(d-1)}dP\\ &-\frac{1}{r_{0}^{(d-1)\omega_{q}+d-3}}d\alpha-\frac{2}{(d-2)r_{0}^{d-4}}da.\\ \end{aligned}\label{eqn:ne10}%
\end{equation}
Using Eq. $\left( \ref{eqn:pp1}\right) $, it is easy to get
\begin{equation}
\begin{aligned} &f\left(M+dM,Q+dQ,P+dP,a+da,\alpha+d\text{\ensuremath{\alpha}},r_{0}+dr_{0}\right)\\ &=\delta+\frac{(r_{0}^{3-d}-r_{h}^{3-d})16\pi q}{r_{0}^{d-3}\Omega_{d-2}\sqrt{2(d-3)(d-2)}}dQ\\ &-\frac{16\pi p^{r}}{r_{0}^{d-3}(d-2)\Omega_{d-2}}-\frac{2r_{0}^{2}}{l^{3}}dl\\ &-\frac{1}{r_{0}^{(d-1)\omega_{q}+d-3}}d\alpha-\frac{2}{(d-2)r_{0}^{d-4}}da.\\ \end{aligned}\label{eqn:ne111}%
\end{equation}
When the initial black hole is the extremal black hole, $r_{0}=r_{h}$, $T=0$
and $\delta=0$. Then we can obtain $f_{min}=\delta=0$ and $f_{min}^{\prime}%
=0$. Hence, Eq. $\left( \ref{eqn:ne111}\right) $ is written as
\begin{equation}
f\left( M+dM,Q+dQ,P+dP,a+da,\alpha+d\text{$\alpha$},dr_{0}+r_{0}\right)
<0.\label{eqn:wcc6}%
\end{equation}
When the initial black hole is the near-extremal black hole, $r_{0}$ and
$r_{h}$ do not coincide. In a similar way, the Eq. $\left( \ref{eqn:ne10}%
\right) $ can be expanded near the minimum point by using the relation
$r_{h}=r_{0}+\epsilon$, which is
\begin{equation}
\begin{aligned} &f\left(M+dM,Q+dQ,P+dP,a+da,\alpha+d\text{\ensuremath{\alpha}},r_{0}+dr_{0}\right)\\ &=\delta_{\epsilon}+\frac{[\frac{(d-3)\epsilon}{r_{0}^{d-2}}+O(\epsilon)^{2}]16\pi q}{r_{0}^{d-3}\Omega_{d-2}\sqrt{2(d-3)(d-2)}}dQ\\ &-\frac{16\pi p^{r}}{r_{0}^{d-3}(d-2)\Omega_{d-2}}-\frac{2r_{0}^{2}}{l^{3}}dl\\ &-\frac{1}{r_{0}^{(d-1)\omega_{q}+d-3}}d\alpha-\frac{2}{(d-2)r_{0}^{d-4}}da.\\ \end{aligned}\label{eqn:ne112}%
\end{equation}
We set $dQ\sim\epsilon$, $d\alpha\sim\epsilon$, $da\sim\epsilon$. Thus, Eq.
$\left( \ref{eqn:ne112}\right) $ is modified as
\begin{equation}
\begin{aligned} &f\left(M+dM,Q+dQ,P+dP,a+da,\alpha+d\text{\ensuremath{\alpha}},r_{0}+dr_{0}\right)\\ &=\delta_{\epsilon}-\frac{16\pi p^{r}}{r_{0}^{d-3}(d-2)\Omega_{d-2}}-\frac{2r_{0}^{2}\epsilon}{l^{3}}-\frac{\epsilon}{r_{0}^{(d-1)\omega_{q}+d-3}}-\frac{2\epsilon}{(d-2)r_{0}^{d-4}}+O(\epsilon^{2}).\\ \end{aligned}\label{eqn:ne1122}%
\end{equation}
Correspondingly, at the minimum point, we have
\begin{equation}
\begin{aligned} &f\left(M+dM,Q+dQ,P+dP,a+da,\alpha+d\text{\ensuremath{\alpha}},r_{0}+dr_{0}\right)\leq0.\\ \end{aligned}\label{eqn:wc77}%
\end{equation}
Therefore, it is obviously that the horizon stably exists at the final state
of the near-extremal black hole.

\section{The scalar field}

\label{sec:C}

\subsection{ Solution to Charged Scalar Field Equation}

\label{sec:Ca} In order to investigate the scattering of the nonminimally
coupled massive scalar field with RN-AdS black hole with a cloud of strings in
d-dimensional spacetime, the amount of conserved quintessence taken into the
black hole is given as the fluxes of the scattered external field. The action
of the charged scalar field in the fixed gravitational and electromagnetic
fields is
\begin{equation}
S_{\varPsi}=-\frac{1}{2}\int d^{D}x\sqrt{-g}(\mathbf{\mathcal{D}}_{\mu
}\varPsi\mathbf{\mathcal{D}}^{\mu}\varPsi^{*}+(\mu^{2}+\zeta
R)\varPsi\varPsi^{*}),
\end{equation}
where the spacetime dimension is assumed to be $D\geq4$. Owing to a scalar
field with electric charge $q$, we consider the covariant derivative
$\mathcal{\mathbf{\mathcal{D}}}_{\mu}=\partial_{\mu}-iqA_{\mu}$. The scalar
field has the mass $\mu$ and nonminimal coupling $\zeta$ with the curvature. There are two field equations, including the complex
conjugate
\begin{equation}
\frac{1}{\sqrt{-g}}\mathbf{\mathcal{D}}_{\mu}(\sqrt{-gg}^{\mu\nu
}\mathbf{\mathcal{D}}_{\nu}\phi)-(\mu^{2}+\zeta R)\varPsi=0, \frac{1}%
{\sqrt{-g}}\mathbf{\mathcal{D}}_{\mu}^{*}(\sqrt{-gg}^{\mu\nu}%
\mathbf{\mathcal{D}}_{\nu}^{*}\phi^{*})-(\mu^{2}+\zeta R)\varPsi^{*}=0.
\end{equation}
The determinant of the metric is simply noted as
\begin{equation}
\sqrt{-g}=r^{d-2}\prod_{j=0}^{d-3}\sin^{d-2-j}\theta_{j}.
\end{equation}
Then the separable equation with respect to $\varPsi$ is obtained as
\begin{equation}
\frac{1}{\sqrt{-g}}\partial_{\mu}(\sqrt{-gg}^{\mu\nu}\partial_{\nu}%
\phi)-2iqA_{0g}g^{00}\partial_{0}\varPsi-q^{2}g^{00}(A_{0})^{2}\varPsi-(\mu
^{2}+\zeta R)\varPsi=0.
\end{equation}
The solution to the scalar field is
\begin{equation}
\varPsi(t,r,\phi,\varTheta)=e^{-i\omega t}R(r)Y_{lm}(\varTheta_{1}%
,\varTheta_{2},...\varTheta_{d-2}),
\end{equation}
where $Y_{lm}(\varTheta_{1},\varTheta_{2},...\varTheta_{d-2})$ is the
hyperspherical harmonics on a $(d-2)$-dimensional sphere. At the outer
horizon, the radial solution of the scalar field is \cite{Gwak:2019asi}
\begin{equation}
R(r)=e^{\pm i(\omega_{q}-q\phi)r^{*}}.\label{eqn:ssf6}%
\end{equation}
The negative sign in the Eq. $\left( \ref{eqn:ssf6}\right) $ selected to
represent the scalar field entering the outer horizon under scalar field
scattering. Thus two solutions of the scalar field is represented as
\begin{equation}
\varPsi=e^{-i\omega_{q}t}e^{-i(\omega-q\phi)r^{*}}Y_{lm}(\varTheta_{1}%
,\varTheta_{2},...\varTheta_{d-2}), \varPsi^{*}=e^{i\omega_{q}t}%
e^{i(\omega_{q}-q\phi)r^{*}}Y_{lm}^{*}(\varTheta_{1},\varTheta_{2}%
,...\varTheta_{d-2}).\label{eqn:ssf7}%
\end{equation}
According to these solutions, it can be deduced that the relationship between
the black hole conserved and the scalar field, considering the $PV$term. When
entering a black hole, the energy and charge of the scalar field change as
much as the changes in the black hole. These transfer fluxes entering the
black hole can be obtained by the energy of the momentum tensor of the scalar
field
\begin{equation}
\begin{aligned} T_{\nu}^{\mu}=\frac{1}{2}\mathbf{\mathcal{D}}^{\mu}\partial_{\nu}\varPsi^{*}+\frac{1}{2}\varPsi\mathbf{\mathcal{D}}^{*\mu}\varPsi^{*}\partial_{\nu}-\delta_{\nu}^{\mu}(\frac{1}{2}\mathbf{\mathcal{D}}_{\mu}\varPsi\mathbf{\mathcal{D}}^{*\mu}\varPsi^{*}-\frac{1}{2}(\mu^{2}+\zeta R)\varPsi\varPsi^{*})). \end{aligned}\label{eqn:ssf8}%
\end{equation}
The energy flux is the component $T_{\nu}^{\mu}$ integrated by a solid angle
on an $S^{d-2}$ sphere at the outer horizon. Then, fluxes of energy and
electric charge are
\begin{equation}
\begin{aligned} &\frac{dE}{dt}=\int T_{t}^{r}\sqrt{-g}d\varOmega_{d-2}=\omega_{q}(\omega_{q}-\sqrt{\frac{d-2}{2(d-3)}}\frac{q^{2}}{r_{h}^{d-3}})r_{h}^{d-2},\\ &\frac{de}{dt}=\frac{q}{\omega_{q}}\frac{dE}{dt}=q(\omega_{q}-\sqrt{\frac{d-2}{2(d-3)}}\frac{q^{2}}{r_{h}^{d-3}})r_{h}^{d-2}.\\ \label{eqn:ssf9} \end{aligned}
\end{equation}
The fluxes in the above formulas will change the corresponding properties of
the black hole during the infinitesimal time interval $dt$.

\subsection{The first and second laws of thermodynamics}

\label{sec:Cb} In this section, we will discuss issues related to
thermodynamics under scalar field scattering. When the change in the enthalpy
is connected to the change in internal energy \cite{Gwak:2019asi}, the charge
flux corresponds to the change in that of the black hole. Moreover, the
changes in internal energy and charge are given as
\begin{equation}
dU=(\frac{dE}{dt})dt,dQ=(\frac{de}{dt})dt.\label{eqn:ssf9}%
\end{equation}
with
\begin{equation}
\begin{aligned} &dU=d(M-PV)=\omega_{q}(\omega_{q}-\sqrt{\frac{d-2}{2(d-3)}}\frac{q^{2}}{r_{h}^{d-3}})r_{h}^{d-2}dt,\\ &dQ=q(\omega_{q}-\sqrt{\frac{d-2}{2(d-3)}}\frac{q^{2}}{r_{h}^{d-3}})r_{h}^{d-2}.\\ \label{eqn:ssf10} \end{aligned}
\end{equation}

The location of the outer horizon is of great significance in the analysis
process, and the outer horizon $r_{h}$ is located at the point satisfying
$f(M,Q,P,a,\alpha,r_{h})=0$. Assuming that the initial state of the black hole
is represented by $(M,Q,P,a,\alpha,r_{h})$, and the final state is represented
by $(M+dM,Q+dQ,P+dP,a+da,\alpha+d\alpha,r_{h}+dr_{h})$. The functions
$f(M,Q,P,a,\alpha,r_{h})$ and $f(M+dM,Q+dQ,P+dP,a+da,\alpha+d\alpha
,r_{h}+dr_{h})$ satisfy
\begin{equation}
f(M,Q,P,a,\alpha,r_{h})=f\left( M+dM,Q+dQ,P+dP,a+da,\alpha+d\alpha
,r_{h}+dr_{h}\right) =0.
\end{equation}
The relation between the functions $f(M,Q,P,a,\alpha, r_{h})$ and
$f(M+dM,Q+dQ,P+dP,a+da,\alpha+d\alpha, r_{h}+dr_{h})$is
\begin{equation}
\begin{aligned} &f\left(M+dM,Q+dQ,P+dP,a+da,\alpha+d\alpha,r_{h}+dr_{h}\right)=f(r)\\ &+\frac{\partial f}{\partial M}|_{r=r_{h}}dM+\frac{\partial f}{\partial Q}|_{r=r_{h}}dQ+\frac{\partial f}{\partial r}|_{r=r_{h}}dr_{h}+\\ &\frac{\partial f}{\partial P}|_{r=r_{h}}dP+\frac{\partial f}{\partial a}|_{r=r_{h}}da+\frac{\partial f}{\partial\alpha}|_{r=r_{h}}d\text{\ensuremath{\alpha}},\\ \end{aligned}
\end{equation}
where
\begin{equation}
\begin{aligned} &\frac{\partial f}{\partial M}|_{r=r_{h}}=-\frac{16\pi}{r_{h}^{d-3}(d-2)\Omega_{d-2}},\\ &\frac{\partial f}{\partial Q}|_{r=r_{h}}=\frac{16\pi q}{r_{h}^{2\text{(}d-3)}\Omega_{d-2}\sqrt{2(d-3)(d-2)}},\\ &\frac{\partial f}{\partial P}|_{r=r_{h}}=\frac{16\text{\ensuremath{\pi}}r_{h}^{2}}{(d-2)(d-1)},\\ &\frac{\partial f}{\partial r}|_{r=r_{h}}=4\pi T,\\ &\frac{\partial f}{\partial\alpha}|_{r=r_{h}}=-\frac{1}{r_{h}^{(d-1)\omega_{q}+d-3}},\\ &\frac{\partial f}{\partial a}|_{r=r_{h}}=\frac{-2}{(d-2)r_{h}^{d-4}}.\\ \end{aligned}
\end{equation}
Then, we can figure out
\begin{equation}
\begin{aligned} &dr_{h}=\frac{16\pi r_{h}}{(\frac{16\pi Pr_{h}}{d-2}-4\pi T])\varOmega}[\frac{2q^{2}\omega_{q}}{r_{h}^{d-3}\sqrt{2(d-2)(d-3)}}-\frac{\omega_{q}^{2}}{d-2}-\frac{q^{4}}{2(d-3)r_{h}^{2(d-3)}}]dt\\ &-\frac{2}{(\frac{16\pi Pr_{h}}{d-2}-4\pi T)(d-2)r_{h}^{d-4}}da-\frac{1}{(\frac{16\pi Pr_{h}}{d-2}-4\pi T)r_{h}^{(d-1)\omega_{q}+d-3}}d\alpha.\\ \end{aligned}
\end{equation}
Then the variation of the entropy and volume are obtained
\begin{equation}
dS=\frac{4\pi r_{h}^{d-2}(d-2)[\frac{2q^{2}\omega_{q}}{r_{h}^{d-3}\sqrt{2(d-2)(d-3)}}-\frac{\omega_{q}^{2}}{d-2}-\frac{q^{4}}{2(d-3)r_{h}^{2(d-3)}}]dt}{\frac{16\pi Pr_{h}}{d-2}-4\pi T}-\frac{\frac{r_{h}\varOmega_{d-2}}{2}da+\frac{\varOmega_{d-2}(d-2)}{4r_{h}^{(d-1)\omega_{q}}}d\alpha}{\frac{16\pi Pr_{h}}{d-2}-4\pi T}.\label{eqn:ds22}%
\end{equation}
and
\begin{equation}
dV=\frac{16\pi r_{h}^{d-1}[\frac{2q^{2}\omega_{q}}{r_{h}^{d-3}\sqrt{2(d-2)(d-3)}}-\frac{\omega_{q}^{2}}{d-2}-\frac{q^{4}}{2(d-3)r_{h}^{2(d-3)}}]dt}{\frac{16\pi Pr_{h}}{d-2}-4\pi T}-\frac{\frac{2r_{h}^{2}\varOmega_{d-2}}{(d-2)}da+\frac{\varOmega_{d-2}}{r_{h}^{(d-1)\omega_{q}-1}}d\alpha}{\frac{16\pi Pr_{h}}{d-2}-4\pi T}.\label{eqn:dv22}%
\end{equation}
Using Eqs. $\left( \ref{eqn:ds22}\right) $ and $\left( \ref{eqn:dv22}\right)
$, we yield
\begin{equation}
\begin{aligned} &\frac{[4\pi r_{h}^{d-2}T(d-2)-16\pi r_{h}^{d-1}P][\frac{2q^{2}\omega_{q}}{r_{h}^{d-3}\sqrt{2(d-2)(d-3)}}-\frac{\omega_{q}^{2}}{d-2}-\frac{q^{4}}{2(d-3)r_{h}^{2(d-3)}}]dt}{\frac{16\pi Pr_{h}}{d-2}-4\pi T}\\ &-\frac{T(d-2)r_{h}\varOmega_{d-2}-4r_{h}^{2}P\varOmega_{d-2}}{(d-2)(\frac{32\pi Pr_{h}}{d-2}-8\pi T)}da-\frac{T\varOmega_{d-2}(d-2)-4P\varOmega_{d-2}r_{h}}{(\frac{64\pi Pr_{h}}{(d-2)}-16\pi T)r_{h}^{(d-1)\omega_{q}}}d\alpha.\\ \end{aligned}
\end{equation}
The generalized first law in the extended phase space which account for the
cosmological constant effect, cloud of strings and the quintessence
contributions is then expressed as
\begin{equation}
dM=TdS+VdP+\phi dQ+\mathcal{A}da+\mathcal{Q}d\alpha.\label{eqn:fcc2}%
\end{equation}
Thus, the first law of thermodynamics is recovered by the scattering of the
scalar field. Then the second law of thermodynamics is validated in the
extended phase space. When it is the extremal black hole, the temperature is
zero. Then Eq. $\left( \ref{eqn:ds22}\right) $ is modified as
\begin{equation}
dS=\frac{4\pi r_{h}^{d-2}(d-2)[\frac{2q^{2}\omega_{q}}{r_{h}^{d-3}\sqrt{2(d-2)(d-3)}}-\frac{\omega_{q}^{2}}{d-2}-\frac{q^{4}}{2(d-3)r_{h}^{2(d-3)}%
}]dt}{\frac{16\pi Pr_{h}}{d-2}}-\frac{\frac{r_{h}\varOmega_{d-2}}{2}%
da+\frac{\varOmega_{d-2}(d-2)}{4r_{h}^{(d-1)\omega_{q}}}d\alpha}{\frac{16\pi
Pr_{h}}{d-2}}.\label{eqn:ds222}%
\end{equation}
which is positive when $d\alpha>0$ and $da>0$, and the reverse is uncertain.
Therefore, the second law of thermodynamics can be indefinite for the extremal
black hole in the extended phase space.

\begin{table}
\caption{The relation between $dS$, $Q$ and $r_{h}$ for $d = 5$ in the
extended phase space via scalar field scattering.}%
\label{tab:Bdsa1}%
\begin{tabular}
[c]%
{p{0.6in}|p{0.6in}|p{0.65in}|p{0.6in}|p{0.6in}|p{0.65in}|p{0.6in}|p{0.6in}|p{0.65in}}%
\hline
\multicolumn{3}{c|}{$a=0.01$} & \multicolumn{3}{|c|}{$a=10$} &
\multicolumn{3}{|c}{$a=20$}\\\hline
$Q$ & $r_{h}$ & $dS$ & $Q$ & $r_{h}$ & $dS$ & $Q$ & $r_{h}$ & $dS$\\\hline
0.640747 & 1.43509 & 0.2762150 & 0.965855 & 1.752080 & 0.9524490 & 1.33654 &
2.01130 & 2.6726000\\
0.64 & 1.40537 & 0.2427220 & 0.96 & 1.674810 & 0.7045120 & 0.99 & 1.32019 &
0.1864200\\
0.6 & 1.19173 & 0.0931272 & 0.9 & 1.468580 & 0.3134130 & 0.9 & 1.20661 &
0.1101180\\
0.55 & 1.04937 & 0.0464416 & 0.8 & 1.265150 & 0.1320450 & 0.8 & 1.08407 &
0.0595115\\
0.5 & 0.93212 & 0.0248162 & 0.7 & 1.097070 & 0.0594952 & 0.7 & 0.96294 &
0.0306009\\
0.45 & 0.82613 & 0.0133363 & 0.6 & 0.941706 & 0.0260095 & 0.6 & 0.84134 &
0.0146652\\
0.4 & 0.72648 & 0.0070078 & 0.5 & 0.791284 & 0.0104770 & 0.5 & 0.71771 &
0.0063719\\
0.35 & 0.63079 & 0.0035247 & 0.4 & 0.641677 & 0.0036787 & 0.4 & 0.59042 &
0.0024033\\
0.3 & 0.53766 & 0.0016586 & 0.3 & 0.490067 & 0.0010301 & 0.3 & 0.45777 &
0.0007265\\
0.2 & 0.35598 & 0.0002610 & 0.2 & 0.334134 & 0.0001891 & 0.2 & 0.31756 &
0.0001458\\
0.1 & 0.17741 & 0.0000140 & 0.1 & 0.171632 & 0.0000119 & 0.1 & 0.16667 &
0.0000102\\\hline
\end{tabular}
\end{table}

\begin{table}
\caption{The relation between $dS$, $Q$ and $r_{h}$ for $d = 6$ in the
extended phase space via scalar field scattering.}%
\label{tab:Bdsa2}%
\begin{tabular}
[c]%
{p{0.6in}|p{0.6in}|p{0.65in}|p{0.6in}|p{0.6in}|p{0.65in}|p{0.6in}|p{0.6in}|p{0.65in}}%
\hline
\multicolumn{3}{c|}{$a=0.01$} & \multicolumn{3}{|c|}{$a=10$} &
\multicolumn{3}{|c}{$a=20$}\\\hline
0.751095 & 1.26380 & 0.2052720 & 1.09831 & 1.43654 & 0.4682340 & 1.48119 &
1.57562 & 0.8792900\\
0.75 & 1.24297 & 0.1781660 & 0.99 & 1.23344 & 0.1334240 & 0.99 & 1.07695 &
0.0447711\\
0.7 & 1.10288 & 0.0689191 & 0.9 & 1.12473 & 0.0667108 & 0.9 & 1.01152 &
0.0286431\\
0.65 & 1.02277 & 0.0394930 & 0.8 & 1.02853 & 0.0350302 & 0.8 & 0.93810 &
0.0169569\\
0.6 & 0.95369 & 0.0240479 & 0.7 & 0.93622 & 0.0182371 & 0.7 & 0.86279 &
0.0096341\\
0.55 & 0.88937 & 0.0148948 & 0.6 & 0.84395 & 0.0091214 & 0.6 & 0.78436 &
0.0051753\\
0.5 & 0.82727 & 0.0092044 & 0.5 & 0.74901 & 0.0042539 & 0.5 & 0.70140 &
0.0025733\\
0.4 & 0.70438 & 0.0033197 & 0.4 & 0.64873 & 0.0017768 & 0.4 & 0.61198 &
0.0011440\\
0.3 & 0.57726 & 0.0010185 & 0.3 & 0.53976 & 0.0006191 & 0.3 & 0.51317 &
0.0004255\\
0.2 & 0.43871 & 0.0002224 & 0.2 & 0.41656 & 0.0001529 & 0.2 & 0.39967 &
0.0001132\\
0.1 & 0.27580 & 0.0000198 & 0.1 & 0.26667 & 0.0000156 & 0.1 & 0.25906 &
0.0000127\\\hline
\end{tabular}
\end{table}

\begin{table}
\caption{The relation between $dS$, $Q$ and $r_{h}$ for $d = 7$ in the
extended phase space via scalar field scattering.}%
\label{tab:Bdsa3}%
\begin{tabular}
[c]%
{p{0.6in}|p{0.6in}|p{0.65in}|p{0.6in}|p{0.6in}|p{0.65in}|p{0.6in}|p{0.6in}|p{0.65in}}%
\hline
\multicolumn{3}{c|}{$a=0.01$} & \multicolumn{3}{|c|}{$a=10$} &
\multicolumn{3}{|c}{$a=20$}\\\hline
$Q$ & $r_{h}$ & $dS$ & $Q$ & $r_{h}$ & $dS$ & $Q$ & $r_{h}$ & $dS$\\\hline
0.818145 & 1.16885 & 0.1640370 & 1.17528 & 1.28263 & 0.3158390 & 1.56137 &
1.373450 & 0.5141460\\
0.8 & 1.10675 & 0.0928402 & 0.99 & 1.07803 & 0.0550888 & 0.99 & 0.990496 &
0.0220919\\
0.75 & 1.03766 & 0.0496371 & 0.9 & 1.01703 & 0.0324783 & 0.9 & 0.943946 &
0.0146366\\
0.7 & 0.98534 & 0.0308256 & 0.8 & 0.95202 & 0.0182544 & 0.8 & 0.890724 &
0.0090351\\
0.65 & 0.93833 & 0.0199918 & 0.7 & 0.88699 & 0.0100934 & 0.7 & 0.835066 &
0.0053799\\
0.6 & 0.89360 & 0.0131606 & 0.6 & 0.81988 & 0.0053710 & 0.6 & 0.775936 &
0.0030504\\
0.5 & 0.80578 & 0.0056642 & 0.5 & 0.74876 & 0.0026852 & 0.5 & 0.711979 &
0.0016162\\
0.4 & 0.71509 & 0.0022778 & 0.4 & 0.67125 & 0.0012179 & 0.4 & 0.641222 &
0.0007755\\
0.3 & 0.61620 & 0.0007898 & 0.3 & 0.58382 & 0.0004702 & 0.3 & 0.560427 &
0.0003172\\
0.2 & 0.50162 & 0.0002017 & 0.2 & 0.47991 & 0.0001335 & 0.2 & 0.463348 &
0.0000959\\
0.1 & 0.35416 & 0.0000229 & 0.1 & 0.34299 & 0.0000172 & 0.1 & 0.333862 &
0.0000134\\\hline
\end{tabular}
\end{table}\begin{table}
\caption{The relation between $dS$, $Q$ and $r_{h}$ for $d=8$ in the extended
phase space via scalar field scattering.}%
\label{tab:Bdsa4}%
\begin{tabular}
[c]%
{p{0.6in}|p{0.6in}|p{0.65in}|p{0.6in}|p{0.6in}|p{0.65in}|p{0.6in}|p{0.6in}|p{0.65in}}%
\hline
\multicolumn{3}{c|}{$a=0.01$} & \multicolumn{3}{|c|}{$a=10$} &
\multicolumn{3}{|c}{$a=20$}\\\hline
$Q$ & $r_{h}$ & $dS$ & $Q$ & $r_{h}$ & $dS$ & $Q$ & $r_{h}$ & $dS$\\\hline
0.861699 & 1.111510 & 0.1368470 & 1.22409 & 1.194940 & 0.2407210 & 1.6107 &
1.261110 & 0.36552000\\
0.8 & 1.017820 & 0.0469615 & 0.99 & 1.014820 & 0.0342930 & 0.99 & 0.951314 &
0.01439230\\
0.75 & 0.977116 & 0.0297433 & 0.9 & 0.969496 & 0.0210418 & 0.9 & 0.914867 &
0.00972157\\
0.7 & 0.940827 & 0.0198000 & 0.8 & 0.919793 & 0.0122545 & 0.8 & 0.872725 &
0.00614304\\
0.65 & 0.906358 & 0.0134389 & 0.7 & 0.869011 & 0.0070074 & 0.7 & 0.828122 &
0.00375529\\
0.6 & 0.872532 & 0.0091714 & 0.6 & 0.815649 & 0.0038632 & 0.6 & 0.780130 &
0.00219467\\
0.5 & 0.804095 & 0.0042006 & 0.5 & 0.758080 & 0.0020099 & 0.5 & 0.727477 &
0.00120488\\
0.4 & 0.731161 & 0.0017970 & 0.4 & 0.694122 & 0.0009557 & 0.4 & 0.668245 &
0.00060337\\
0.3 & 0.649196 & 0.0006688 & 0.3 & 0.620283 & 0.0003914 & 0.3 & 0.599186 &
0.00026037\\
0.2 & 0.550705 & 0.0001872 & 0.2 & 0.529782 & 0.0001204 & 0.2 & 0.513820 &
0.00008474\\
0.1 & 0.416850 & 0.0000247 & 0.1 & 0.404567 & 0.0000176 & 0.1 & 0.394645 &
0.00001337\\\hline
\end{tabular}
\end{table}

\begin{figure}[tbh]
\begin{center}
\subfigure[{$d=5$ .}]{
\includegraphics[width=0.45\textwidth]{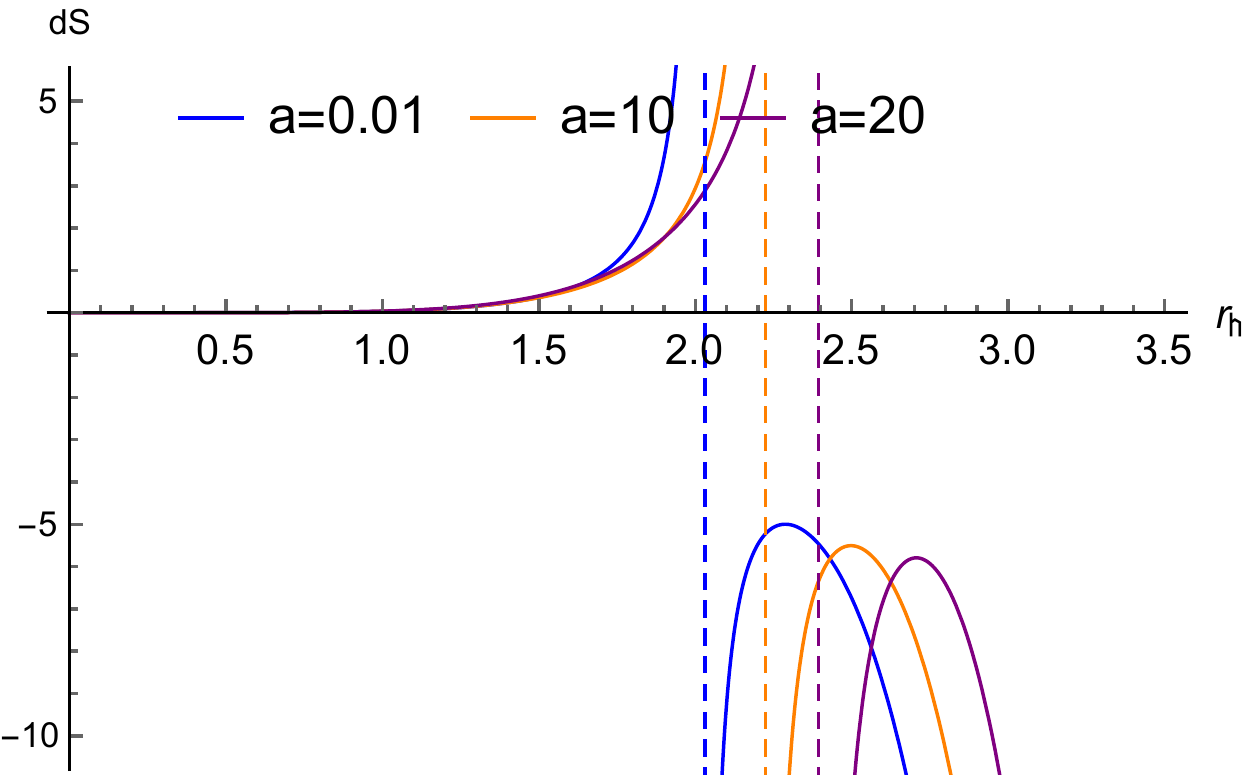}\label{Bdsa5}}
\subfigure[{$d=6$ .}]{
\includegraphics[width=0.45\textwidth]{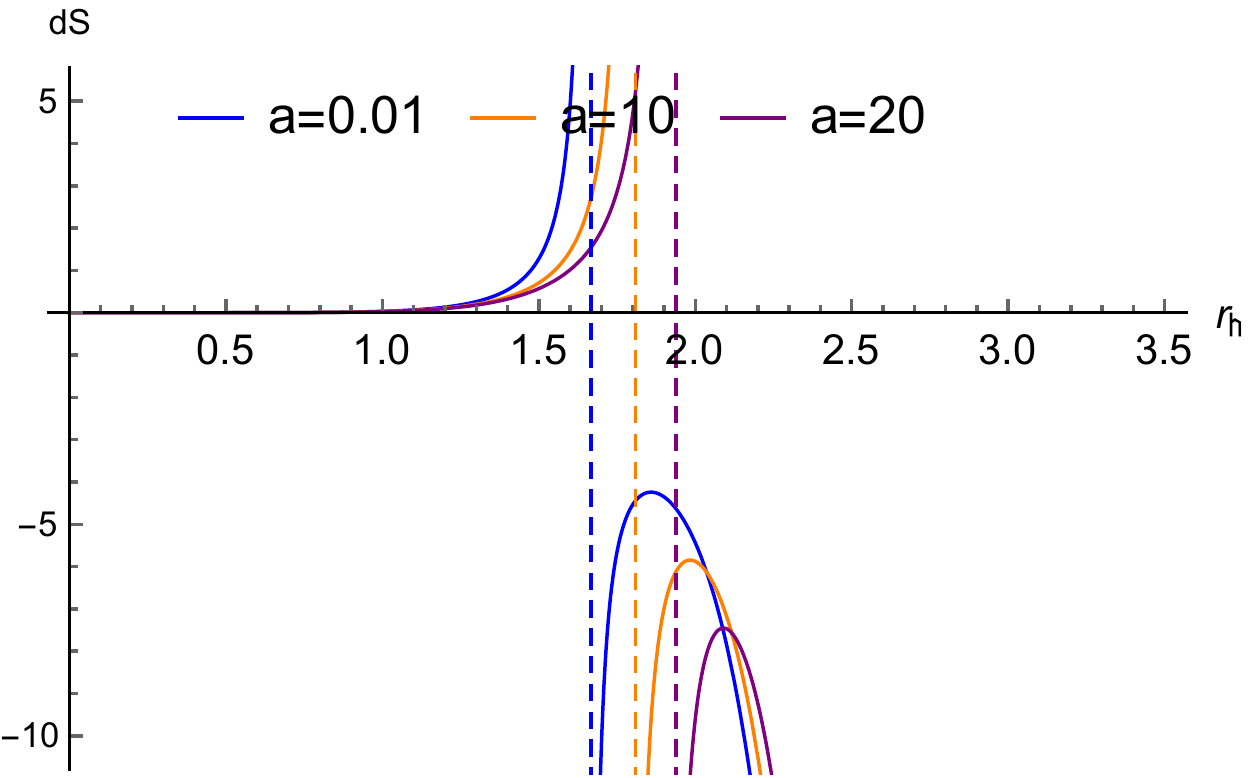}\label{Bdsa6}}
\subfigure[{$d=7$ .}]{
\includegraphics[width=0.45\textwidth]{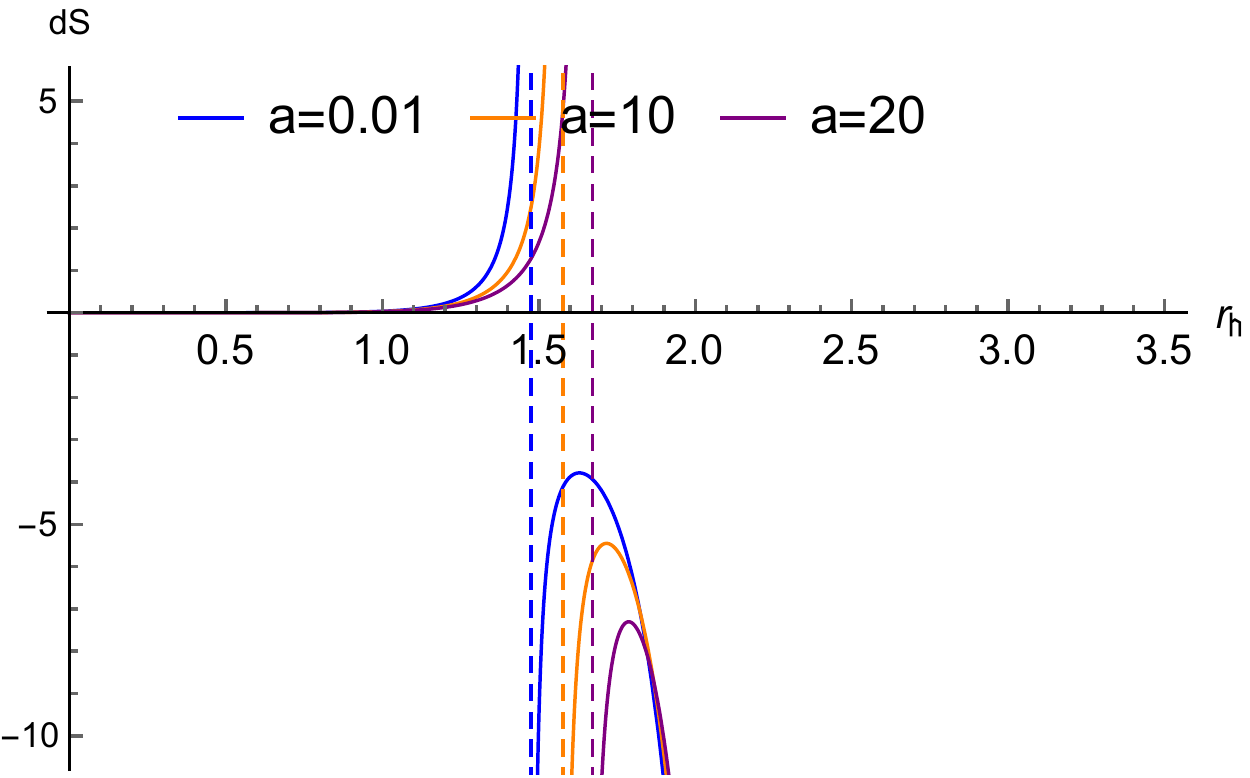}\label{Bdsa7}}
\subfigure[{$d=8$ .}]{
\includegraphics[width=0.45\textwidth]{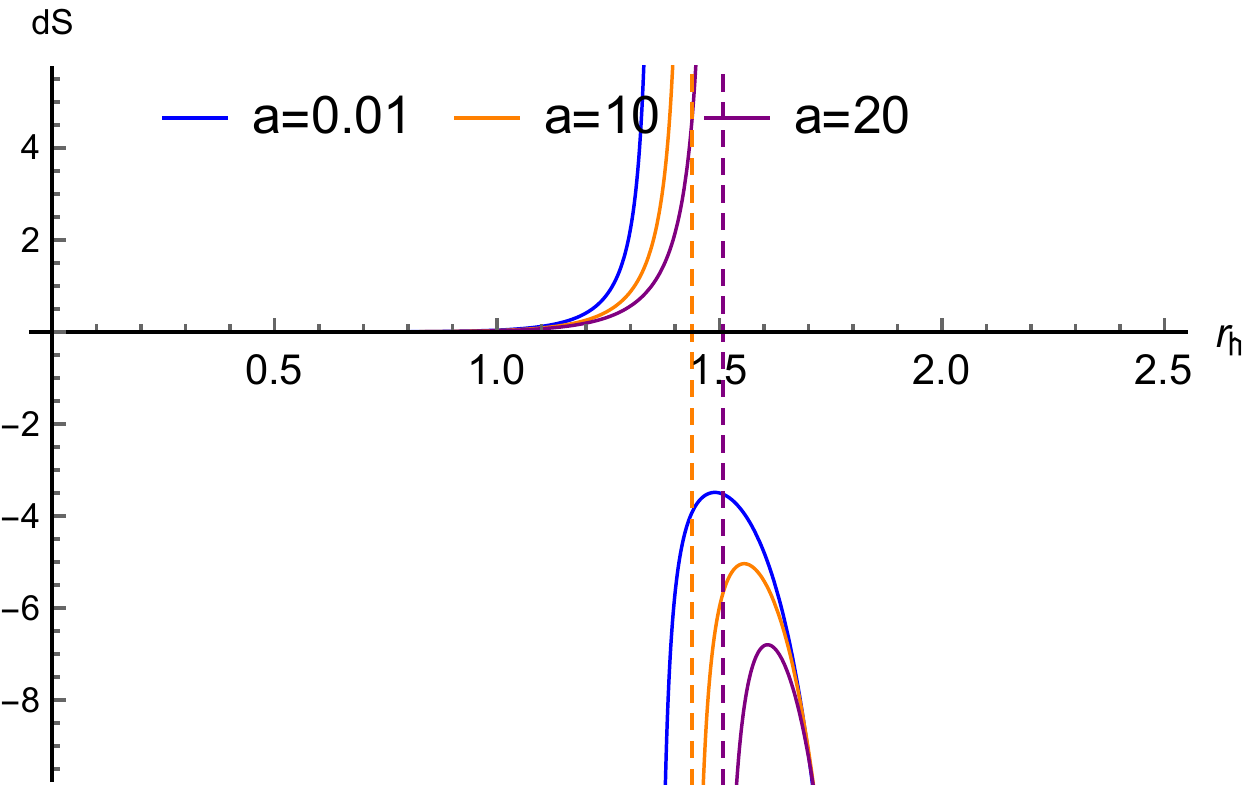}\label{Bdsa8}}
\end{center}
\caption{The relationship between $dS$,$Q$ and $r_{h}$ for $a=0.1,10,20$. }%
\label{fig:Bds1}%
\end{figure}

\begin{table}
\caption{The relation between $dS$, $Q$ and $r_{h}$ for $d = 5$ in the
extended phase space via scalar field scattering.}%
\label{tab:Bdspha1}%
\begin{tabular}
[c]%
{p{0.6in}|p{0.6in}|p{0.65in}|p{0.6in}|p{0.6in}|p{0.65in}|p{0.6in}|p{0.6in}|p{0.65in}}%
\hline
\multicolumn{3}{c|}{$\alpha=0.01$} & \multicolumn{3}{|c|}{$\alpha=10$} &
\multicolumn{3}{|c}{$\alpha=20$}\\\hline
$Q$ & $r_{h}$ & $dS$ & $Q$ & $r_{h}$ & $dS$ & $Q$ & $r_{h}$ & $dS$\\\hline
0.640747 & 1.435090 & 0.2762150 & 0.99 & 1.25816 & 0.8322130 & 0.99 &
1.113580 & 1.7461700\\
0.6 & 1.191730 & 0.0931272 & 0.9 & 1.18899 & 0.3740940 & 0.9 & 1.058110 &
0.4420350\\
0.55 & 1.049370 & 0.0464415 & 0.8 & 1.10560 & 0.1700950 & 0.8 & 0.991187 &
0.4420350\\
0.5 & 0.932126 & 0.0248162 & 0.7 & 1.01377 & 0.0789687 & 0.7 & 0.917336 &
0.0689460\\
0.45 & 0.826131 & 0.0133363 & 0.6 & 0.91142 & 0.0355355 & 0.6 & 0.834588 &
0.0297781\\
0.4 & 0.726485 & 0.0070078 & 0.5 & 0.79599 & 0.0147432 & 0.5 & 0.740149 &
0.0123187\\
0.35 & 0.630789 & 0.0035247 & 0.4 & 0.66473 & 0.0052727 & 0.4 & 0.630155 &
0.0045427\\
0.3 & 0.537666 & 0.0016586 & 0.3 & 0.51596 & 0.0014571 & 0.3 & 0.500041 &
0.0013274\\
0.2 & 0.355981 & 0.0002609 & 0.2 & 0.35138 & 0.0002507 & 0.2 & 0.347263 &
0.0002419\\
0.1 & 0.177411 & 0.0000140 & 0.1 & 0.17711 & 0.0000139 & 0.1 & 0.176826 &
0.0000138\\\hline
\end{tabular}
\end{table}

\begin{table}
\caption{The relation between $dS$, $Q$ and $r_{h}$ for $d = 6$ in the
extended phase space via scalar field scattering.}%
\label{tab:Bdspha2}%
\begin{tabular}
[c]%
{p{0.6in}|p{0.6in}|p{0.65in}|p{0.6in}|p{0.6in}|p{0.65in}|p{0.6in}|p{0.6in}|p{0.65in}}%
\hline
\multicolumn{3}{c|}{$\alpha=0.01$} & \multicolumn{3}{|c|}{$\alpha=10$} &
\multicolumn{3}{|c}{$\alpha=20$}\\\hline
$Q$ & $r_{h}$ & $dS$ & $Q$ & $r_{h}$ & $dS$ & $Q$ & $r_{h}$ & $dS$\\\hline
0.751095 & 1.26380 & 0.2052720 & 0.99 & 1.04990 & 0.0990892 & 0.99 &
0.968419 & 0.0815352\\
0.7 & 1.10288 & 0.0689191 & 0.9 & 1.00783 & 0.0636439 & 0.9 & 0.933293 &
0.0507251\\
0.6 & 0.95369 & 0.0240479 & 0.8 & 0.95630 & 0.0381356 & 0.8 & 0.890286 &
0.0298474\\
0.55 & 0.88937 & 0.0148948 & 0.7 & 0.89852 & 0.0220132 & 0.7 & 0.842006 &
0.0172073\\
0.5 & 0.82727 & 0.0092043 & 0.6 & 0.83282 & 0.0119845 & 0.6 & 0.786858 &
0.0095171\\
0.45 & 0.76594 & 0.0056019 & 0.5 & 0.75695 & 0.0059689 & 0.5 & 0.722502 &
0.0049009\\
0.4 & 0.70438 & 0.0033196 & 0.4 & 0.66796 & 0.0025935 & 0.4 & 0.645455 &
0.0022377\\
0.3 & 0.57726 & 0.0010184 & 0.3 & 0.56214 & 0.0009055 & 0.3 & 0.550610 &
0.0008287\\
0.2 & 0.43871 & 0.0002224 & 0.2 & 0.43457 & 0.0002136 & 0.2 & 0.430850 &
0.0002062\\
0.1 & 0.27580 & 0.0000198 & 0.1 & 0.27538 & 0.0000196 & 0.1 & 0.274968 &
0.0000195\\\hline
\end{tabular}
\end{table}

\begin{table}
\caption{The relation between $dS$, $Q$ and $r_{h}$ for $d = 7$ in the
extended phase space via scalar field scattering.}%
\label{tab:Bdspha3}%
\begin{tabular}
[c]%
{p{0.6in}|p{0.6in}|p{0.65in}|p{0.6in}|p{0.6in}|p{0.65in}|p{0.6in}|p{0.6in}|p{0.65in}}%
\hline
\multicolumn{3}{c|}{$\alpha=0.01$} & \multicolumn{3}{|c|}{$\alpha=10$} &
\multicolumn{3}{|c}{$\alpha=20$}\\\hline
$Q$ & $r_{h}$ & $dS$ & $Q$ & $r_{h}$ & $dS$ & $Q$ & $r_{h}$ & $dS$\\\hline
0.818145 & 1.16885 & 0.1640370 & 0.99 & 0.975201 & 0.0429284 & 0.99 &
0.918038 & 0.0318422\\
0.8 & 1.12844 & 0.1131090 & 0.9 & 0.944555 & 0.0298752 & 0.9 & 0.891935 &
0.0221392\\
0.75 & 1.03766 & 0.0496371 & 0.8 & 0.906704 & 0.0194035 & 0.8 & 0.859722 &
0.0145037\\
0.7 & 0.98534 & 0.0308256 & 0.7 & 0.863856 & 0.0120788 & 0.7 & 0.823226 &
0.0092066\\
0.6 & 0.89360 & 0.0131606 & 0.6 & 0.814597 & 0.0070827 & 0.6 & 0.781100 &
0.0055698\\
0.5 & 0.80578 & 0.0056642 & 0.5 & 0.756955 & 0.0038133 & 0.5 & 0.731328 &
0.0031311\\
0.4 & 0.71509 & 0.0022778 & 0.4 & 0.688128 & 0.0018105 & 0.4 & 0.670800 &
0.0015682\\
0.3 & 0.61620 & 0.0007898 & 0.3 & 0.604011 & 0.0007059 & 0.3 & 0.594569 &
0.0006477\\
0.2 & 0.50162 & 0.0002017 & 0.2 & 0.497826 & 0.0001937 & 0.2 & 0.494408 &
0.0001869\\
0.1 & 0.35416 & 0.0000230 & 0.1 & 0.353678 & 0.3536780 & 0.1 & 0.353200 &
0.0000226\\\hline
\end{tabular}
\end{table}\begin{table}
\caption{The relation between $dS$, $Q$ and $r_{h}$ for $d = 8$ in the
extended phase space via scalar field scattering.}%
\label{tab:Bdspha4}%
\begin{tabular}
[c]%
{p{0.6in}|p{0.6in}|p{0.65in}|p{0.6in}|p{0.6in}|p{0.65in}|p{0.6in}|p{0.6in}|p{0.65in}}%
\hline
\multicolumn{3}{c|}{$\alpha=0.01$} & \multicolumn{3}{|c|}{$\alpha=10$} &
\multicolumn{3}{|c}{$\alpha=20$}\\\hline
$Q$ & $r_{h}$ & $dS$ & $Q$ & $r_{h}$ & $dS$ & $Q$ & $r_{h}$ & $dS$\\\hline
0.861699 & 1.111510 & 0.1368470 & 0.99 & 0.941476 & 0.0269757 & 0.99 &
0.897107 & 0.0193602\\
0.8 & 1.017820 & 0.0469615 & 0.9 & 0.917179 & 0.0194393 & 0.9 & 0.876162 &
0.0140900\\
0.75 & 0.977116 & 0.0297433 & 0.8 & 0.887017 & 0.0131023 & 0.8 & 0.850188 &
0.0096751\\
0.7 & 0.940827 & 0.0198000 & 0.7 & 0.852669 & 0.0084608 & 0.7 & 0.820590 &
0.0064206\\
0.6 & 0.872532 & 0.0091714 & 0.6 & 0.812908 & 0.0051536 & 0.6 & 0.786200 &
0.0040569\\
0.5 & 0.804095 & 0.0042005 & 0.5 & 0.765975 & 0.0028935 & 0.5 & 0.745248 &
0.0023849\\
0.4 & 0.731161 & 0.0017969 & 0.4 & 0.709280 & 0.0014438 & 0.4 & 0.694940 &
0.0012554\\
0.3 & 0.649196 & 0.0006688 & 0.3 & 0.638772 & 0.0005998 & 0.3 & 0.630644 &
0.0005514\\
0.2 & 0.550705 & 0.0001872 & 0.2 & 0.547191 & 0.0001798 & 0.2 & 0.544033 &
0.0001734\\
0.1 & 0.416850 & 0.0000246 & 0.1 & 0.416333 & 0.0000245 & 0.1 & 0.415826 &
0.0000243\\\hline
\end{tabular}
\end{table}\begin{figure}[tbh]
\begin{center}
\subfigure[{$d=5$ .}]{
\includegraphics[width=0.45\textwidth]{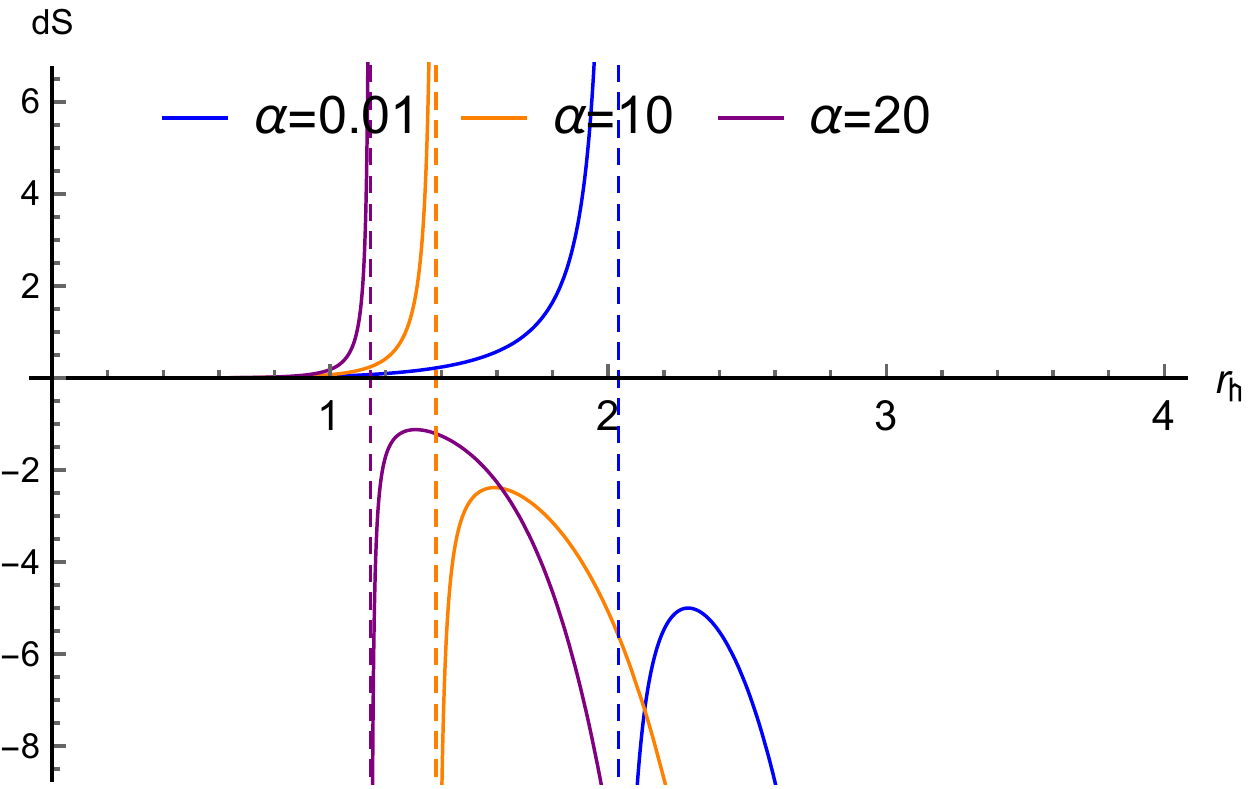}\label{Bdsph5}}
\subfigure[{$d=6$ .}]{
\includegraphics[width=0.45\textwidth]{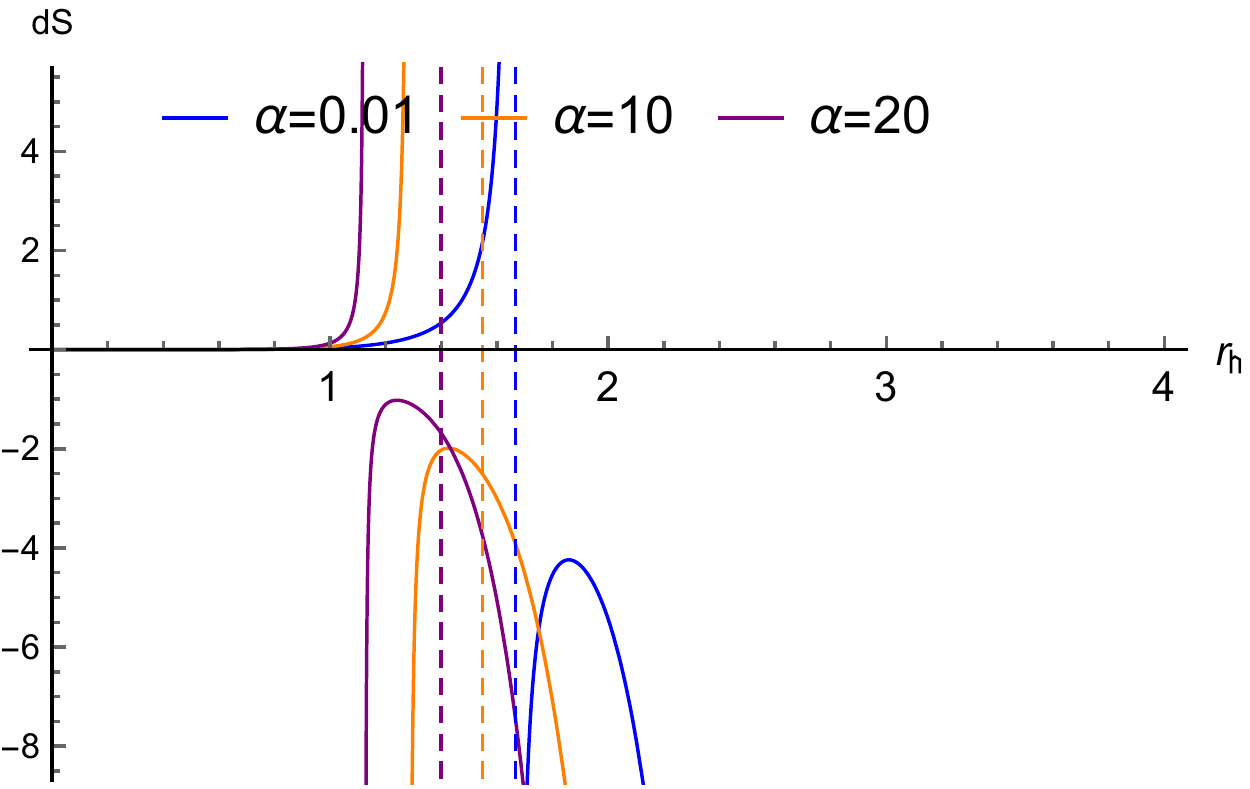}\label{Bdsph6}}
\subfigure[{$d=7$ .}]{
\includegraphics[width=0.45\textwidth]{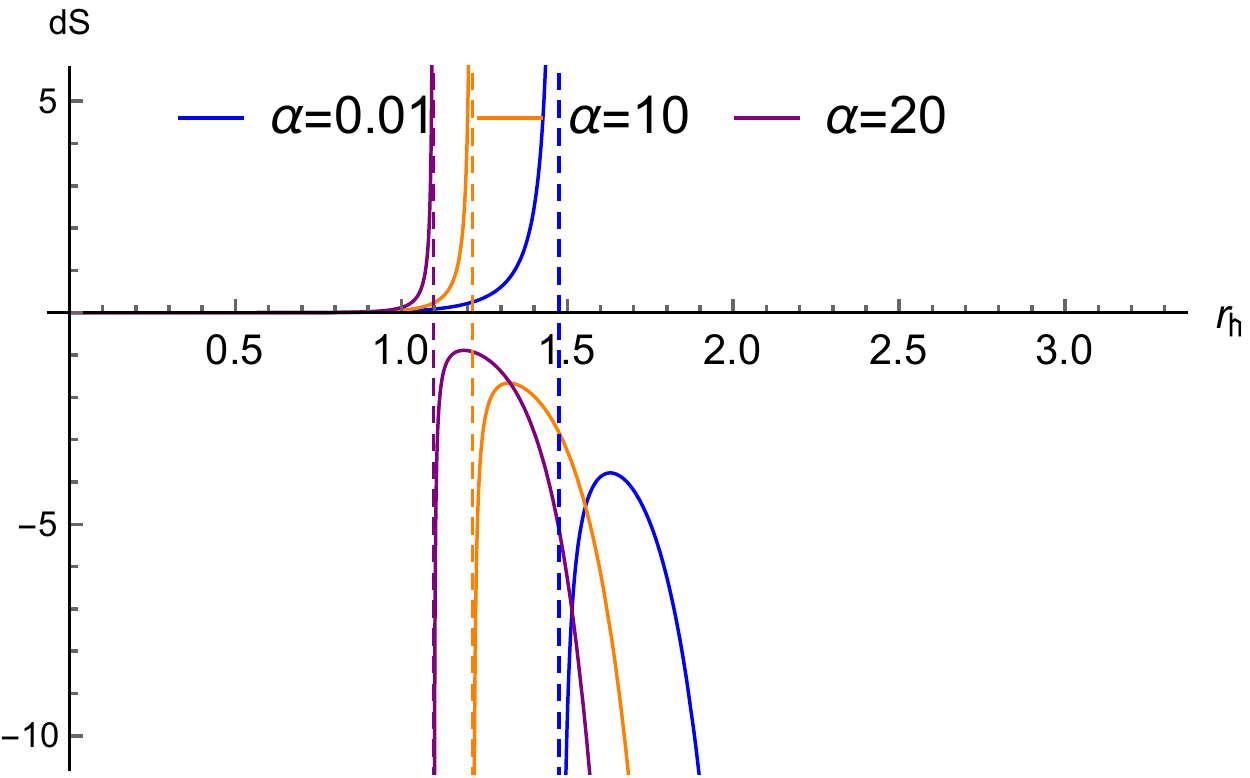}\label{Bdsph7}}
\subfigure[{$d=8$ .}]{
\includegraphics[width=0.45\textwidth]{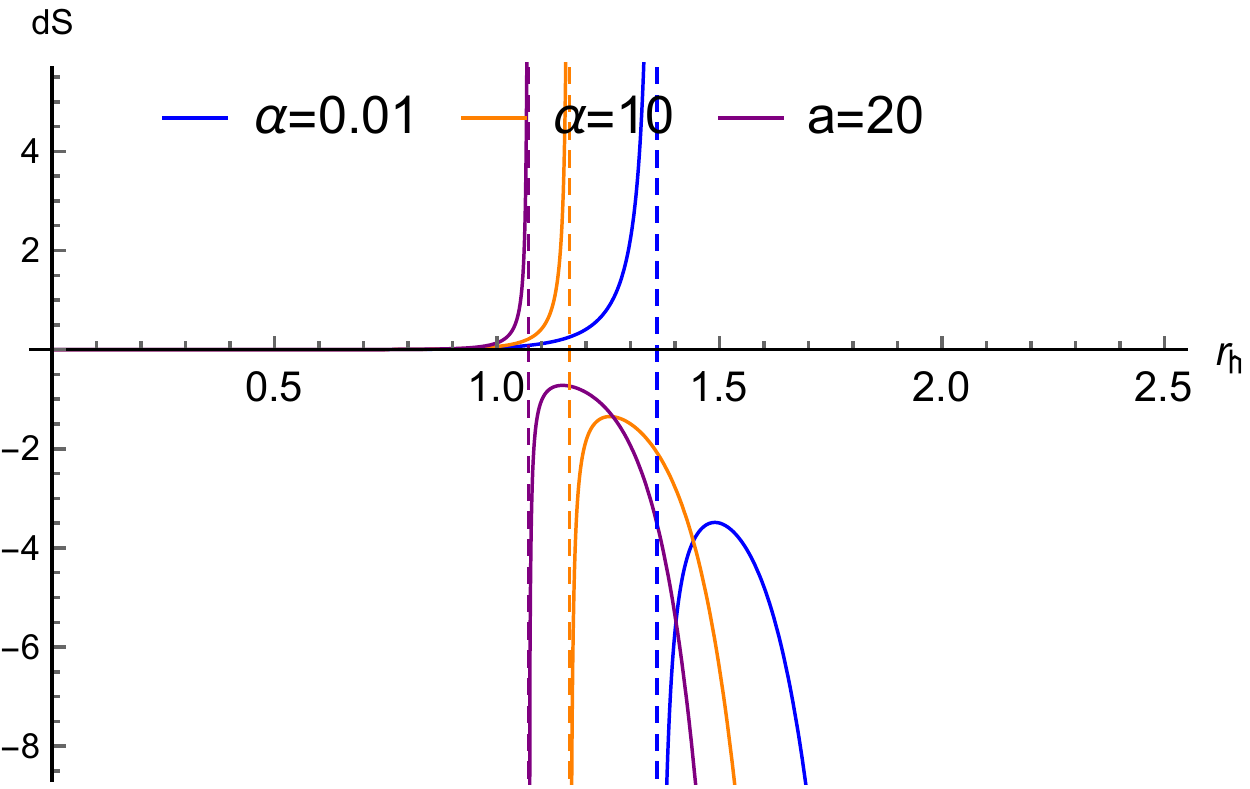}\label{Bdsph8}}
\end{center}
\caption{The relationship between $dS$,$Q$ and $r_{h}$ for $\alpha=0.1,10,20$.
}%
\label{fig:Bds2}%
\end{figure}

\begin{table}
\caption{The relation between $dS$, $Q$ and $r_{h}$ for $d = 5$ in the
extended phase space via scalar field scattering.}%
\label{tab:Bdsda1}%
\begin{tabular}
[c]%
{p{0.6in}|p{0.6in}|p{0.65in}|p{0.6in}|p{0.6in}|p{0.65in}|p{0.6in}|p{0.6in}|p{0.65in}}%
\hline
\multicolumn{3}{c|}{$d\alpha=0.6,da=0.9$} & \multicolumn{3}{|c|}{$d\alpha
=0.6,da=-0.9$} & \multicolumn{3}{|c}{$d\alpha=-0.6,da=0.9$}\\\hline
$Q$ & $r_{h}$ & $dS$ & $Q$ & $r_{h}$ & $dS$ & $Q$ & $r_{h}$ & $dS$\\\hline
0.796654 & 1.69771 & 1.373840 & 0.796654 & 1.69771 & 0.803655 & 0.796654 &
1.69771 & -0.26956\\
0.7 & 1.29939 & 0.175067 & 0.7 & 1.29939 & 0.078073 & 0.7 & 1.29939 &
0.011301\\
0.6 & 1.08142 & 0.058755 & 0.6 & 1.08142 & 0.017997 & 0.6 & 1.08142 &
0.011090\\
0.5 & 0.88942 & 0.020155 & 0.5 & 0.88942 & 0.002618 & 0.5 & 0.88942 &
0.006282\\
0.4 & 0.70741 & 0.006226 & 0.4 & 0.70741 & -0.00058 & 0.4 & 0.70741 &
0.002819\\
0.3 & 0.52949 & 0.001546 & 0.3 & 0.52949 & -0.00056 & 0.3 & 0.52949 &
0.000955\\
0.2 & 0.35303 & 0.000251 & 0.2 & 0.35303 & -0.00017 & 0.2 & 0.35303 &
0.000199\\
0.1 & 0.17676 & 0.000013 & 0.1 & 0.17676 & -0.00001 & 0.1 & 0.17676 &
0.000013\\\hline
\end{tabular}
\end{table}

\begin{table}
\caption{The relation between $dS$, $Q$ and $r_{h}$ for $d = 6$ in the
extended phase space via scalar field scattering.}%
\label{tab:Bdsda2}%
\begin{tabular}
[c]%
{p{0.6in}|p{0.6in}|p{0.65in}|p{0.6in}|p{0.6in}|p{0.65in}|p{0.6in}|p{0.6in}|p{0.65in}}%
\hline
\multicolumn{3}{c|}{$d\alpha=0.6,da=0.9$} & \multicolumn{3}{|c|}{$d\alpha
=0.6,da=-0.9$} & \multicolumn{3}{|c}{$d\alpha=-0.6,da=0.9$}\\\hline
$Q$ & $r_{h}$ & $dS$ & $Q$ & $r_{h}$ & $dS$ & $Q$ & $r_{h}$ & $dS$\\\hline
0.978391 & 1.46517 & 2.1220000 & 0.978391 & 1.46517 & 1.6944400 & 0.978391 &
1.465170 & 0.328897\\
0.9 & 1.26723 & 0.4825120 & 0.9 & 1.26723 & 0.3713990 & 0.9 & 1.267230 &
0.181025\\
0.8 & 1.13797 & 0.1971970 & 0.8 & 1.13797 & 0.1423440 & 0.8 & 1.137970 &
0.0894198\\
0.7 & 1.02504 & 0.0866350 & 0.7 & 1.02504 & 0.0567645 & 0.7 & 1.025040 &
0.0437401\\
0.6 & 0.91634 & 0.0369707 & 0.6 & 0.91634 & 0.0207463 & 0.6 & 0.916339 &
0.0203260\\
0.5 & 0.80685 & 0.0145217 & 0.5 & 0.806853 & 0.0061859 & 0.5 & 0.806853 &
0.0086836\\
0.4 & 0.69303 & 0.0049729 & 0.4 & 0.693037 & 0.0011404 & 0.4 & 0.693037 &
0.0032719\\
0.3 & 0.57118 & 0.0013726 & 0.3 & 0.571182 & -0.000073 & 0.3 & 0.571182 &
0.0010134\\
0.2 & 0.43577 & 0.0002610 & 0.2 & 0.435772 & -0.000112 & 0.2 & 0.435772 &
0.0002198\\
0.1 & 0.27477 & 0.0000205 & 0.1 & 0.274769 & -0.000017 & 0.1 & 0.274769 &
0.0000194\\\hline
\end{tabular}
\end{table}

\begin{table}
\caption{The relation between $dS$, $Q$ and $r_{h}$ for $d = 7$ in the
extended phase space via scalar field scattering.}%
\label{tab:Bdsda3}%
\begin{tabular}
[c]%
{p{0.5in}|p{0.6in}|p{0.7in}|p{0.5in}|p{0.6in}|p{0.7in}|p{0.5in}|p{0.6in}|p{0.7in}}%
\hline
\multicolumn{3}{c|}{$d\alpha=0.6,da=0.9$} & \multicolumn{3}{|c|}{$d\alpha
=0.6,da=-0.9$} & \multicolumn{3}{|c}{$d\alpha=-0.6,da=0.9$}\\\hline
$Q$ & $r_{h}$ & $dS$ & $Q$ & $r_{h}$ & $dS$ & $Q$ & $r_{h}$ & $dS$\\\hline
1.09785 & 1.329700 & 1.0543800 & 1.09785 & 1.329700 & 0.72938600 & 1.09785 &
1.329700 & -0.6389280\\
0.9 & 1.094730 & 0.0862086 & 0.9 & 1.094730 & 0.03887250 & 0.9 & 1.094730 &
-0.0271017\\
0.8 & 1.019380 & 0.0427440 & 0.8 & 1.019380 & 0.01428640 & 0.8 & 1.019380 &
-0.0084702\\
0.7 & 0.945735 & 0.0214957 & 0.7 & 0.945735 & 0.00425739 & 0.7 & 0.945735 &
-0.0014881\\
0.6 & 0.870696 & 0.0105465 & 0.6 & 0.870696 & 0.00039618 & 0.6 & 0.870696 &
0.00082362\\
0.5 & 0.791846 & 0.0048885 & 0.5 & 0.791846 & -0.0007313 & 0.5 & 0.791846 &
0.00120608\\
0.4 & 0.706553 & 0.0020574 & 0.4 & 0.706553 & -0.0007421 & 0.4 & 0.706553 &
0.00089463\\
0.3 & 0.611095 & 0.0007353 & 0.3 & 0.611095 & -0.0004288 & 0.3 & 0.611095 &
0.00046471\\
0.2 & 0.498774 & 0.0001918 & 0.2 & 0.498774 & -0.0001517 & 0.2 & 0.498774 &
0.00015644\\
0.1 & 0.352893 & 0.0000222 & 0.1 & 0.352893 & -0.0000209 & 0.1 & 0.352893 &
0.00002114\\\hline
\end{tabular}
\end{table}

\begin{table}
\caption{The relation between $dS$, $Q$ and $r_{h}$ for $d = 8$ in the
extended phase space via scalar field scattering.}%
\label{tab:Bdsda4}%
\begin{tabular}
[c]%
{p{0.5in}|p{0.6in}|p{0.65in}|p{0.5in}|p{0.6in}|p{0.75in}|p{0.5in}|p{0.6in}|p{0.8in}}%
\hline
\multicolumn{3}{c|}{$d\alpha=0.6,da=0.9$} & \multicolumn{3}{|c|}{$d\alpha
=0.6,da=-0.9$} & \multicolumn{3}{|c}{$d\alpha=-0.6,da=0.9$}\\\hline
$Q$ & $r_{h}$ & $dS$ & $Q$ & $r_{h}$ & $dS$ & $Q$ & $r_{h}$ & $dS$\\\hline
1.17861 & 1.244110 & 0.9120490 & 1.17861 & 1.244110 & 0.65663100 & 1.17861 &
1.244110 & -0.6105110\\
0.99 & 1.075100 & 0.0921938 & 0.99 & 1.075100 & 0.04653060 & 0.99 & 1.075100 &
-0.0389781\\
0.9 & 1.024280 & 0.0510932 & 0.9 & 1.024280 & 0.02101080 & 0.9 & 1.024280 &
-0.0167391\\
0.8 & 0.969409 & 0.0273018 & 0.8 & 0.969409 & 0.00799241 & 0.8 & 0.969409 &
-0.0057605\\
0.7 & 0.913819 & 0.0145162 & 0.7 & 0.913819 & 0.00223971 & 0.7 & 0.913819 &
-0.0011299\\
0.6 & 0.855703 & 0.0074858 & 0.6 & 0.855703 & -0.0000564 & 0.6 & 0.855703 &
0.00056517\\
0.5 & 0.793252 & 0.0036501 & 0.5 & 0.793252 & -0.0007068 & 0.5 & 0.793252 &
0.00091316\\
0.4 & 0.724135 & 0.0016255 & 0.4 & 0.724135 & -0.0006499 & 0.4 & 0.724135 &
0.00071939\\
0.3 & 0.644708 & 0.0006217 & 0.3 & 0.644708 & -0.0003811 & 0.3 & 0.644708 &
0.00039836\\
0.2 & 0.547972 & 0.0001775 & 0.2 & 0.547972 & -0.0001433 & 0.2 & 0.547972 &
0.00014579\\
0.1 & 0.415445 & 0.0000238 & 0.1 & 0.415445 & -0.0000225 & 0.1 & 0.415445 &
0.00002262\\\hline
\end{tabular}
\end{table}

\begin{figure}[tbh]
\begin{center}
\subfigure[{$d=5$ .}]{
\includegraphics[width=0.45\textwidth]{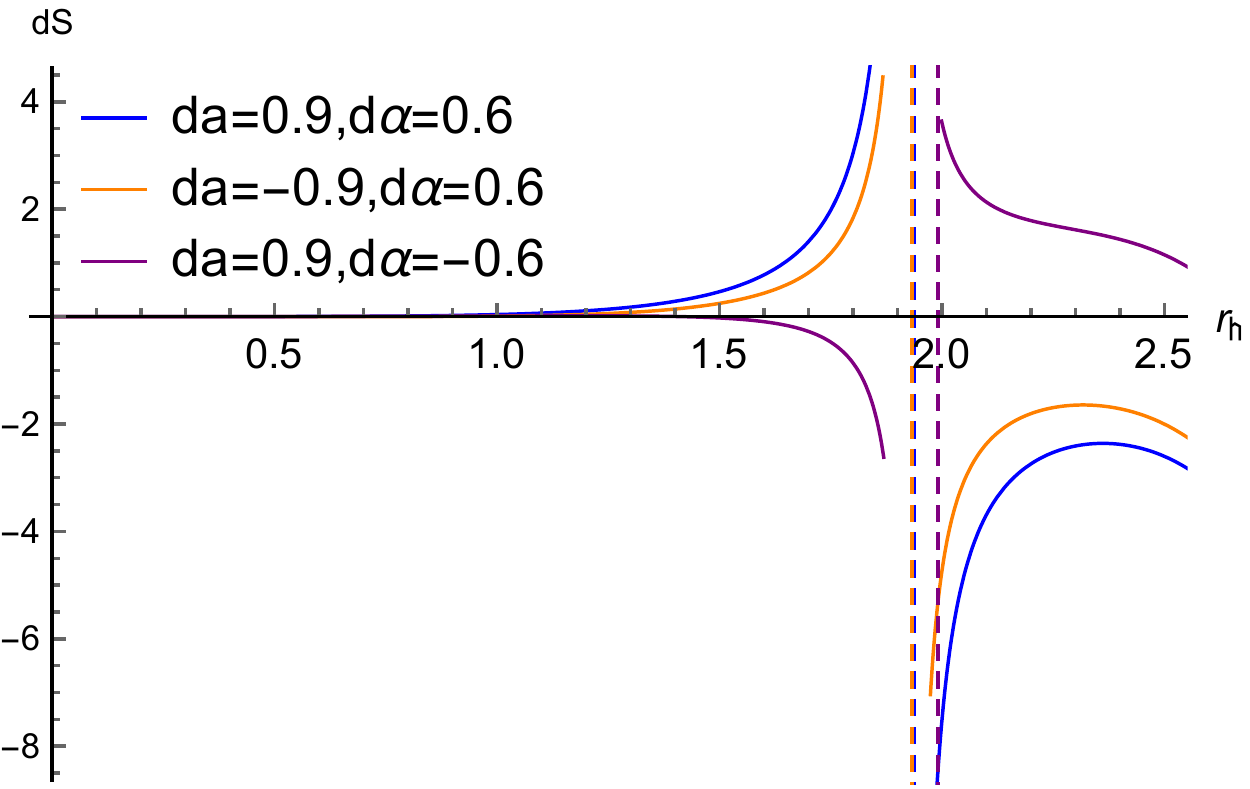}\label{Bdsda5}}
\subfigure[{$d=6$ .}]{
\includegraphics[width=0.45\textwidth]{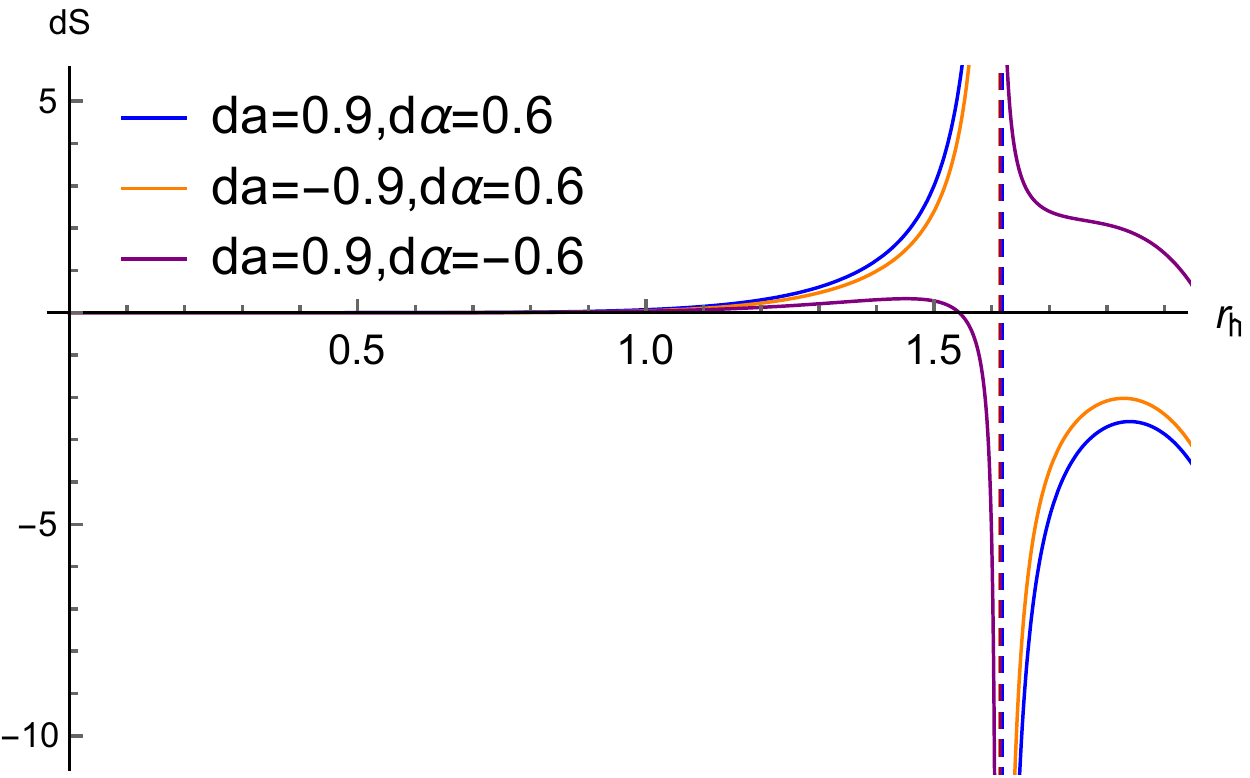}\label{Bdsda6}}
\subfigure[{$d=7$ .}]{
\includegraphics[width=0.45\textwidth]{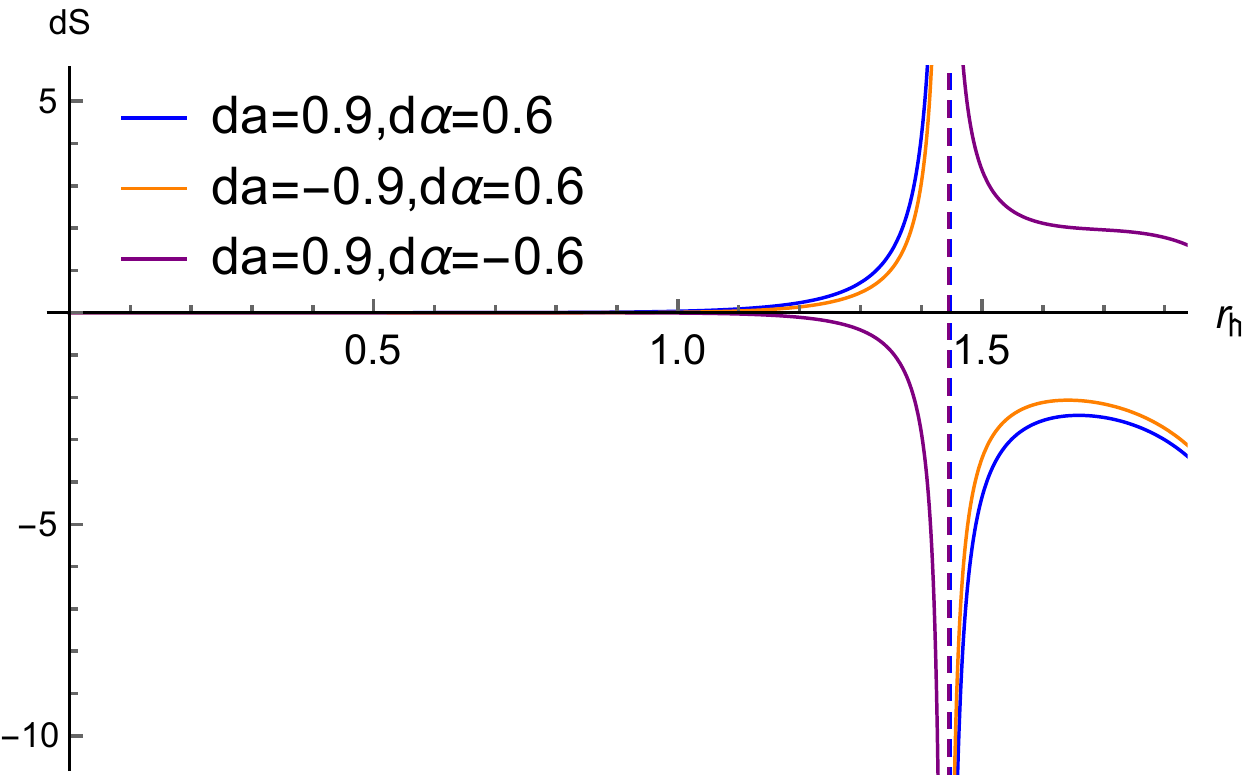}\label{Bdsda7}}
\subfigure[{$d=8$ .}]{
\includegraphics[width=0.45\textwidth]{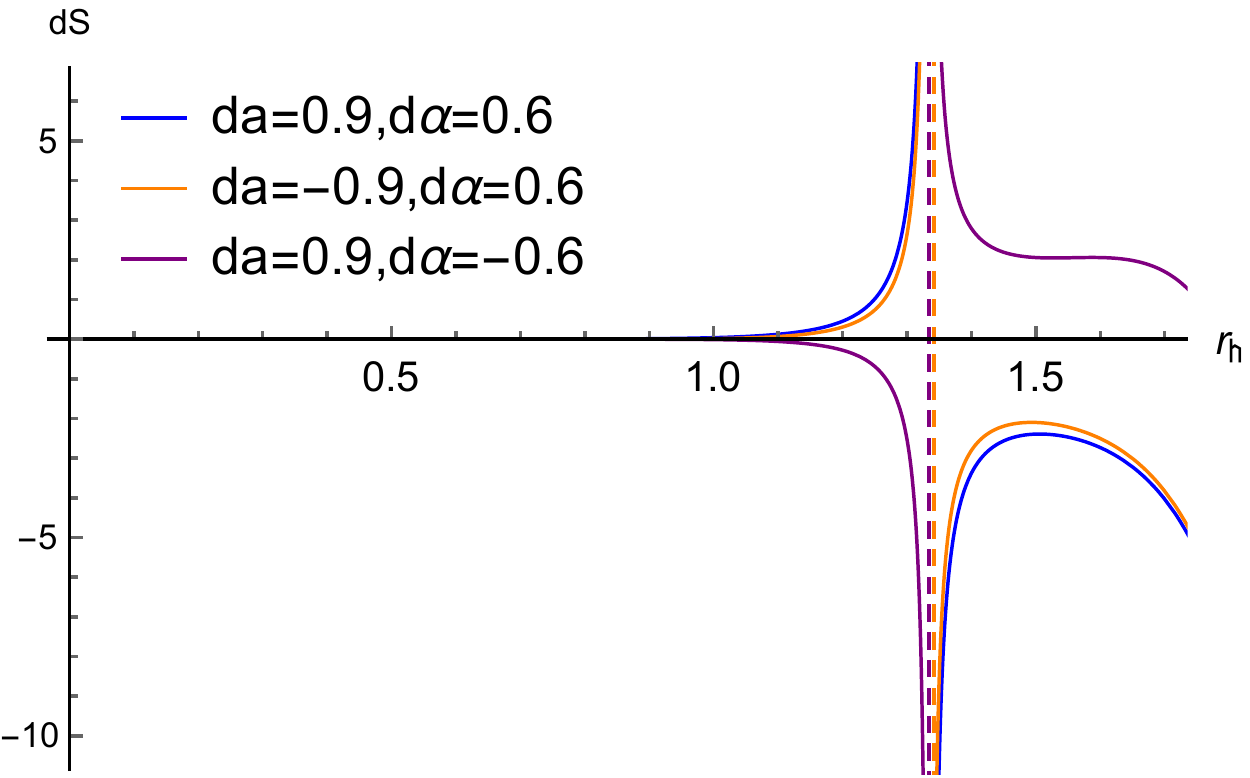}\label{Bdsda8}}
\end{center}
\caption{The relationship between $dS$,$Q$ and $r_{h}$ for $da$ and $d\alpha$.
}%
\label{fig:Bds3}%
\end{figure}\begin{figure}[tbh]
\begin{center}
\subfigure[{$d=5$ .}]{
\includegraphics[width=0.45\textwidth]{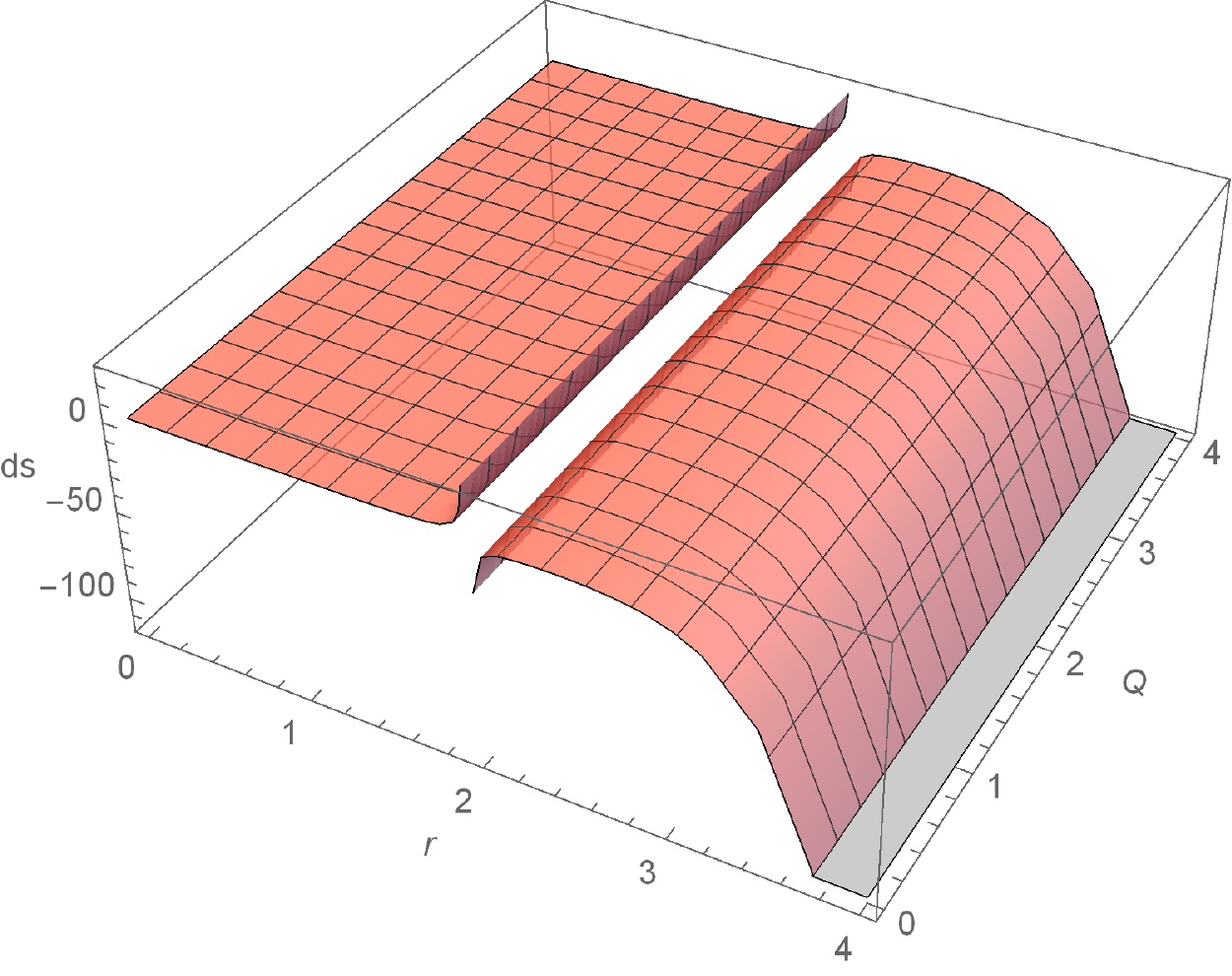}\label{fig:3DBds5}}
\subfigure[{$d=6$ .}]{
\includegraphics[width=0.45\textwidth]{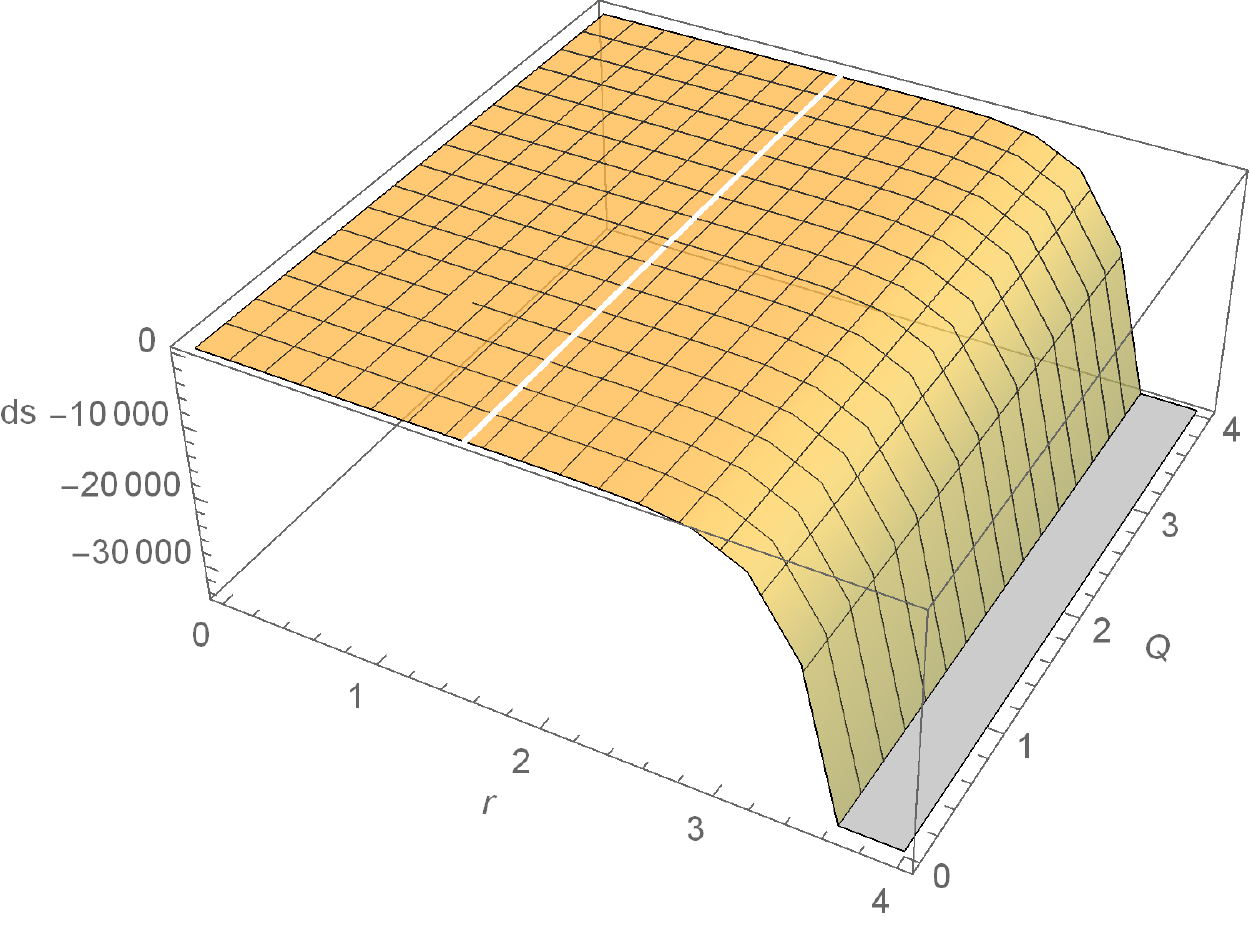}\label{fig:3DBds6}}
\subfigure[{$d=7$ .}]{
\includegraphics[width=0.45\textwidth]{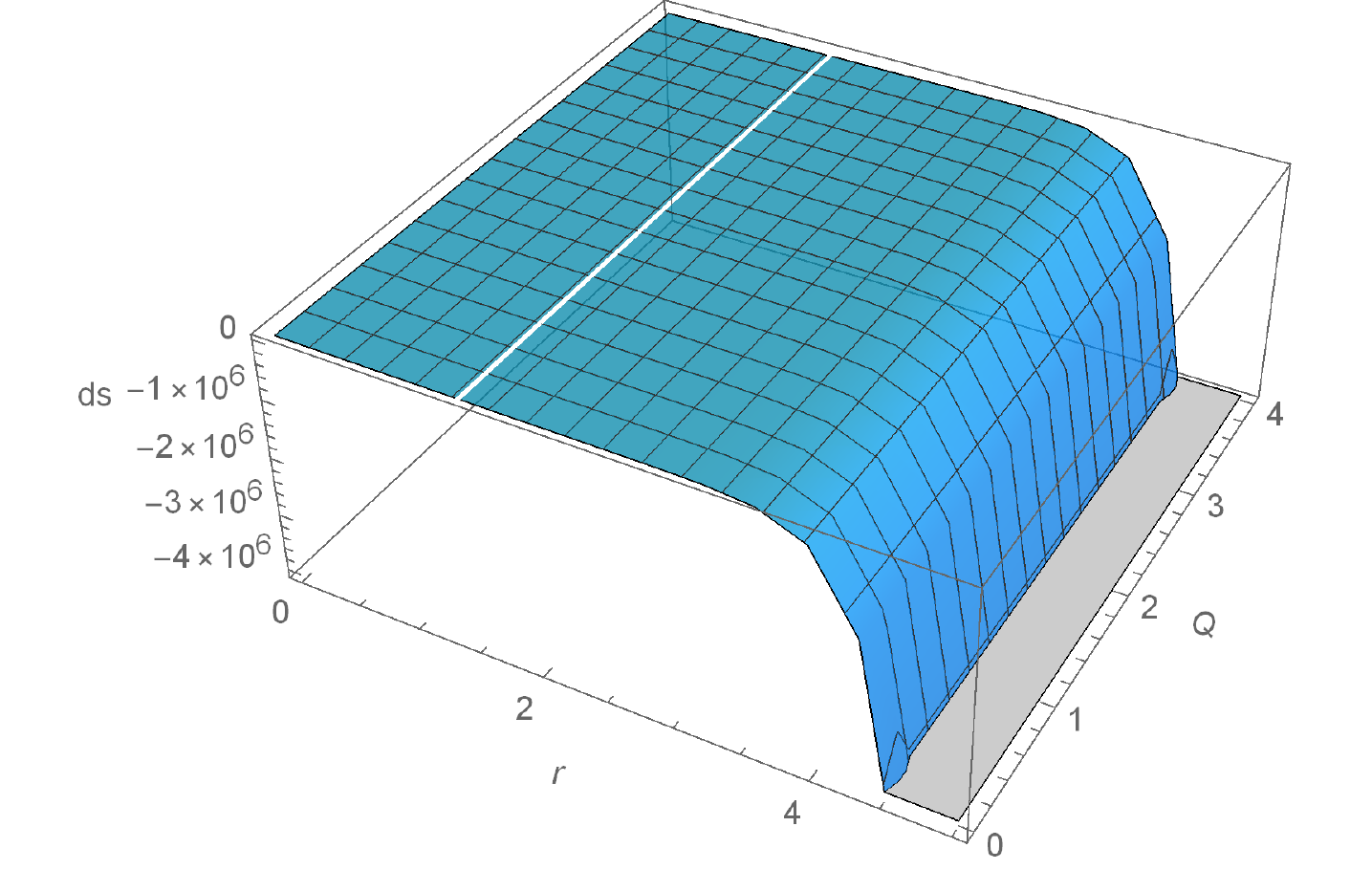}\label{fig:3DBds7}}
\subfigure[{$d=8$ .}]{
\includegraphics[width=0.45\textwidth]{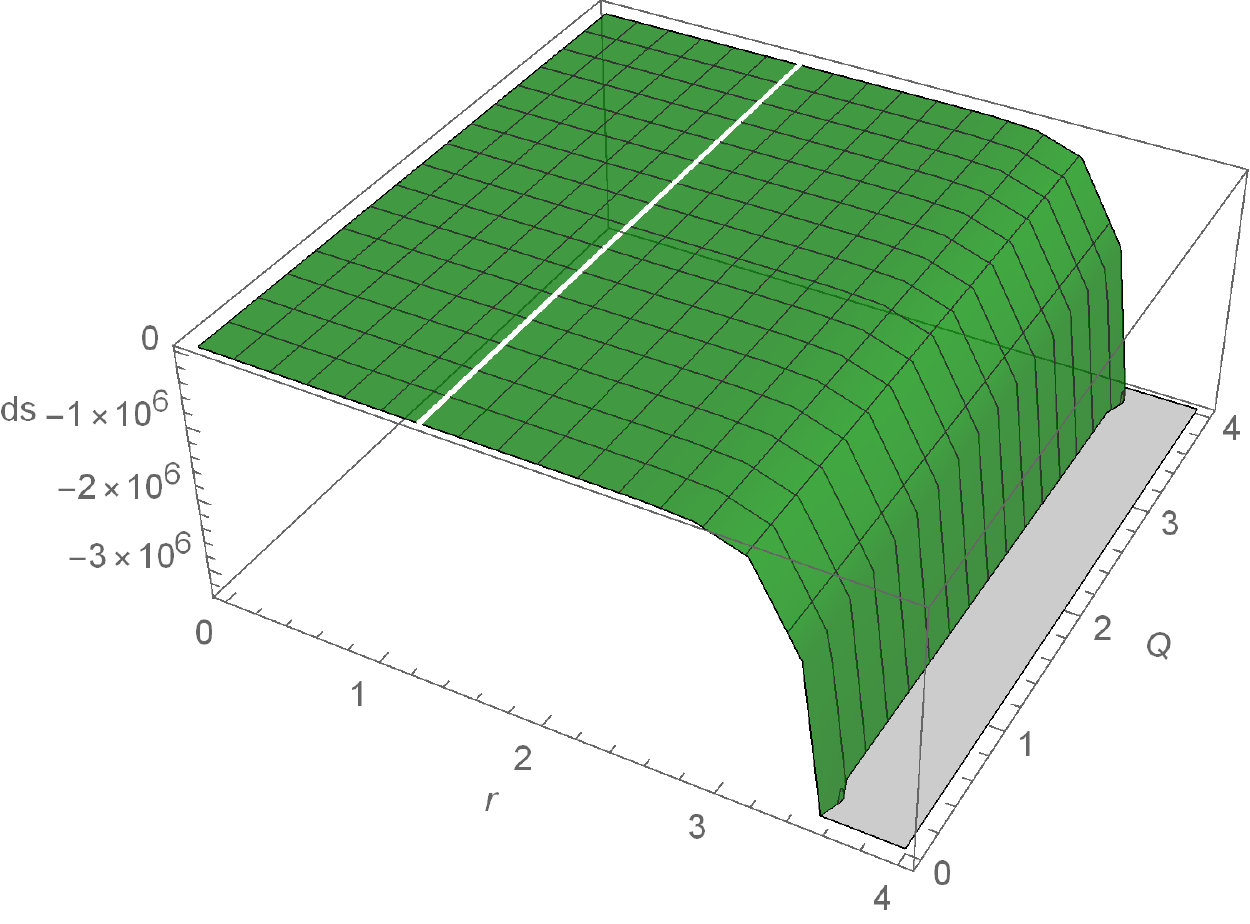}\label{fig:3DBds8}}
\end{center}
\caption{The relationship between $dS$,$Q$ and $r_{h}$. }%
\label{fig:Bds4}%
\end{figure}\begin{figure}[tbh]
\begin{center}
\subfigure[{$d=4$ .}]{
\includegraphics[width=0.45\textwidth]{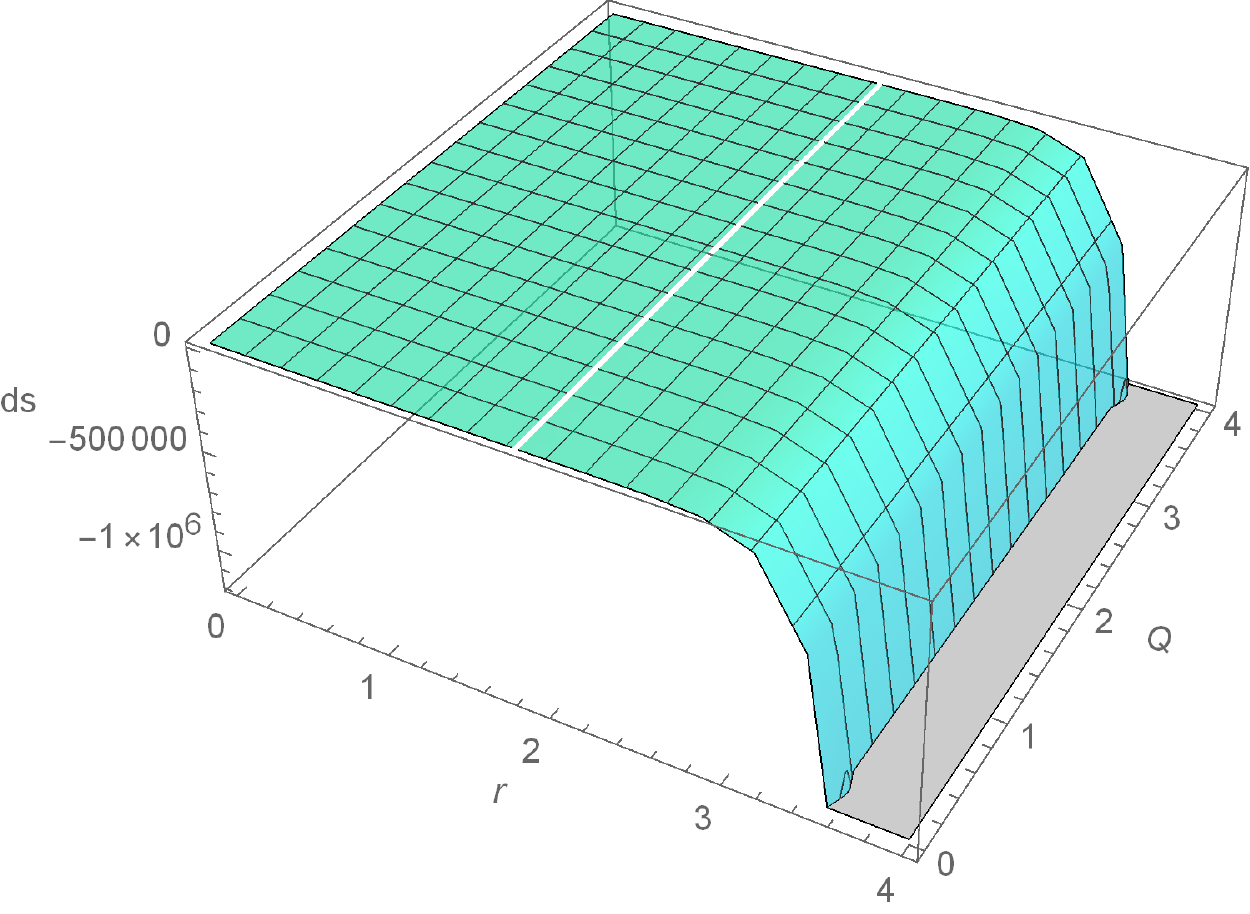}\label{fig:3DBds4}}
\end{center}
\caption{The relationship between $dS$,$Q$ and $r_{h}$. }%
\label{fig:Bds5}%
\end{figure}

Then we focus on the near-extremal black hole. In the process of exploring the
change of entropy, we set $M=1, l=1,\Omega_{d-2}=1, P=1,\omega_{q}%
=\text{-}\frac{d-1}{d-2}$ and $dt=0.0001$, using different charge values in
various dimensions to investigate.  We analysed the effect of parameter $a$ on
$dS$ in the beginning. From Table \ref{tab:Bdsa1}, Table \ref{tab:Bdsa2},
Table \ref{sec:Bb} and Table \ref{tab:Bdsa4}, it is evident that the event
horizon of the black hole and the variation of entropy decreases when the
charge of the black hole decreases, which is the same as the conclusion
obtained at the section \ref{fig:Bds2}. Besides, as the value of $a$
decreases, the value of the critical horizon become smaller. And as the values
of $d$ decrease, the values of the divergent point become greater. From Fig.
$\left( \ref{fig:Bds1}\right) $, we conclude that there are regions of $dS$
which are positive and negative. Similar to that in the particle absorption
section, we will also investigate how $d$ and $\alpha$ affect the value of
$dS_{h}$. By observing Table \ref{tab:Bdspha1}, Table \ref{tab:Bdspha2}, Table
\ref{tab:Bdspha3} and Table \ref{tab:Bdspha4}, we found the event horizon of
the black hole and the variation of entropy decrease when the charge of the
black hole decreases. Fig. $\left( \ref{fig:Bds2}\right) $ has shown that the
values of the divergent point decreases as $d$ or $\alpha$ increases. At the
same time , there's always a region where entropy is less than zero. So far,
the second law of thermodynamics has been violated, irrespective of the values
of $a$ and $\alpha$. In order to intuitively understand the changes in entropy
associated with $dS$ and $d\alpha$, we list different tables and graphs of
functions. In the Table \ref{tab:Bdsda1}, Table \ref{tab:Bdsda2}, Table
\ref{tab:Bdsda3} and Table \ref{tab:Bdsda4}, the influence of $d\alpha$ on the
change of entropy is more obvious. From Fig. $\left( \ref{fig:Bds3}\right) $,
it is obviously that there is indeed a phase change point causes a positive or
negative change in the value of $dS$. It's worth mentioning that the change of
$d\alpha$ has a different effect on the changes of entropy than the section
\ref{sec:B}. In order to explore the difference of entropy change in high and
low dimensional cases, we plot Fig. $\left( \ref{fig:Bds4}\right) $, and
compare which with Fig. $\left( \ref{fig:Bds5}\right) $, then the conclusion
obtained is the same as when the particle is absorbed. That is indeed a phase
change point that divides $dS$ into positive and negative values independent
of dimension $d$. From the above discussion, it can be concluded that the
second law of thermodynamics is not always valid for the near-extremal black
hole in the extended phase space.

\subsection{Stability of horizon}

\label{sec:Cc} In the extended phase space, the stability of horizon also
tests through checking the sign of the minimum value of the function $f(r)$ in
the initial state. Assuming that there is a minimum value of $f(r)$ and the
minimum value is less than zero. For the extremal black hole, $\delta= 0$. For
the near-extremal black hole, $\delta$ is a small quantity. After the flux of
the scalar field enters the black hole, the sign of the minimum value in the
final state can be obtained in term of the initial state. Assuming
$(M,Q,P,r_{0},a,\alpha)$ and $(M+dM,Q+dQ,P+dP,r_{0}+dr_{0},a+da,\alpha
+d\alpha)$ represent the initial state and the finial state, respectively. At
$r=r_{0}+dr_{0}$, $f(M+dM,Q+dQ,P+dP,r_{0}+dr_{0},a+da,\alpha+d\alpha)$ is
written as
\begin{equation}
\begin{aligned} &f\left(M+dM,Q+dQ,P+dP,a+da,\alpha+d\text{\ensuremath{\alpha}},dr_{0}+r_{0}\right)\\ &=\delta+\frac{\partial f}{\partial M}|_{r=r_{0}}dM+\frac{\partial f}{\partial Q}|_{r=r_{0}}dQ+\frac{\partial f}{\partial P}|_{r=r_{0}}dP\\ &+\frac{\partial f}{\partial a}|_{r=r_{0}}da+\frac{\partial f}{\partial\alpha}|_{r=r_{0}}d\text{\ensuremath{\alpha}}+\frac{\partial f}{\partial r}|_{r=r_{0}}dr\\ &=\delta+\delta_{1}+\delta_{2}.\\ \end{aligned}\label{eqn:wccc1}%
\end{equation}
where
\begin{equation}
\begin{aligned} &\frac{\partial f}{\partial r}|_{r=r_{0}}=0,\\ &\frac{\partial f}{\partial M}|_{r=r_{o}}=-\frac{16\pi}{r_{0}^{d-3}(d-2)\Omega_{d-2}},\\ &\frac{\partial f}{\partial Q}|_{r=r_{0}}=\frac{16\pi q}{r_{0}^{2\text{(}d-3)}\Omega_{d-2}\sqrt{2(d-3)(d-2)}},\frac{\partial f}{\partial P}|_{r=r_{0}}=\frac{16\text{\ensuremath{\pi}}r_{0}^{2}}{(d-2)(d-1)},\\ &\frac{\partial f}{\partial\alpha}|_{r=r_{o}}=-\frac{1}{r_{0}^{(d-1)\omega_{q}+d-3}},\\ &\frac{\partial f}{\partial a}|_{r=r_{0}}=\frac{-2}{(d-2)r_{0}^{d-4}}.\\ \end{aligned}\label{eqn:wccc2}%
\end{equation}
Inserting Eq. $\left( \ref{eqn:wccc1}\right) $ into Eq. $\left(
\ref{eqn:wccc2}\right) $ yields
\begin{equation}
\begin{aligned} &f\left(M+dM,Q+dQ,P+dP,a+da,\alpha+d\text{\ensuremath{\alpha}},r_{0}+dr_{0}\right)\\ &=\delta+\frac{16\pi r_{h}^{d-2}P}{r_{0}^{d-3}}[\omega_{q}-\sqrt{\frac{d-2}{2(d-3)}}\frac{q^{2}}{r_{h}^{d-3}}]\{\frac{16\pi Pr_{h}}{(d-2)4\pi T-16\pi Pr_{h})}[\frac{q^{2}}{r_{h}^{d-3}}-\frac{\omega_{q}}{d-2}]\\ &+[\frac{q^{2}}{r_{0}^{d-3}}-\frac{\omega_{q}}{d-2}]\}dt+\frac{16\pi}{(d-2)(d-1)}(r_{0}^{2}-\frac{r_{h}^{d-1}}{r_{0}^{d-3}})dP\\ &+\frac{2}{d-2}(\frac{16\pi Pr_{h}}{16\pi Pr_{h}-(d-2)4\pi T}\frac{r_{h}^{d-3}}{r_{0}^{d-3}}r_{h}^{4-d}-r_{0}^{4-d})da\\ &+(\frac{16\pi Pr_{h}}{16\pi Pr_{h}-(d-2)4\pi T}\frac{r_{h}^{d-3}}{r_{0}^{d-3}}\frac{1}{r_{h}^{(d-1)\omega_{q}+d-3}}-\frac{1}{r_{0}^{(d-1)\omega_{q}+d-3}})d\alpha.\\ \end{aligned}
\end{equation}

Then, we have
\begin{equation}
\begin{aligned} \delta&=0,\\ \delta_{1}&=\frac{16\pi r_{h}^{d-2}P}{r_{0}^{d-3}}[\omega_{q}-\sqrt{\frac{d-2}{2(d-3)}}\frac{q^{2}}{r_{h}^{d-3}}]\{\frac{16\pi Pr_{h}}{(d-2)4\pi T-16\pi Pr_{h})}[\frac{q^{2}}{r_{h}^{d-3}}-\frac{\omega_{q}}{d-2}\\ &+[\frac{q^{2}}{r_{0}^{d-3}}-\frac{\omega_{q}}{d-2}]\}dt\\ \delta_{2}&=\frac{2}{d-2}(\frac{16\pi Pr_{h}}{16\pi Pr_{h}-(d-2)4\pi T}\frac{r_{h}^{d-3}}{r_{0}^{d-3}}r_{h}^{4-d}-r_{0}^{4-d})da+\frac{16\pi}{(d-2)(d-1)}(r_{0}^{2}-\frac{r_{h}^{d-1}}{r_{0}^{d-3}})dP\\ &+(\frac{16\pi Pr_{h}}{16\pi Pr_{h}-(d-2)4\pi T}\frac{r_{h}^{d-3}}{r_{0}^{d-3}}\frac{1}{r_{h}^{(d-1)\omega_{q}+d-3}}-\frac{1}{r_{0}^{(d-1)\omega_{q}+d-3}})d\alpha.\\ \end{aligned}\label{eqn:wccc3}%
\end{equation}
In the extremal black hole, $r_{0} = r_{h}, T = 0$ and $d f_{min} = 0$. Hence,
Eq. $\left( \ref{eqn:wccc1}\right) $ is written as
\begin{equation}
\begin{aligned} &\delta=0,\\ &\delta_{1}=0,\\ &\delta_{2}=0.\\ \end{aligned}
\end{equation}
Thus, we have
\begin{equation}
f\left( M+dM,Q+dQ,P+dP,a+da,\alpha+d\text{$\alpha$},dr_{0}+r_{0}\right)
=\delta+\delta_{1}+\delta_{2}=0.\label{eqn:wcc4}%
\end{equation}
Therefore, the scattering of the scalar field doesn't cause the horizon
changes in the minimum value of $f(r)$. This proves that the extremal black
hole is still hold and the horizon is still exists at the final state. For the
near-extremal black hole, $r_{0}$ and $r_{h}$ are very close. To calculate the
value of Eq. $\left( \ref{eqn:wccc3}\right) $, we can suppose that
$r_{h=}r_{0}+\epsilon$, where $0<\epsilon\ll1$. In this situation, Eq. $\left(
\ref{eqn:wccc3}\right) $ is written as
\begin{equation}
\begin{aligned} \delta&<0,\\ \delta_{1}&=\{16\pi Pr_{0}(\omega_{q}-q^{2}\sqrt{\frac{d-2}{2(d-3)}}\frac{1}{r_{0}^{d-3}})(\frac{q^{2}r_{0}+q^{2}}{r_{0}^{d-3}}-\frac{\omega_{q}+\omega_{q}r_{0}}{d-2})+O(\epsilon)+O(\epsilon)^{2}\}dt,\\ \delta_{2}&=\frac{2}{d-2}(\frac{16\pi P(r_{0}+\epsilon)}{16\pi P(r_{0}+\epsilon)-(d-2)4\pi T}\frac{r_{0}+(d-3)\epsilon+O(\epsilon)^{2}}{r_{0}^{d-3}})da\\ &-\frac{16\pi}{(d-2)(d-1)}r_{0}^{2}(d-1)\epsilon dP\\ &+\{\frac{16\pi P}{16\pi P(r_{0}+\epsilon)-(d-2)4\pi T}[\frac{1}{r_{0}^{(d-1)\omega_{q}d-4}}+\frac{(2-d)\omega_{q}\epsilon}{r_{0}^{(d-1)\omega_{q}+d-3}}+O(\epsilon)^{2}]-\frac{1}{r_{0}^{(d-1)\omega_{q}+d-3}}\}d\alpha,\\ \end{aligned}
\end{equation}
where $dt$ is an infinitesimal scale and is set as $dt\sim\epsilon$. If the
initial black hole is near extremal, we have $dP\sim\epsilon,d\alpha
\sim\epsilon,da\sim\epsilon$. So we have
\begin{equation}
\delta<0,\delta_{1}+\delta_{2}\ll\delta,
\end{equation}
and
\begin{equation}
f\left( M+dM,Q+dQ,P+dP,a+da,\alpha+d\text{$\alpha$},dr_{0}+r_{0}\right)
\thickapprox\delta<0.
\end{equation}
Therefore, the event horizon exists and the black hole isn't overcharged in
the finial state. The weak cosmic censorship conjecture is valid in the
near-extremal black hole.

\subsection{A new assumption: $dE = dM$}

\label{sec:Cd} In the previous subsection, we found that the second law of
thermodynamics may be violated. It is believed that this assumption of
violation of the second law is not physical but is an absurd conclusion of a
false assumption that scalar field scattering changes the internal energy of a
black hole. In this subsection, we assume that after the scalar field
scattering, the black hole's enthalpy changes. When the energy flux is assumed
to the enthalpy of the black hole
\begin{equation}
dE=dM,\label{eqn:new1}%
\end{equation}
where the variation of the charge of the black hole $dQ$ is the same as the
variation of the electric charge flux of the scalar field $de$
\begin{equation}
dQ=(\frac{de}{dt})dt.\label{eqn:new2}%
\end{equation}
Thus, we obtain
\begin{equation}
dM=\omega_{q}(\omega_{q}+q\phi)r_{h}^{d-2}dt, dQ=q(\omega_{q}+q\phi
)r_{h}^{d-2}dt.\label{eqn:new3}%
\end{equation}
As a charged particle dropped into the black hole, the configurations of the
black hole will be changed. This progress will lead to a shift for the
horizon, The relation between the functions $f(r)$ and
$f(M+dM,Q+dQ,P+dP,a+da,\alpha+d\alpha, r_{h}+dr_{h})$ is
\begin{equation}
\begin{aligned} &f\left(M+dM,Q+dQ,P+dP,a+da,\alpha+d\alpha,r_{h}+dr_{h}\right)=f(r)\\ &+\frac{\partial f}{\partial M}|_{r=r_{h}}dM+\frac{\partial f}{\partial Q}|_{r=r_{h}}dQ+\frac{\partial f}{\partial r}|_{r=r_{h}}dr_{h}+\\ &\frac{\partial f}{\partial P}|_{r=r_{h}}dP+\frac{\partial f}{\partial a}|_{r=r_{h}}da+\frac{\partial f}{\partial\alpha}|_{r=r_{h}}d\text{\ensuremath{\alpha}}. \label{eqn:new4} \end{aligned}
\end{equation}
Substituting Eq. $\left( \ref{eqn:new3}\right) $ into Eq. $\left(
\ref{eqn:new4}\right) $, we can obtain the value of the $dr_{h}$, which is
\begin{equation}
\begin{aligned} &dr_{h}=\frac{-4r_{h}}{T\varOmega_{d-2}}[\frac{2q^{2}\omega_{q}}{r_{h}^{d-3}\sqrt{2(d-2)(d-3)}}-\frac{\omega_{q}^{2}}{d-2}-\frac{q^{4}}{2(d-3)r_{h}^{2(d-3)}}]dt\\ &-\frac{4r_{h}^{2}}{T(d-2)(d-1)}dP+\frac{1}{4\pi Tr_{h}^{(d-1)\omega_{q}+d-3}}d\alpha+\frac{2}{4\pi T(d-2)r_{h}^{d-4}}da. \end{aligned}\label{eqn:ne55}%
\end{equation}
With the aid of [$dS=\frac{\Omega_{d-2}(d-2)r_{h}^{d-3}}{4}dr_{h}$], the variation
of entropy is given by
\begin{equation}
\begin{aligned} &dS=\frac{(d-2)r_{h}^{d-2}}{-T}[\frac{2q^{2}\omega_{q}}{r_{h}^{d-3}\sqrt{2(d-2)(d-3)}}-\frac{\omega_{q}^{2}}{d-2}-\frac{q^{4}}{2(d-3)r_{h}^{2(d-3)}}]dt\\ &-\frac{r_{h}^{d-1}\varOmega_{d-2}}{T(d-1)}dP+\frac{\varOmega_{d-2}(d-2)}{16\pi Tr_{h}^{(d-1)\omega_{q}}}d\alpha+\frac{\varOmega_{d-2}(d-2)r_{h}}{8\pi T(d-2)}da.\\ \end{aligned}\label{eqn:new6}%
\end{equation}
Using Eq. $\left( \ref{eqn:new6}\right) $, it is easy to get
\begin{equation}
\begin{aligned} &TdS-VdP\\ &=\frac{4Pr_{h}^{d-1}-(d-2)Tr_{h}^{d-2}}{T}[\frac{2q^{2}\omega_{q}}{r_{h}^{d-3}\sqrt{2(d-2)(d-3)}}-\frac{\omega_{q}^{2}}{d-2}-\frac{q^{4}}{2(d-3)r_{h}^{2(d-3)}}]dt\\ &+\frac{\varOmega_{d-2}[4Pr_{h}^{d}-(d-2)Tr_{h}^{d-1}]}{T(d-2)(d-1)}dP-\frac{\varOmega_{d-2}[4Pr_{h}^{d-2}-(d-2)Tr_{h}^{d-3}]}{16\pi Tr_{h}^{(d-1)\omega_{q}+d-3}}d\alpha\\ &-\frac{\varOmega_{d-2}[4Pr_{h}^{d-2}-(d-2)Tr_{h}^{d-3}]}{8\pi T(d-2)r_{h}^{d-4}}da.\\ \end{aligned}\label{eqn:new7}%
\end{equation}
Then, the Eq. $\left( \ref{eqn:new3}\right) $ reduces to
\begin{equation}
dM=TdS+VdP+\phi dQ+\mathcal{A}da+\mathcal{Q}d\alpha.\label{eqn:new8}%
\end{equation}

Obviously, the Eq. $\left( \ref{eqn:new8}\right) $ is exactly same as Eq.
$\left( \ref{eqn:fcc2}\right) $. This means that the first law of black hole
thermodynamics still holds. Next, we will continue to check the second law of
black hole thermodynamics when a charged particle is captured by the black
hole. As the black hole entropy increases in a clockwise direction will not be
less than zero, we can examine the second law of thermodynamics black hole by
studying the change in entropy. For the extremal black hole, the temperature
is zero. Then, combining this condition and the black hole mass and the
variation of entropy finally reads
\begin{equation}
dS_{extremal}\rightarrow\infty.\label{eqn:new9}%
\end{equation}
It is true from Eq. $\left( \ref{eqn:new9}\right) $ that the second law of
black hole thermodynamics is still hold for the extremal black holes. In
addition, the temperatures of the non-extremal black hole is greater than
zero, so the variation of entropy $dS$ always has a positive value under
certain conditions, which means the second law of black hole thermodynamics
dose not violate for the non-extremal black holes. Next, we will further check
the stability of horizon of the black hole. In a similar way, $f\left(
M+dM,Q+dQ,P+dP,a+da,\alpha+d\text{$\alpha$},r_{0}+dr_{0}\right) $ is rewritten
as
\begin{equation}
\begin{aligned} &f\left(M+dM,Q+dQ,P+dP,a+da,\alpha+d\text{\ensuremath{\alpha}},r_{0}+dr_{0}\right)\\ &=\delta+\frac{16r_{h}^{d-2}}{r_{0}^{d-3}\varOmega_{d-2}}[\frac{q^{2}\omega_{q}}{r_{0}^{d-3}\sqrt{2(d-2)(d-3)}}-\frac{\omega_{q}^{2}}{d-2}-\frac{q^{4}}{r_{0}^{d-3}r_{h}^{d-3}2(d-3)}+\frac{\omega_{q}q^{2}}{r_{h}^{d-3}\sqrt{2(d-3)(d-2)}}]dt\\ &-\frac{16\pi p^{r}}{r_{0}^{d-3}(d-2)\Omega_{d-2}}+\frac{16\text{\ensuremath{\pi}}r_{0}^{2}}{(d-2)(d-1)}dP\\ &-\frac{1}{r_{0}^{(d-1)\omega_{q}+d-3}}d\alpha-\frac{2}{(d-2)r_{0}^{d-4}}da.\\ \end{aligned}\label{eqn:neee10}%
\end{equation}
Therefore, at the minimum point, we have
\begin{equation}
\begin{aligned} &\delta=0,\\ &\delta_{1}=+\frac{16r_{h}^{d-2}}{r_{0}^{d-3}\varOmega_{d-2}}[\frac{q^{2}\omega_{q}}{r_{0}^{d-3}\sqrt{2(d-2)(d-3)}}-\frac{\omega_{q}^{2}}{d-2}-\frac{q^{4}}{r_{0}^{d-3}r_{h}^{(d-3)}2(d-3)}+\frac{\omega_{q}q^{2}}{r_{h}^{d-3}\sqrt{2(d-3)(d-2)}}]dt.\\ &\delta_{2}=-\frac{16\pi p^{r}}{r_{0}^{d-3}(d-2)\Omega_{d-2}}-\frac{2r_{0}^{2}}{l^{3}}dl -\frac{1}{r_{0}^{(d-1)\omega_{q}+d-3}}d\alpha-\frac{2}{(d-2)r_{0}^{d-4}}da.\\ \label{eqn:zn3} \end{aligned}
\end{equation}
In the extremal black hole, $r_{0} = r_{h}$, $T = 0$, and $d f_{min} = 0$.
Hence,
\begin{equation}
f\left( M+dM,Q+dQ,P+dP,a+da,\alpha+d\text{$\alpha$},r_{0}+dr_{0}\right)
<0.\label{eqn:wcc444}%
\end{equation}
Therefore, the event horizon exists in the extremal black hole. For the
near-extremal black hole, the location $r_{0}$ is no longer equal to the event
horizon $r_{h}$, which leads to that the condition is not available. To
calculate the value of Eq. $\left( \ref{eqn:zn3}\right) $, we can suppose that
$r_{h=}r_{0}+\epsilon$, where $0<\epsilon\ll1$. Using the same method above we
can get
\begin{equation}
\begin{aligned} &\delta=0,\\ &\delta_{1}=\{\frac{16\pi}{\Omega_{d-2}}[\frac{q^{2}r_{0}\omega_{q}+q^{2}\omega_{q}}{r_{0}^{d-3}\sqrt{2(d-3)(d-2)}}-\frac{\omega_{q}^{2}r_{0}}{d-2}-\frac{q^{4}}{2r_{0}^{d-3}(d-3)}]+O(\epsilon)+O(\epsilon)^{2}\}dt,\\ &\delta_{2}=-\frac{16\pi p^{r}}{r_{0}^{d-3}(d-2)\Omega_{d-2}}-\frac{2r_{0}^{2}}{l^{3}}dl -\frac{1}{r_{0}^{(d-1)\omega_{q}+d-3}}d\alpha-\frac{2}{(d-2)r_{0}^{d-4}}da.\\ \label{eqn:zn4} \end{aligned}
\end{equation}
If the initial black hole is near extremal, we have $dl\sim\epsilon
,d\alpha\sim\epsilon,da\sim\epsilon$,$dt\sim\epsilon$. So we haave
\begin{equation}
\delta<0,\delta_{1}+\delta_{2}\ll\delta.
\end{equation}
and
\begin{equation}
f\left( M+dM,Q+dQ,P+dP,a+da,\alpha+d\text{$\alpha$},r_{0}+dr_{0}\right)
\thickapprox\delta<0.
\end{equation}
Therefore, the event horizon exists and the black hole isn't overcharged in
the finial state. The weak cosmic censorship conjecture is valid in the
near-extremal black hole.

\section{Discussion and Conclusion}

\label{sec:D} This paper investigated the first and second laws of
thermodynamics and the stability of the horizon of a charged AdS black hole
with cloud of strings and quintessence present in d-dimensional spacetime via
particle absorption and scalar field scattering in the extended phase space.
Our research was based on two assumptions in two cases, i.e., the energy of
the particle is related to the internal energy or enthalpy of the black hole
in the case of particle absorption, and the energy flux of the scalar field is
combined with the internal energy or enthalpy of the black hole under the
scalar field scattering.

At first, we reviewed the thermodynamics of the black hole by considering the
cosmological constant as the function of thermodynamic pressure $P$, and
treating the state parameters of cloud of strings and quintessence as
variables. Then we studied the absorption of scalar particle and fermion, and
found they finally simplified to the same relation $p^{r}=\omega-q\phi$ by
deriving the Hamilton Jacobi equation. Furthermore, we tested the validity of
the first and second laws of thermodynamics and the stability of the horizon
under the assumption that the energy of particle $E$ changes the internal
energy of the black hole $dU$. The first law of thermodynamics is recovered,
and the second law of thermodynamics is indefinite. The WCCC is valid all the
time for extremal and near-extremal black holes, which means the horizons
stable exist.

During the discussion of the second law, we mainly studied the change of the
black hole entropy under different circumstances dimensions after fixing the
variables. With the variation of the charge of the black hole, we found that
there was always a phase transition point, which divides the variation of
entropy into positive and negative region. The variation of entropy is
negative for the extremal and near-extremal black holes, while positive for
the far-extremal black holes. Therefore, it is concluded that in the extended
phase space, the second law is violated for the extremal and near-extremal
black holes. In addition, we compared the entropy changes of black holes in
high and low dimensions, and found that the value of the phase change point
increases with the decreases of dimension. While for the stability of
horizons, we checked the sign of the minimum value of $f(r)$, and found it
never greater than zero. Therefore, neither extremal black holes nor
near-extremal black holes will be overcharged.

Furthermore, another assumption was considered, namely the energy of particle
$E$ changes the enthalpy of the black hole $dM$. In this case, we found that
the first law of thermodynamics and the stability of horizons results were
same with the results obtained by the former $E=dU$. Moreover, the increment
of the black hole's entropy is always positive after particle absorption.
Therefore, the second law of thermodynamics holds. The results are concluding
in Table \ref{tab:12wccc}.

\begin{table}%
\begin{tabular}
[c]{|p{0.7in}|p{2.7in}|p{2.7in}|}\hline
& \multicolumn{2}{l|}{Particle absorption}\\\hline
& E=dU. & E=dM.\\\hline
1st law & $dM=TdS+VdP+\phi dQ+\mathcal{A}da+\mathcal{Q}d\alpha$. &
$dM=TdS+VdP+\phi dQ+\mathcal{A}da+\mathcal{Q}d\alpha$.\\\hline
2nd law & Indefinite. & Satisfied.\\\hline%
\begin{tabular}
[c]{@{}l@{}}%
The\\
stability\\
of horizon
\end{tabular}
&
\begin{tabular}
[c]{@{}l@{}}%
The horizon still exists for the\\
extremal and near-extremal\\
black holes.
\end{tabular}
&
\begin{tabular}
[c]{@{}l@{}}%
The horizon still exists for the\\
extremal and near-extremal\\
black holes.
\end{tabular}
\\\hline
\end{tabular}
\caption{{\footnotesize {}{}{}{}Results for the first and second laws of
thermodynamics and the the stability of horizons, which are tested for
d-dimensional charged AdS black holes with cloud of strings and quintessence
via particle absorption.}}%
\label{tab:12wccc}%
\end{table}

\begin{table}%
\begin{tabular}
[c]{|p{0.7in}|p{2.7in}|p{2.7in}|}\hline
& \multicolumn{2}{l|}{Scalar field scattering}\\\hline
& dE=dU. & dE=dM.\\\hline
1st law & $dM=TdS+VdP+\phi dQ+\mathcal{A}da+\mathcal{Q}d\alpha$. &
$dM=TdS+VdP+\phi dQ+\mathcal{A}da+\mathcal{Q}d\alpha$.\\\hline
2nd law & Indefinite. & Satisfied.\\\hline%
\begin{tabular}
[c]{@{}l@{}}%
The\\
stability\\
of horizon
\end{tabular}
&
\begin{tabular}
[c]{@{}l@{}}%
Satisfied for the extremal and\\
near-extremal black holes. The\\
extremal/near-extremal black hole stays\\
extremal/near-extremal after the scalar\\
field scattering.
\end{tabular}
&
\begin{tabular}
[c]{@{}l@{}}%
Satisfied for the extremal and\\
near-extremal black holes. The\\
extremal/near-extremal black hole stays\\
extremal/near-extremal after the scalar\\
field scattering.
\end{tabular}
\\\hline
\end{tabular}
\caption{{\footnotesize {}{}{}{}Results for the first and second laws of
thermodynamics and the the stability of horizons, which are tested for
d-dimensional charged AdS black holes with cloud of strings and quintessence
via scalar field scattering.}}%
\label{tab:13wccc}%
\end{table}

In the section \ref{sec:C} , at first the variations of the energy and charge
of the black hole in an infinitesimal time interval after scalar field
scattering were calculated. Then we recovered the first law of thermodynamics
and discussed the validity of the second law of thermodynamics. Using the same
research methods as the particle absorption part, we also found that there was
always a phase transition point. Then, we further calculated and discussed the
stability of the horizon via checking the sign of the minimum value of $f(r)$.
Moreover, the thermodynamics and the stability of the horizon were also
discussed under two assumptions, i.e., the energy flux of the scalar field
$dE$ changes the internal energy of the black hole $dU$ and the energy flux of
the scalar field $dE$ changes the enthalpy of the black hole $dM$. Our results
are summarized in Table \ref{tab:13wccc}. \begin{table}[tbh]
\begin{centering}
\begin{tabular}{|p{3.0in}|p{3.0in}|}
\hline
Types of black holes & 1st law  \tabularnewline
\hline
RN-AdS BH  & $dM=TdS+VdP+\text{\ensuremath{\phi}}dQ$\tabularnewline
\hline
RN-AdS BH with cloud of strings  & $dM=TdS+VdP+\text{\ensuremath{\phi}}dQ-\frac{r_{h}}{2}da$\tabularnewline
\hline
RN-AdS BH with quintessence  & $dM=TdS+VdP+\text{\ensuremath{\phi}}dQ-\frac{1}{2r_{h}^{3\omega_{q}}}d\alpha$\tabularnewline
\hline
RN-AdS BH with cloud of strings and quintessence  & $dM=TdS+VdP+\text{\ensuremath{\phi}}dQ-\frac{1}{2r_{h}^{3}\omega_{q}}d\alpha-\frac{r_{h}}{2}da$\tabularnewline
\hline
d-dimensional RN-AdS BH with cloud of strings and quintessence  & $dM=TdS+VdP+\phi dQ-\frac{\varOmega_{d-2}r_{h}}{8\pi}da+\frac{(2-d)\Omega_{d-2}}{16\pi r_{h}^{(d-1)\omega_{q}}}d\alpha$\tabularnewline
\hline
\end{tabular}
\par\end{centering}
\caption{{\footnotesize {}{}{}{}Results for the first law of thermodynamic
under different conditions. }}%
\label{tab:1st}%
\end{table}

As shown in Refs. \cite{Gwak:2019asi,Gwak:2017kkt}, the RN-AdS black hole is
studied in d-dimensional space via scalar field scattering and particle
absorption, respectively. When the dimension is reduced to four, the laws of
thermodynamics and the overcharging problem of the charged AdS black hole with
cloud of strings and quintessence are investigated by particle absorption in
\cite{Liang:2020uul}, and studied under scalar field in \cite{Liang:2020hjz}.
When only quintessence is considered without cloud of strings, the
thermodynamics and the stability of horizon for RN-AdS black hole with
quintessence are investigated by particle absorption in Ref. \cite{He:2019fti}%
, and are studied under scalar field scattering in Ref.
\cite{intro-Hong:2019yiz}. The results of the first thermodynamic law under
different conditions are summarized in Table \ref{tab:1st}.

In Refs. \cite{Gwak:2017kkt,Liang:2020uul,He:2019fti}, the energy of the
particle is assumed to correspond to internal energy of the black hole, i.e.,
$E=dU$ in the extended phase space. In Refs.
\cite{Gwak:2019asi,Liang:2020hjz,intro-Hong:2019yiz}, the energy flux of the
field is assumed to correspond to internal energy of the black hole, i.e.,
$dE=dU$ in the extended phase space. Under this assumption, the second law of
thermodynamics for black holes is violated in extended phase space. In Refs.
\cite{Hu:2019lcy,Li:2020dnc,Liang:2021voh}, another assumption is proposed. In
this assumption, the energy(energy flux) is assumed to change the enthalpy of
the black hole instead of the internal energy of the black hole, i.e.,
$E=dM$($dE=dM$). Under this assumption, the second law of thermodynamics of
the black hole is valid. Besides, the first law of thermodynamics and the
stability of the horizon under this assumption have the same results as the
previous one. The results of the black hole under two assumptions are same in
normal phase space, since the mass can be regarded as the internal energy,
i.e., $M=U$.

\begin{acknowledgments}
We are grateful to Wei Hong, Peng Wang, Haitang Yang, Jun Tao, Deyou Chen and
Xiaobo Guo for useful discussions. This work is supported in part by NSFC
(Grant No. 11747171), Natural Science Foundation of Chengdu University of TCM
(Grants nos. ZRYY1729 and ZRYY1921), Discipline Talent Promotion Program of
/Xinglin Scholars(Grant no.QNXZ2018050) and the key fund project for Education
Department of Sichuan (Grantno. 18ZA0173).
\end{acknowledgments}

\end{document}